\documentclass[11pt]{article}
\pdfoutput=1

\usepackage{stolenstyle}
\usepackage{amsmath,epsf,amssymb,latexsym,amsthm,setspace,bbm,array,pifont,enumerate,mathrsfs,comment}
\usepackage[table]{xcolor}
\usepackage{longtable}

\DeclareFontFamily{U}{rsf}{}
\DeclareFontShape{U}{rsf}{m}{n}{
  <5> <6> rsfs5 <7> <8> <9> rsfs7 <10-> rsfs10}{}
\DeclareMathAlphabet\Scr{U}{rsf}{m}{n}

\def\rep#1{{{\boldsymbol{#1}}}}
\def\brep#1{{{\overline{\boldsymbol{#1}}}}}

\def\cDb{{\overline{\cD}}}

\def\GUL{\GU(1)_{\text{L}}}
\def\GUR{\GU(1)_{\text{R}}}

\def\C{{\mathbb C}}

\def\P{{\mathbb P}}
\def\R{{\mathbb R}}
\def\Z{{\mathbb Z}}

\def\Img{\operatorname{Im}}

\def\Pic{\operatorname{Pic}}
\def\Rea{\operatorname{Re}}

\def\Sym{\operatorname{Sym}}

\def\tr{\operatorname{tr}}

\def\SO{\operatorname{SO}}
\def\SL{\operatorname{SL}}

\def\SU{\operatorname{SU}}
\def\GU{\operatorname{U{}}}

\def\GE{\operatorname{E}}

\def\so{\operatorname{\mathfrak{so}}}

\def\su{\operatorname{\mathfrak{su}}}
\def\Lu{\operatorname{\mathfrak{u}}}
\def\Le{\operatorname{\mathfrak{e}}}
\def\Lg{\operatorname{\mathfrak{g}}}

\def\p{\partial}
\def\pb{\bar{\partial}}

\def\la{\langle}
\def\ra{\rangle}

\def\ff#1#2{{\textstyle\frac{#1}{#2}}}

\def\cA{{\cal A}}
\def\cB{{\cal B}}
\def\cC{{\cal C}}
\def\cD{{\cal D}}
\def\cE{{\cal E}}
\def\cF{{\cal F}}
\def\cG{{\cal G}}
\def\cH{{\cal H}}
\def\cI{{\cal I}}

\def\cL{{\cal L}}
\def\cM{{\cal M}}
\def\cN{{\cal N}}
\def\cO{{\cal O}}

\def\cU{{\cal U}}
\def\cV{{\cal V}}
\def\cW{{\cal W}}

\newcommand\thetab{\overline{\theta}}

\newcommand\Lambdab{\overline{\Lambda}}

\newcommand\Omegab{\overline{\Omega}}

\newcommand\zb{\overline{z}}

\theoremstyle{definition}

\usepackage{upgreek}

\usepackage{tikz}
\usetikzlibrary[shapes.geometric]
\usetikzlibrary{positioning} 
\usetikzlibrary{calc,intersections,through,backgrounds}
\usetikzlibrary{cd}
\tikzset{>=stealth}
\tikzset{every picture/.style={very thick}}

\def\ii{\mathrm{i}}
\def\dd{\mathrm{d}}
\def\ee{\mathrm{e}}
\def\sleft{\text{\tiny{L}}}
\def\sright{\text{\tiny{R}}}
\def\cir{{\mathrm{cir}}}
\def\mm{{\mathrm{min}}}
\def\tq{{\mathrm{q}}}
\def\ty{{\mathrm{y}}}

\def\preprint{ ~~ }

\title{Phases of flux vacua and their quantum symmetries}
\author[a] {Dan Isra\"el,}
\author[b] {Ilarion V.~Melnikov,}
\author[c] {Yann Proto\,}
\affiliation[a] {Sorbonne Universit\'e, CNRS, Laboratoire de Physique Th\'eorique et Hautes {\'E}nergies, LPTHE, F-75005 Paris, France}
\affiliation[b] {Department of Physics and Astronomy,
James Madison University,
Harrisonburg, VA 22807, USA}
\affiliation[c] {Bethe Center for Theoretical Physics, Universit\"at Bonn, Bonn, Germany}

\emailAdd{israel@lpthe.jussieu.fr}
\emailAdd{melnikix@jmu.edu}
\emailAdd{yproto@uni-bonn.de}

\abstract{We present a torsional linear sigma model with (0,2) supersymmetry that realizes, in its large radius phase, a five-dimensional heterotic flux vacuum, and we use the linear sigma model to connect this geometry to a stringy regime realized as an asymmetric orbifold of a Gepner model and a circle.  Along the way we develop a number of tools relevant for more general (0,2) gauged linear sigma model constructions of heterotic vacua.}

\begin{document}

\maketitle

\section{Introduction}
Much of what we know about theories of quantum gravity derives from the study of the landscape of string vacua.  While there are a variety of methods for constructing particular loci in this landscape, it is in general difficult to determine whether two seemingly different constructions yield the same vacuum, or, if there are moduli present, to determine whether two distinct vacua are connected in the moduli space.  The moduli space of Calabi-Yau (CY) compactifications of the type II string with geometry $\R^{1,3} \times X$ provides a clear example of each of these phenomena:  mirror symmetry relates type IIA string theory compactified on a manifold $X$ to type IIB string theory compactified on a mirror manifold $X^\circ$, while the moduli space of compactification on the quintic $3$-fold contains a Gepner point, where the worldsheet conformal field theory is a rational superconformal theory (SCFT).\footnote{A thorough review of these classic results can be found in~\cite{Hori:2003ds,Aspinwall:2009isa}.}  The gauged linear sigma model (GLSM), introduced in~\cite{Witten:1993yc}, remains the best tool for exploring these phenomena. 

In this work we use a linear sigma model approach to study the stringy limit of a heterotic flux vacuum with geometry $\R^{1,4} \times X$, where  the five-dimensional manifold $X$ is a principal $\GU(1)$ bundle over a K3 manifold $M$ whose curvature is supported by non-zero NS-NS flux $H$.  A further compactification on $X\times S^1$ is a non-K\"ahler heterotic flux geometry.   These vacua preserve $8$ spacetime supercharges and are of interest as toy models for the largely uncharted territory of heterotic flux vacua.  They have been explored extensively ever since their identification via string duality~\cite{Dasgupta:1999ss} and explicit geometric analysis~\cite{Goldstein:2002pg,Becker:2006et,Fu:2006vj}. The so-called torsional linear sigma models (TLSMs) developed  to study such vacua were first presented in~\cite{Adams:2006kb}, but to our knowledge this work is the first example to realize the following two goals:
\begin{enumerate}
\item  we present a TLSM with a large radius phase that satisfies all of the necessary conditions that should be obeyed by a heterotic flux vacuum, including bundle stability;
\item we analyze the stringy regime where the base K3 volume becomes of order $\alpha'^2$, and the vacuum is described as a chiral orbifold $ (\cC_{\text{LG}} \times S^1)/\Gamma$, where $\cC_{\text{LG}}$ is a Gepner model realized as the RG fixed point of a Landau--Ginzburg theory, $\Gamma$ is a finite abelian group, and the orbifold acts on the circle, held at special radius, as a chiral shift. 
\end{enumerate}
There are two distinct ways to deform away from the stringy Gepner point: we can either go back to the large radius phase that realizes the heterotic flux vacuum, or we can decompactify the fibered circle.  In the former limit  the K3 geometry decompactifies to a $9$-dimensional heterotic string vacuum, while by taking the latter limit we obtain a compactification of the heterotic string to $6$ dimensions on a string-size K3 realized as a Gepner model.

Along the way we obtain a number of results that are of relevance to more general (0,2) GLSMs and should be contrasted with the tamer (2,2) GLSMs.  Some of these are qualitative and are reflected in the rich structure of the phase diagram, which includes jumping of the central charge in different phases and an elaborate pattern of gauge enhancement in phase limits.  There are also a number of technical developments, such as a GLSM realization of line bundle stability conditions and a new method to construct GLSMs for string vacua by incorporating interacting fermi multiplets that correspond to trivial bundle factors in the geometry.  In the rest of the introduction we will expand on these latter points and explain their relation to the torsional sigma model construction and heterotic flux vacua.

\subsection{Line bundles and gauged linear sigma models}
To get our bearings let us consider a textbook construction of a large class of perturbative $(\GE_8\times\GE_8) \rtimes \Z_2$  heterotic string vacua realized in the RNS formulation as a (0,2) supersymmetric non-linear sigma model (NLSM) with target space $Y$---a Calabi-Yau $3$-fold and $k$ Weyl holomorphic fermions coupled to a connection on a holomorphic rank $k$ vector bundle $\cV \to Y$.   For simplicity we set $c_1(\cV) = 0$. In order for the NLSM to be well-defined the bundle must satisfy $c_2(\cV) = c_2(T_Y)$.   Assuming that $Y$ is smooth and has a large radius limit---these are necessary conditions for the NLSM description to be reliable---a necessary condition for the configuration to correspond to a heterotic vacuum is that $\cV$ is a zero--slope polystable bundle:  $\cV = \oplus_i \cV_i$, and each summand is zero--slope stable.\footnote{A review of bundle stability in heterotic string constructions can be found in~\cite{Anderson:2018pui} and references therein.}  

Suppose now that we also have a GLSM construction of this geometry.  That is, we can choose the data of a two-dimensional gauge theory (we assume the gauge group to be abelian throughout this work), so that for some limit in the GLSM parameter space the IR degrees of freedom are given by the (0,2) NLSM just described.  As described in~\cite{Witten:1993yc,Distler:1993mk,Distler:1995mi} this means that $Y$ is presented as a complete intersection in some toric variety $Z$, while the bundle $\cV$ is the cohomology of a complex
\begin{equation}
\begin{tikzcd}
0 \ar[r]	& \cA \ar[r,"g"]		& \cB \ar[r,"f"] & \cC \ar[r] &0
\end{tikzcd}~,
\label{eq:monad}
\end{equation}
with each term a sum of line bundles over $Y$. The map $f$ depends holomorphically on parameters encoded in the (0,2) superpotential $\cW$ of the GLSM, while the map $g$ depends on the parameters encoded in the chirality conditions of the fermi multiplets of the GLSM---these are (0,2) multiplets with lowest component transforming as a holomorphic Weyl fermion, i.e. a section of the square root of the canonical bundle over the worldsheet.  In a little more detail, $\cW$ depends on three sets of holomorphic parameters: the complexified Fayet-Iliopoulos (FI) parameters $\tau^\alpha$, with $\alpha = 1,\ldots,r$ for a gauge group of rank $r$, the parameters in polynomials in chiral superfields that define $Y \subset Z$ as a complete intersection, and the parameters in  $f$.  Denoting the fermi multiplets by $\Gamma_A$ and the chiral superfields collectively by $\Phi$, we have the conditions\footnote{A detailed presentation of the notation is given below in section~\ref{s:tlsms}.}
\begin{align}
\cDb \Gamma_A  &= E_A(\Phi)~, &   
\cW &\supset \sum_{A} \Gamma_A J_A(\Phi)~,&
\sum_{A} E_A J_A & = 0~.
\label{eq:EJconstraint}
\end{align}
The polynomials $E_A(\Phi)$ and $J_A(\Phi)$ are constrained by gauge invariance and additional symmetries of the GLSM:  in all of our constructions we will insist that the global symmetry contains at least a $\GU(1)_{\sleft} \times\GU(1)_{\sright}$ subgroup, with the second factor being the R-symmetry of the (0,2) supersymmetry algebra.  We denote the set of the GLSM parameters by $\cM_{\text{uv}}$.

Having specified the theory, we have the key question:  does this data define a heterotic vacuum for some choice of parameters in $\cM_{\text{uv}}$?  Assuming an affirmative answer, there are several crude checks that we can make of the proposal.
\begin{enumerate}
\item  We can check that the $\GU(1)_{\sright}$ has the correct 't Hooft anomaly to match the putative right-moving central charge $c_{\sright} = 9$, while assigning integer R-charges to gauge-invariant operators in the NS sector.
\item Similarly, $\GU(1)_{\sleft}$ should have the correct 't Hooft anomaly in a normalization that assigns integer charges to gauge-invariant operators in the NS sector, so that it can be enhanced to the full spacetime gauge symmetry algebra.\footnote{For example, if $r=3$, then to obtain a consistent theory we need to have the enhancement $\Lu(1)_{\sleft} \oplus \so(10) \to \Le_6 \subset \Le_8$.  This requires the $\Lu(1)_{\sleft}$ current to satisfy $J_{\sleft}(z) J_{\sleft}(0) \sim r z^{-2}$.}
\item The elliptic genus of the GLSM, which can be computed using techniques developed in~\cite{Benini:2013xpa}, should match that of the NLSM, which can be evaluated as in~\cite{Kawai:1994qy}.
\end{enumerate}
These conditions are necessary if the GLSM description is to be a useful guide to the physics of the (0,2) SCFT of the heterotic vacuum, but they are not sufficient.  A well-known subtlety is the possibility that gauge instanton corrections in the GLSM destabilize the putative vacuum at a generic point in $p \in \cM_{\text{uv}}$.   Not much is known about the end-point of the RG flow in this situation, but it is plausible that at least in some examples the flow can be interpreted as a motion on $\cM_{\text{uv}}$, and the fixed points are parameterized by some locus $\cM'_{\text{uv}} \subset \cM_{\text{uv}}$.  It may be that every point in $\cM'_{\text{uv}}$ corresponds to some infinite distance singularity, e.g. a decompactification limit of $Y$---exactly the sort of destabilization discussed long ago in~\cite{Dine:1986zy,Dine:1987bq} in the NLSM context.  Remarkably, there are arguments that for a large class of (0,2) GLSMs such a destabilization does not occur~\cite{Silverstein:1995re,Beasley:2003fx,Bertolini:2014dna}, and we assume that our GLSM for $\cV \to Y$ belongs to this protected class. 

Even when the necessary conditions are satisfied, and the instanton corrections are absent, there can be further sources of destabilization~\cite{Bertolini:2021hal}.  In the geometric setting just reviewed there is an obvious culprit:  bundle stability.  In a sense, we expect it to have the same consequences as the gauge instanton corrections.  That is, we might consider some subset $\cM_{\text{geo}} \subset \cM_{\text{uv}}$ where for a generic $p \in \cM_{\text{geo}}$ the (0,2) NLSM based on the geometry $\cV \to X$ is a reliable description of the low energy physics.\footnote{As familiar from (2,2) GLSMs this corresponds to a ``smooth large radius phase'', which involves a choice of FI parameters deep in a phase, as well as a generic choice of the $E$ and $J$ couplings.}  If this generic point $p$ corresponds to an unstable bundle, it may be that the RG fixed points are parameterized by $\cM'_{\text{geo}} \subset \cM_{\text{geo}}$, but a generic point in $\cM'_{\text{geo}}$ may not correspond to a smooth geometry, and, worse, it may well be a singular SCFT. Thus, we come to a crucial---and to our knowledge unanswered---question in (0,2) GLSMs:  \textit{can stability conditions be stated directly in terms of the UV data of the 2d gauge theory?}

In this work we provide an affirmative answer in the special case when the bundle is a sum of holomorphic line bundles  $\cV = \oplus_i \cL_i$ linearly realized in the GLSM. For each $\cL_i$ there is a chiral fermi multiplet $\Gamma_i$  (i.e. $E_i = 0$), with gauge charges $Q_i^\alpha$ and trivial superpotential coupling: $J_i =0$.  In this situation stability conditions are related to redundancies in the GLSM parameter space $\cM_{\text{uv}}$.  Just as in the better-understood (2,2) GLSMs, not all parameters in $\cM_{\text{uv}}$ correspond to marginal deformations of the SCFT describing the RG fixed point.  A better approximation to the space of marginal deformations is the quotient
\begin{align}
\label{eq:Mglsm}
\cM_{\text{glsm}} = \left( \cM_{\text{uv}} \setminus \Delta \right) / \cG~.
\end{align}
$\Delta$ is the locus where the low energy limit is a singular SCFT, and $\cG$ is the group of holomorphic field redefinitions acting on the chiral superfields $\Phi$ and the fermi multiplets $\Gamma_A$.\footnote{This structure is reviewed in~\cite{Melnikov:2019tpl} and references therein, as well as in~\cite{Bertolini:2021hal}.  Its (2,2) antecedents include the LG--minimal model correspondence~\cite{Martinec:1988zu,Vafa:1988uu}, as well as the monomial--divisor mirror map in the context of mirror symmetry~\cite{Aspinwall:1993rj,Cox:2000vi}.  The essential idea is that the D-terms in the (0,2) Lagrangian are presumed to contain only marginally irrelevant deformations.  Thus, for any $p \in \cM_{\text{uv}} \setminus \Delta$, $\cW(p)$ and $\cW(\cG \cdot p)$ define theories in the same basin of attraction with respect to the RG flow.}
It is well-known (or at least noted long ago)~\cite{Kreuzer:2010ph} that $\cG$ acts linearly on the couplings encoded in $E_A$, $J_A$, as well as on the exponentiated FI parameters $q_\alpha = e^{2\pi \ii \tau^\alpha}$.  The last action is peculiar to (0,2) GLSMs versus (2,2) GLSMs, and it arises precisely because the $\Gamma_A$ and $\Phi$ can be redefined independently in (0,2) theories.  

In a (0,2) GLSM with a linearly realized line bundle factor $\cL_i \subset \cV$---it is easy to generalize to a sum of line bundles---there is a field redefinition parameterized by a complex constant $\xi_i$ that rescales $\Gamma_i$ and leaves all other fields invariant, and it induces a rescaling of the $q_\alpha$ while leaving the $E$ and $J$ potentials invariant:
\begin{align}
q_\alpha \mapsto q_\alpha e^{2\pi \ii \xi^i Q^\alpha_i}~.
\end{align}
This generates an equivalence relation on the complexified FI parameters: 
\begin{align}
\label{eq:equivalencelinebundle}
\tau^\alpha \sim \tau^\alpha + \xi^i Q^\alpha_i~.
\end{align}
In a large radius phase this has the following geometric interpretation.  The complexified K\"ahler class on $Y$ can be written as\footnote{This is a relation that holds deep in a large radius phase: already in (2,2) theories the relation between the geometric ($B+\ii J$) and GLSM coordinates $(\tau^\alpha)$ involves a constant shift in $B$, as well as a series of instanton corrections due to point-like instantons of the GLSM (see~\cite{Jockers:2012dk} for a discussion of the precise relation).   We expect that the situation is more complicated in (0,2) theories.}
\begin{align}
B+\ii J & = \sum_{\alpha=1}^r \tau^\alpha \eta_\alpha~,
\end{align}
where the $\eta_\alpha \in H^2(Y,\R)$ are classes pulled back from the ambient toric variety $Z$.  Thus, while the $\tau^\alpha$ parameterize a subset of the complexified K\"ahler cone of $Y$, the equivalence relation implies that the IR fixed points are described by a quotient, and motion in the K\"ahler moduli space in the direction $Q^\alpha_i$ is a classical illusion.

The equivalence relation does not fix a particular representative class, but in the large radius limit the expectation is that $\xi_i$ should be chosen so that $\cL_i$ is stable, which is simply the condition
\begin{align}
\label{eq:stability3fold}
\text{c}_1(\cL_i) \wedge J \wedge J & = 0~,
\end{align}
and it is easy to see that imposing this condition leads to a quadratic equation for $\Img(\xi_i)$.   If a solution exists it may lead to a singular geometry, in which case additional arguments are necessary to make sure that the SCFT is nevertheless smooth.  Away from the large radius limit~(\ref{eq:stability3fold}) will receive corrections, but we do not expect the essential qualitative feature---a quadratic  constraint on the parameters---to be modified.

\subsection{Line bundles and circle bundles}
The same structure applies to the geometry of a $5$-dimensional heterotic flux vacuum with a linear sigma model realization.  As we will now review, the reason for this is easy to understand from the perspective described in detail in our  earlier work with R.~Minasian~\cite{Israel:2023tjw}.

We think of the construction in two steps: first we compactify the heterotic string to $9$ dimensions, and we then further compactify the theory on a K3 manifold $M$.  In order to obtain a supersymmetric vacuum we must turn on a non-trivial connection for some subgroup of the $9$--dimensional spacetime gauge group, and there are two seemingly qualitatively different choices:  we can either turn on a connection for the $\Le_8\oplus \Le_8$ gauge fields, or we can use a $\Lu(1)$ symmetry associated to the circle.  In the former situation the compactification geometry is a familiar flux-free compactification with geometry $M\times S^1$, while in the latter case supersymmetry requires that the $\Lu(1)$ is the left-moving symmetry of the compact boson CFT, and the non-trivial fibration requires an NS $3$-form flux $H$ proportional to $\upomega\wedge \Theta$, where $\upomega$ is the first Chern class of the circle bundle over $M$, and $\Theta$ is the invariant $1$-form associated to the circle fiber.  Now the geometry is $X$---the total space of the circle bundle over $M$.

The two cases are limits of a more general setup, where we fiber the circle and turn on a connection for the $\Le_8\oplus\Le_8$ gauge fields, with the latter associated to some stable vector bundle $\cE \to M$.  The heterotic Bianchi identity for $H$ and flux quantization conditions lead to constraints
\begin{align}
c_2(T_M) - c_2(\cE) & = - r^2 \upomega \wedge \upomega~,  & r^2 \upomega \in H^2(M,\Z)~,
\label{eq:heteroticBianchi}
\end{align}
where $r$ is the circle radius in units of $\alpha'$,\footnote{We work in the conventions where $r = 1$ is the self-dual radius.  It is possible to turn on a Wilson line along the circle direction, which modifies the quantization condition on the circle radius.  Since our primary interest is in the TLSM, we do not consider that possibility in this work.} while spacetime supersymmetry leads to further requirements
\begin{align}
\label{eq:CircleHYM}
 \upomega &\in H^{1,1}(M)~,&
 J \wedge \upomega & = 0~.
\end{align}
If we take $\cE$ to be trivial, the $5$--dimensional gauge symmetry includes the full $\Le_8\oplus\Le_8$, and $X$ is a space with trivial first Pontryagin class.

From this perspective it is not surprising that a heterotic flux vacuum is closely related to choosing $\cE = \cL$ to be a line bundle: this is exactly what we might expect from $9$--dimensional T-duality, and this expectation is confirmed in~\cite{Evslin:2008zm,Israel:2023tjw}.  At first glance this raises a puzzle: if we use T-duality to trade a fibered circle for a line bundle $\cL$, how can we preserve the full $\Le_8\oplus\Le_8$ gauge symmetry?  The answer is very pretty and stringy: although the circle is no longer fibered in the T-dual description, it sits at a specific radius $r^\circ$ and comes with a Wilson line characterized by a lattice vector $\kappa \in \Gamma_8+\Gamma_8$, and these are just right to recover the full $\Le_8\oplus \Le_8$ gauge symmetry, but now some of the worldsheet holomorphic currents explicitly involve circle winding modes.  Unlike in the circle-fibered description, where the radius is quantized, $r^\circ$ can be made as large as we like, and in the decompactification limit the holomorphic currents correspond to roots $\alpha \in \Gamma_8+\Gamma_8$ that are orthogonal to $\kappa$.  We will see a similar feature emerge in our TLSM construction, but now without having to perform a T-duality.  Instead, in the LGO phase we find that there is a modulus that corresponds to a decompactification of a circle, leading to a $6$--dimensional heterotic compactification on a stringy K3 SCFT.

From~(\ref{eq:CircleHYM}) we see that the TLSM construction presents a puzzle essentially identical to bundle stability:  supersymmetry requires that the curvature of the circle bundle is anti-self-dual (ASD), but there is no such constraint on the data of the classical TLSM Lagrangian.  Fortunately, the resolution is also similar, but with an important twist.  The key feature of a TLSM is that its fermion content is gauge--anomalous, with the anomaly encoded in a symmetric matrix $\cA^{\alpha\beta}$.  Nevertheless, the full theory is well-defined because it contains a so-called shift multiplet $\Omega$, which transforms inhomogeneously under the gauge symmetry and has Green-Schwarz--like couplings to the gauge field:  thus, the classical action is gauge-dependent in such a way as to cancel the gauge variation of the measure of the chiral fermions.  As we discuss in more detail below, this is possible when the anomaly matrix takes the form
\begin{align}
\label{eq:anommatrix}
\cA = -2 r^2w w^{\text{t}}~,
\end{align}
where $w$ is an $r$--dimensional vector with integer coefficients that encodes the circle fibration in the geometric phase via
\begin{align}
\upomega = \sum_\alpha w^\alpha \eta_\alpha~.
\end{align} 
When this holds there is a field redefinition that induces an identification on the complexified FI parameters:
\begin{align}
\label{eq:equivalencecirclebundle}
\tau^\alpha\sim\tau^\alpha + \xi w^\alpha~.
\end{align}
This is the TLSM analogue of the identification~(\ref{eq:equivalencelinebundle}) described for the GLSM with line bundle factors.  Incidentally this shows that a TLSM that admits a large radius phase with a smooth heterotic flux geometry must have gauge group rank $r \ge 2$.  This mild restriction on the model's simplicity is dwarfed by the complications due to the stringent conditions for gauge anomaly cancellation, as we know review.

\subsection{Linear representations of bundle data}
Anyone who has attempted to construct a linear sigma model representation for a particular bundle $\cV \to X$ has encountered the vexing issue that the linear model's consistency conditions are stronger than those of the non-linear model.   For instance, we might consider the simple (0,2) non-linear sigma model with target space $\P^1$ and take the bundle to be trivial: this is a well-defined non-linear sigma model because $c_2(T_{\P^1}) = 0$, so that it is possible to choose the gauge bundle to be trivial.  It is very easy to write down a classical (0,2) GLSM that realizes a $\P^1$ target space:  we take two chiral superfields $\Phi_{1},\Phi_{2}$, and couple them to a single $\GU(1)$ gauge field with charge $+1$.   In the large radius phase the D-term constraint leads to
\begin{align}
|\phi_1|^2 + |\phi_2|^2 = \rho \gg 0~,
\end{align}
so that, modulo gauge equivalence the classical vacua of the theory are parameterized by a $\P^1$.  Unfortunately this is an anomalous gauge theory with $\cA = -2$.

This is a typical situation in (0,2) GLSMs: the linear realization of the desired data has a gauge anomaly, and the same issue comes up in the TLSM, where instead of seeking gauge-anomaly-free matter content with $\cA = 0$, we want to choose it to satisfy~(\ref{eq:anommatrix}).    

In constructing our TLSMs we found an elegant trick that allows for extra wiggle room: it may be possible to realize the desired geometry and bundle by including GLSM fermi couplings that lead to trivial bundle factors in a geometric phase, and we believe it will be of some use in constructing many anomaly--free GLSMs and TLSMs.  The idea is simple:  rather than finding a GLSM with a large radius phase that directly realizes the desired bundle $\cV \to X$, it may instead be easier to find a realization of a GLSM for the bundle $\cV\oplus \cO^{\oplus n} \to X$, where the trivial factor only emerges at low energy.

To illustrate the idea, consider the following family of (0,2) GLSMs, with $4$ chiral multiplets and $3$ chiral fermi multiplets coupled to a single $\GU(1)$ gauge field with charges specified by two non-negative integers $d$ and $k$
\begin{align}
\arraycolsep=4pt
\begin{array}{ccccccc}
\Phi_1 & \Phi_2 & \Phi_3 	& \Phi_4  & \Gamma_1 & \Gamma_2 & \Gamma_3  \\
1	   &1		&1		&-k-1	&1		    &k			 &-d
\end{array}
\end{align}
and superpotential
\begin{align}
\cW & = \Gamma_3 \left(\Phi_1^d + \Phi_2^d + \Phi_3^d\right) + \Phi_4 \left(\Gamma_1 \Phi_1^k + \Gamma_2\Phi_2\right)~.
\end{align}
The gauge anomaly is given by
\begin{align}
\cA = 1 + k^2+d^2 - 3 -(k+1)^2 = d^2 -2k -3~,
\end{align}
and the general solution to $\cA = 0$ can be expressed n in terms of a non-negative integer $\ell$:
\begin{align}
d &= 2\ell +3~,& k & = 2\ell^2 + 6\ell + 3~.
\end{align}
When the FI parameter is positive, the target space geometry is a degree $d=2\ell+3$ hypersurface in $\P^2$, i.e. an algebraic curve $C$ of genus~\cite{Griffiths:1978pa}
\begin{align}
g(C) & = \frac{(d-1)(d-2)}{2} = (\ell+1)(2\ell+1)~,
\end{align}
while the bundle $\cV$ is a monad determined by the exact sequence
\begin{equation}
\begin{tikzcd}
0 \ar[r]	& \cV \ar[r]	& \cO_C(1)\oplus\cO_C(k) \ar[r,"(\phi_1^k~\phi_2)"] 	& \cO_C(k+1) \ar[r] &0
\end{tikzcd}~,
\end{equation}
where $\cO_C(k)$ denote the restriction of the sheaf $\cO(k)$ over $\P^2$ to the curve $C$.  Since $\cV$ is a line bundle with $c_1(\cV) = 0$, it is trivial, and thus we obtain a (0,2) GLSM construction for an algebraic curve of genus $g(C) = (\ell+1)(2\ell+1)$ equipped with a trivial bundle.  

The family includes the elliptic curve at $\ell = 0$, for which we of course also have  the familiar (2,2) GLSM description, but this construction is a little more economical, involving just $3$ fermi multiplets and neither $E$--couplings nor gauge-neutral chiral superfields. 

\subsection{Roadmap}
The rest of this work is organized as follows.  We introduce a class of TLSMs that realize a large radius heterotic flux vacuum in section~\ref{s:tlsms}.  Next, in section~\ref{s:phases} we study the phase diagram of one of those TLSMs, and we discuss the collapse of the naive two-dimensional phase diagram to a one-dimensional subspace that interpolates between the large radius flux vacuum phase and a stringy regime $ (\cC_{\text{LG}} \times S^1)/\Gamma$.  Section~\ref{s:dressedellgen} is devoted to the computation of the dressed elliptic genus in different phases.  Among other things, this provides evidence that the large radius and stringy phases of our TLSM are indeed connected.  In section~\ref{s:lgo} we turn to a detailed analysis of the stringy phase and argue that in fact this entire phase collapses to a single point:  the asymmetric orbifold of a product of a Gepner model and a ( (0,2) supersymmetric)  compact boson.  We wrap up with a discussion and outlook in section~\ref{s:discussion}.  Several appendices contain technical details of some of the computations.

\subsection*{Acknowledgements}
It is a pleasure to thank K.~Hori, J.~Knapp, G.~Lockhart, R.~Minasian, E.~Morgensztern, R.~Plesser, and R.~Singhal for useful discussions.  We are also grateful to the organizers of the GLSM26 conference and the Mainz Institute for Theoretical Physics (MITP) of the Cluster of Excellence PRISMA+ (Project ID 390831469) for hospitality and support. The work of YP work has received funding from the European Research Council (ERC) under the Horizon Europe (grant agreement No. 101078365) research and innovation program. IVM's work was in part supported by the Humboldt Research Award and the Jean d'Alembert Program at the University of Paris--Saclay, as well as the Education Leave program at James Madison University.  In addition, he is grateful for hospitality and support from LPTHE at Sorbonne University, as well as from the Max Planck Institute for Gravitational Physics (AEI).

\section{A class of torsional linear sigma models} \label{s:tlsms}
In this section we introduce a class of gauged linear sigma models with a large radius phase realizing a heterotic flux vacuum $X$.

\subsection{The (0,2) Lagrangian}
We begin by reviewing the structure of the TLSM. Throughout this paper we adopt the (0,2) superspace conventions of~\cite{Melnikov:2019tpl}. We work on a flat worldsheet of Euclidean signature, with superspace coordinates $(z;\bar{z},\theta,\bar{\theta})$.

Torsional linear sigma models, introduced in~\cite{Adams:2006kb}, are two-dimensional abelian gauge theories with (0,2) supersymmetry. They are constructed so that, in an appropriate large radius phase, the linear model describes the total space of a principal torus bundle over a K\"ahler base. We focus on the simplest setting, in which the total space $X$ is a circle fibration over a K3 surface $M$. The TLSM is then obtained by supplementing a conventional (0,2) linear model, whose large radius phase describes a non-linear sigma model with target $M$, by a chiral multiplet transforming linearly under the GLSM gauge symmetries. In the IR, this multiplet gives rise to a principal $\C^\ast$-bundle over the K3 sigma model.

The matter content of the K3 gauged linear sigma model consists of bosonic (0,2) chiral multiplets $\Phi_i$, $i=1,\dots,N_\Phi$, and fermi multiplets $\Gamma_A$, $A=1,\dots,N_\Gamma$, all of which are annihilated by the superspace derivative
\begin{align}
\cDb & = \p_{\thetab} + \theta \pb_{\zb}~.
\end{align}
These superfields are coupled to (0,2) vector multiplets for an abelian gauge group $\GU(1)^{r}$, whose gauge-invariant field strengths are denoted by $\Upsilon_\alpha$, with $\alpha=1,\dots,r$. The complexified FI parameters are
\begin{align}
\tau^\alpha=\frac{\vartheta^\alpha}{2\pi}+\ii\rho^\alpha~.
\end{align}
In the large radius phase of the GLSM, they determine the complexified K\"ahler class of the target K3 surface. 

The left- and right-moving fermions of a (0,2) model generally belong to different representations of the gauge group. Under a super-gauge transformation parameterized by chiral superfields $\Lambda_\alpha$, the matter multiplets transform as
\begin{align}
\label{eq:gaugetransform1}
\delta_\Lambda \Gamma_A &= Q^\alpha_A\Lambda_\alpha\Gamma_A~,&
\delta_\Lambda \Phi_i &= Q^\alpha_i\Lambda_\alpha\Phi_i~,
\end{align}
where $Q^\alpha_A$ and $Q^\alpha_i$ respectively denote the gauge charges of the fermi and chiral multiplets. Consequently, the fermionic path-integral measure may fail to be gauge invariant. The local gauge anomaly is encoded by the symmetric matrix
\begin{align}
\mathcal{A}^{\alpha\beta}= \sum_{A=1}^{N_\Gamma} Q^\alpha_A Q^\beta_A - \sum_{i=1}^{N_\Phi} Q^\alpha_i Q^\beta_i~.
\end{align}
For an ordinary GLSM describing a Calabi--Yau compactification, quantum consistency requires the anomaly matrix $\mathcal{A}$ to vanish identically. In a torsional model, this condition is relaxed to a slightly weaker version which we will describe below.

The TLSM contains an extra superfield, which we call a \emph{shift multiplet} and denote by $\Omega$. This is a chiral superfield which transforms by linear shifts under super-gauge symmetries:
\begin{align}
\label{eq:gaugetransform2}
\delta_\Lambda \Omega = w^\alpha\Lambda_\alpha~,
\end{align}
where the shift charges $w^\alpha$ are integers. The gauge-invariant kinetic term for $\Omega$ is determined by a constant parameter $r$, which will be identified with the radius of the circle fibration measured in units of $\alpha'$. The shift multiplet enters the (0,2) superpotential
\begin{align}
\mathcal{W} = \sum_{A=1}^{N_\Gamma} \Gamma_A J_A(\Phi)+\sum_{\alpha=1}^r\left(\tau^\alpha-\tfrac{1}{2} r^2 w^{\alpha}\Omega \right) \Upsilon_\alpha
\label{eq:TLSMsuperpotential}
\end{align}
through an axial coupling to the  gauge curvatures. The coefficient of this coupling is completely fixed in terms of the radius $r$, with important consequences for the theory. First, the kinetic term for $\Omega$ includes couplings of the scalar component $\omega$ to the scalars $D_\alpha$ in the vector multiplets, as well as couplings of the right-moving fermion in $\Omega$ to the left-moving gauginos (the lowest components of $\Upsilon_\alpha$). After including the superpotential contributions, these interaction terms disappear from the Lagrangian: $\Rea(\omega)$ becomes a free real scalar, while only $\Img(\omega)$ remains coupled to the linear model via the abelian vector fields. Moreover, large gauge transformations lead to integrality conditions on the radius. The partition function must stay invariant under gauge transformations parametrized by $\Lambda_\alpha\in2\pi\ii\Z$. As reviewed in appendix~\ref{app:shiftmultiplet}, this holds provided
\begin{align}
r^2 w^\alpha\in\Z~.
\end{align}
Large gauge symmetries shift $\Img(\omega)$ by integral periods, while leaving every other field invariant. After quotienting by the gauge symmetries, this scalar field therefore takes values in $\R/\operatorname{Span}_\Z\{ w^1,\dots,w^r\}$. Thus, provided the shift charges are non-trivial, $\Img(\omega)$ is compact, and the shift multiplet parameterizes a target $\C^\ast=\R\times S^1$. 

The central feature of the TLSM is the failure of gauge invariance, in the classical theory, due to the axial coupling of the shift multiplet. Its classical gauge variation has precisely the form of a local gauge anomaly, and can therefore cancel the one-loop anomaly in the quantum theory. Gauge invariance is restored provided
\begin{align}
\mathcal{A}^{\alpha\beta} = - 2r^2w^\alpha w^\beta ~.
\label{eq:anomalycancelation}
\end{align}
This anomaly cancelation condition admits a natural geometric interpretation in the large radius phase of the GLSM.~ In the IR non-linear sigma model each abelian vector multiplet is associated a divisor class $\eta_\alpha$ in the K3 Picard lattice $\operatorname{Pic}(M)=H^2(M,\Z)\cap H^{1,1}(M)$. Their intersection pairing defines the symmetric matrix
\begin{align}
D_{\alpha\beta} = \eta_\alpha.\eta_\beta~.
\end{align}
The cohomology class $\upomega = w^\alpha\eta_\alpha$ determined by the shift charges is identified with the curvature of a principal circle bundle over $M$, whose self-intersection is computed by $w^{\text{t}}D w$. Spacetime supersymmetry requires this curvature class to be anti-self-dual; in the GLSM this translates into the condition
\begin{align}
w^{\text{t}}D \rho =0
\end{align}
on the FI parameters. The anomaly matrix satisfies
\begin{align}
\tfrac{1}{2}\tr(\mathcal{A}D) &= c_2(T_M)-c_2(\mathcal{E})~,
\end{align}
where $T_M$ denotes the holomorphic tangent bundle of $M$, and $\mathcal{E}$ is the holomorphic bundle supporting the massless left-moving fermions. Thus, the anomaly cancellation condition precisely encodes the heterotic Bianchi identity~\eqref{eq:heteroticBianchi}. Notice, however, that the GLSM anomaly constraints are stronger than the topological requirements in supergravity: the traceless part of~\eqref{eq:anomalycancelation} imposes additional conditions on the linear model data that do not have a geometric counterpart.

\subsubsection*{Five-dimensional flux vacua}
Because the shift multiplet contains the free non-compact scalar $\Rea(\omega)$, the IR theory in the large radius phase is a (0,2) non-linear sigma model with six-dimensional target $X\times\R$. The circle fiber of $X$ combines with the non-compact direction to form a (0,2) chiral multiplet, and the resulting SCFT has a right-moving N=2 superconformal algebra with central charge $c_\sright=9$. It is often useful instead to decouple the non-compact direction, consisting of the scalar $\Rea(\omega)$ and a free right-moving Majorana--Weyl fermion, and focus on the compact CFT associated to $X$. This sacrifices the extended worldsheet supersymmetry: the resulting theory has $c_\sright=\frac{15}{2}$ and preserves only (0,1) supersymmetry. Nevertheless, it retains a right-moving N=4 superconformal algebra with $c_\sright=6$, inherited from the K3 base of the five-dimensional space $X$. Its $\Lu(1)_\sright\subset\su(2)_\sright$ current can be identified in the UV with a global symmetry of the K3 GLSM. The corresponding supercharges, however, do not close on the right-moving translation generator of the full SCFT.

Alternatively, one can replace the non-compact scalar by a compact circle, obtaining a theory with six-dimensional target space $X\times S^1$. The resulting geometry is a complex threefold: the fibered circle of $X$, together with the additional circle, forms an elliptic fiber over the K3 base. The resulting compactification manifold is non-K\"ahler, but carries an $\SU(3)$-structure and satisfies the Hull--Strominger equations, at least to first order in the $\alpha'$ expansion~\cite{Melnikov:2014ywa}.

\subsection{Abelian coverings and K3 surfaces} \label{ss:abeliancoverings}
In addition to the shift charges $w$ and the quantized radius $r$, the remaining defining data of the linear model determines, in the large radius phase, the sigma model on the K3 surface $M$ underlying the five-dimensional circle bundle $X$. This data includes the gauge charges of the chiral and fermi multiplets, as well as  the polynomials $J_A(\Phi)$ appearing in the superpotential.\footnote{Compared to the more general situation described in the introduction, we restrict to chiral fermi multiplets, with $E_A=0$, so that the conditions~\eqref{eq:EJconstraint} are automatically satisfied.} In contrast with (2,2) GLSMs for Calabi--Yau hypersurfaces, where the gauge charges are determined by the combinatorics of the underlying toric fan, a given K3 geometry admits many distinct (0,2) realizations. These correspond to different choices of the holomorphic vector bundle $\cE$ supporting the left-moving fermions. What conditions must $\cE$ satisfy for the resulting theory to be consistent? By construction, $\cE$ and $T_M$ must be topologically distinct in order to satisfy the Bianchi identity~\eqref{eq:heteroticBianchi} with a non-trivial circle fiber. Moreover, for the IR fixed point to describe a spacetime supersymmetric heterotic vacuum, $\cE$ must be polystable with vanishing slope, so that it admits a Hermitian--Yang--Mills connection~\cite{Uhlenbeck:1986ym}. On the other hand, the bundles naturally constructed in the GLSM framework are \emph{monad bundles}, obtained as the cohomology of a complex of the form~\eqref{eq:monad}. Stability criteria are known in some cases, notably for cyclic projective varieties with $\operatorname{Pic}(M)\simeq\Z$~\cite{Hoppe:1984}. Generalizations to larger Picard groups are more subtle: stability depends on the K\"ahler class, and constructing stable bundles remains challenging.

The construction of stable monad bundles can be considerably simplified when the K3 surface $M$ is an abelian branched cover of a smooth projective surface $S$. Schematically, this means that $M$ is obtained by gluing several copies of $S$ along a branch divisor $C\subset S$. Under suitable assumptions on the branching locus~\cite{Pardini:1991}, this yields a smooth $M$ equipped with a natural projection
\begin{align}
\pi:M\to S~.
\end{align}
For specific choices of branching data the pullback $\pi^\ast$ induces an isomorphism of Picard groups, $\operatorname{Pic}(S)\simeq\operatorname{Pic}(M)$. In this case, a holomorphic vector bundle $\cE\to S$ satisfying mild conditions\footnote{A sufficient condition is that the vector bundle $\cE$ satisfies, for all $\ell=1,\dots,\operatorname{rank}(\cE)-1$, the vanishing condition $H^0(S,\Lambda^\ell\cE\otimes\cL)=0$ for all $\cL\in\operatorname{Pic}(S)$ with $\mu(\cL)\leq -\ell\mu(\cE)$, where $\mu$ denotes the slope. The slope is defined with respect to a polarization $H$ on $S$, and correspondingly the pullback bundle is stable with respect to the polarization $\pi^\ast H$.} pulls back to a stable bundle $\pi^\ast\cE$ on $M$. This reduces the stability problem to one on $S$, where it is often much easier to solve. Stable bundles of this type on K3 surfaces were recently constructed in~\cite{Festi:2025}.

The branched covering construction is particularly promising for realizing stable bundles in gauged linear sigma models. Consider a K3 surface $M$ obtained as a branched cover of a smooth K\"ahler surface $S$, with branch divisor consisting of curves $C_a\subset S$, with $a=1,\dots,N$. Let $f_a$ be a section whose zero locus is $C_a$. The cover can then be described algebraically by
\begin{align}
\xi_a^{k_a}=f_a~,
\end{align}
for integers $k_a>1$, where $\xi_a$ is the fiber coordinate of a line bundle $\cL_a\to S$ satisfying $[C_a]=k_a c_1(\cL_a)$. The resulting branched covering is a K3 surface when $\sum_a (k_a-1)c_1(\cL_a)=-K_S$, and is smooth for sufficiently generic branch divisors; it is enough that each $C_a$ be smooth and that distinct components intersect transversely. This construction has a direct GLSM realization, with chiral superfields describing the toric coordinates on $\operatorname{tot}(\cL_1\oplus\dots\oplus\cL_N\to S)$ and fermi superfields whose superpotential couplings $J_A$ impose the defining equations. The resulting linear model has a manifest discrete symmetry $\Gamma=\Z_{k_1}\times\dots\times\Z_{k_N}$ whose generators act by multiplying the corresponding fiber
coordinate $\xi_a$ by a $k_a$-th root of unity. Geometrically, these deck transformations permute the sheets of the cover, and the quotient by this action recovers the base surface, $M/\Gamma\simeq S$. We denote the degree of the covering by 
\begin{align}
k = k_1\times\dots\times k_N~,
\end{align}
and refer to $M$ as a $\Gamma$-covering of $S$.

\subsubsection*{Branched coverings of $\P^1\times\P^1$}
To construct a TLSM whose large radius phase realizes the flux vacua $X$, the toric ambient variety containing the K3 surface $M$ must have at least two independent divisor classes whose pullbacks define $\eta_\alpha\in\operatorname{Pic}(M)$. Indeed, the K3 surface is endowed with two-forms $J$ and $\upomega$ in $H^{1,1}(M)$, obtained as linear combinations of the $\eta_\alpha$ with coefficients given respectively by the FI parameters $\rho^\alpha$ and shift charges $w^\alpha$. The conditions $J.J>0$, ensuring a finite K3 volume, and $J.\upomega=0$, imposed by spacetime supersymmetry, therefore require at least two independent divisor classes. Thus, $\operatorname{Pic}(M)$ must have rank at least two, which in the GLSM requires a gauge group of rank at least two. Since the anomaly cancellation condition~\eqref{eq:anomalycancelation} becomes increasingly restrictive with the gauge-group rank, we focus on the minimal case of linear models with $\GU(1)\times\GU(1)$ gauge group.

Given our previous bundle stability considerations, a natural class of candidates for such $\GU(1)\times\GU(1)$ GLSMs is provided by those whose large radius phases realize branched coverings of a surface $S$ with $H^2(S,\Z)=\Z^2$. The simplest choice\footnote{Other choices of $S$ are certainly possible---the main advantage of $\P^1\times\P^1$ is that the corresponding linear model will admit a LG orbifold phase. This is why we focus on it here.} is
\begin{align}
S\simeq \P^1\times \P^1~,
\end{align}
for which the covering data have a particularly simple description: the sections $f_a$ are bihomogeneous polynomials in the $\P^1\times\P^1$ coordinates. Out of the many presentations of K3 as a branched cover of $\P^1\times\P^1$~\cite{Hayashi:2022}, we consider the three following constructions, with covering degree $k\in\{2,3,4\}$:
\begin{itemize}
\item[-] for $k=2$, the surface $M$ is a double cover branched along a genus-nine curve;
\item[-] for $k=3$, the surface $M$ is a triple cover branched along a genus-four curve;
\item[-] for $k=4$, the surface $M$ is a biquadratic cover branched along two genus-one curves intersecting transversely at eight points.
\end{itemize}
For a generic choice of branching locus, the K3 surface $M$ has Picard rank two, with $\operatorname{Pic}(M)=\operatorname{Span}_\Z\{\eta_1,\eta_2\}$, where $\eta_\alpha$ are the pullbacks of the two hyperplane classes on $\P^1\times\P^1$. Their intersection pairing is
\begin{align}
D = \begin{pmatrix}
0 & k \\
k & 0
\end{pmatrix}~.
\end{align}
The corresponding linear models are specified by the following $\GU(1)\times\GU(1)$ gauge charges and superpotential couplings:
\begin{align*}
\arraycolsep=4pt
k&=2:&
&\begin{array}{cccccc}
\Phi_1 & \Phi_2 & \Phi_3 & \Phi_4 & \Phi_5 & \Gamma_5 \\
1 & 1 & 0 & 0 & 2 & -4 \\
0 & 0 & 1 & 1 & 2 & -4 \\
\end{array}&
&\begin{array}{l}
J_5=\Phi_5^2-f(\Phi_1,\dots,\Phi_4)~,
\end{array}
\\\\
k&=3:&
&\begin{array}{cccccc}
\Phi_1 & \Phi_2 & \Phi_3 & \Phi_4 & \Phi_5 & \Gamma_5 \\
1 & 1 & 0 & 0 & 1 & -3 \\
0 & 0 & 1 & 1 & 1 & -3 \\
\end{array}&
&\begin{array}{l}
J_5=\Phi_5^3-f(\Phi_1,\dots,\Phi_4)~,
\end{array}
\\\\
k&=4:&
&\begin{array}{cccccccc}
\Phi_1 & \Phi_2 & \Phi_3 & \Phi_4 & \Phi_5 & \Phi_6 & \Gamma_5 & \Gamma_6 \\
1 & 1 & 0 & 0 & 1 & 1 & -2 & -2 \\
0 & 0 & 1 & 1 & 1 & 1 & -2 & -2 \\
\end{array}&
&\begin{array}{l}
J_5=\Phi_5^2-f_1(\Phi_1,\dots,\Phi_4)~,\\[2pt]
J_6=\Phi_6^2-f_2(\Phi_1,\dots,\Phi_4)~.
\end{array}
\end{align*}
For $k=2,3$, the resulting K3 surface is a toric hypersurface, while for $k=4$ it is a complete intersection in the ambient toric variety.

As indicated by the labels of the fermi multiplets, this field content must be supplemented by additional superfields: on its own, it does not define a consistent supersymmetric gauge theory, as the charged matter induces a local gauge anomaly. Although our ultimate goal is to construct a TLSM with a non-trivial anomaly, cancelled by the classical variation of the shift multiplet $\Omega$, the anomaly matrix of the field content above has rank two. It therefore cannot be cancelled by a single shift multiplet, which requires the anomaly matrix to have the factorized form~\eqref{eq:anomalycancelation}. We must consequently introduce additional chiral and fermi multiplets to cancel part of the anomaly. These additional fields will encode the holomorphic vector bundle $\cE$.

\subsection{Bundle data}
Every abelian covering constructed above naturally inherits the divisor class
\begin{align}
\upomega = \eta_1-\eta_2~,
\label{eq:ASDclassGLSM}
\end{align}
with self-intersection $\upomega.\upomega = -2k$. When the two $\P^1$ factors are taken to have equal K\"ahler parameters, this class is anti-self-dual and can therefore serve as the curvature class of the circle bundle $X\to M$. In the linear model, the circle fiber is encoded by a shift multiplet of charges
\begin{align}
w = \begin{pmatrix}
1 \\-1\\
\end{pmatrix}~.
\end{align}
Consequently, the anomaly matrix of the linear model must be proportional to $ww^{\text{t}}$.

The matter content introduced above does not satisfy this condition. We therefore supplement it with additional superfields whose contribution to the gauge anomaly has the required form. A minimal choice consists of two additional chiral multiplets and four additional fermi multiplets with the following gauge charges and superpotential couplings:
\begin{align*}
\arraycolsep=4pt
k&=2:&
&\begin{array}{cccccc}
\Phi_6 & \Phi_7 & \Gamma_1 & \Gamma_2 & \Gamma_3 & \Gamma_4 \\
-12 & 0 & 1 & 11 & 0 & 0 \\
0 & -12 & 0 & 0 & 1 & 11 \\
\end{array}&
&\begin{array}{lll}
J_1=\Phi_6 F_1~,&\quad&J_2=\Phi_6 F_2~,\\
J_3=\Phi_7 F_3~,&\quad&J_4=\Phi_7 F_4~,\\
\end{array}
\\\\
k&=3:&
&\begin{array}{cccccc}
\Phi_6 & \Phi_7 & \Gamma_1 & \Gamma_2 & \Gamma_3 & \Gamma_4 \\
-8 & 0 & 1 & 7 & 0 & 0 \\
0 & -8 & 0 & 0 & 1 & 7 \\
\end{array}&
&\begin{array}{lll}
J_1=\Phi_6 F_1~,&\quad&J_2=\Phi_6 F_2~,\\
J_3=\Phi_7 F_3~,&\quad&J_4=\Phi_7 F_4~,\\
\end{array}
\\\\
k&=4:&
&\begin{array}{cccccc}
\Phi_7 & \Phi_8 & \Gamma_1 & \Gamma_2 & \Gamma_3 & \Gamma_4 \\
-6 & 0 & 1 & 5 & 0 & 0 \\
0 & -6 & 0 & 0 & 1 & 5 \\
\end{array}&
&\begin{array}{lll}
J_1=\Phi_7 F_1~,&\quad&J_2=\Phi_7 F_2~,\\
J_3=\Phi_8 F_3~,&\quad&J_4=\Phi_8 F_4~.\\
\end{array}
\end{align*}
The complete linear model has an anomaly matrix of the factorized form~\eqref{eq:anomalycancelation}, with
\begin{align}
r^2=\frac{12}{k}~.
\label{eq:quantizedradius}
\end{align}
Coupling this theory to a shift multiplet of the corresponding radius therefore gives an non-anomalous TLSM.

It remains to identify the holomorphic bundle encoded by the additional fermi multiplets. Their superpotential couplings are specified by four polynomials $F_1,\dots,F_4$. The polynomials $F_1$ and $F_2$ are homogeneous in $\Phi_1,\Phi_2$ of degrees $\frac{24}{k}-1$ and $1$, respectively, while $F_3$ and $F_4$ are homogeneous in $\Phi_3,\Phi_4$ with the same respective degrees. The resulting bundle decomposes as $\mathcal{E}=\cE_1\oplus\cE_2$, where the two factors are described by the kernel bundles
\begin{align}
\mathcal{E}_1 &= \operatorname{ker}(F_1,F_2)~,&
\mathcal{E}_2 &= \operatorname{ker}(F_3,F_4)~,
\label{eq:kernelbundles}
\end{align}
with $(F_1,F_2)$ and $(F_3,F_4)$ understood, respectively, as maps
\begin{align}
\cO(1,0)\oplus\cO(\tfrac{24}{k}-1,0)&\to\cO(\tfrac{24}{k},0)~,&
\cO(0,1)\oplus\cO(0,\tfrac{24}{k}-1)&\to\cO(0,\tfrac{24}{k})~.
\end{align}
Both factors in~\eqref{eq:kernelbundles} are in fact holomorphically trivial: the sections $(F_2,-F_1)$ and $(F_4,-F_3)$ provide nowhere-vanishing sections for the two line bundles.\footnote{The only possible zero of the linear polynomial $F_2$ occurs at a single point of the corresponding $\P^1$, and for generic choices of couplings $F_1$ and $F_2$ do not vanish simultaneously. The same argument applies to the second factor.} Hence,
\begin{align}
\mathcal{E}\simeq\mathcal{O}\oplus\mathcal{O}~.
\end{align}

The resulting heterotic background is therefore one in which the contribution of the gauge bundle to the Bianchi identity~\eqref{eq:heteroticBianchi} vanishes. The second Chern class of the tangent bundle, $c_2(T_M)=24$, is instead cancelled entirely by the flux contribution. Indeed, for the quantized radius~\eqref{eq:quantizedradius} and the anti-self-dual class~\eqref{eq:ASDclassGLSM}, one finds
\begin{align}
r^2(-\upomega.\upomega) = 24~.
\end{align}
We refer to this as a \emph{purely geometric} flux background: the torsional geometry satisfies the Bianchi identity without the presence of a non-trivial gauge bundle. Equivalently, for the associated five-dimensional space $X$, this condition implies $p_1(X)=0$ in $H^4(X,\Z)$.

In hindsight, it is natural that our construction leads to a trivial bundle: the additional chiral and fermi superfields introduced to encode the bundle data describe a direct sum of two line bundles, both with trivial first Chern class. We will retain this choice in what follows, as it provides a particularly simple realization of a flux vacuum without the additional subtleties associated with a non-trivial gauge bundle. There are, however, many other possibilities. In particular, the same (0,2) construction can accommodate non-trivial line bundles by introducing extra fermi multiplets with opposite charges under the two $\GU(1)$ gauge symmetries. This allows for a variety of flux vacua associated with distinct solutions to the Bianchi identity~\eqref{eq:heteroticBianchi}, and consequently gives rise to five-dimensional theories with different gauge groups.

\subsubsection*{Large radius phase}
It remains to verify that the gauged linear sigma models introduced above does indeed admit a large radius phase describing a K3 branched cover of $\P^1\times\P^1$. We consider the model with degree $k=2$; the analysis for the other two cases is analogous. Let $\phi_i$ denote the scalar components of the chiral superfields. The D-term constraints are
\begin{align}
|\phi_1|^2+|\phi_2|^2+2|\phi_5|^2-12|\phi_6|^2&=\rho^1~,&
|\phi_3|^2+|\phi_4|^2+2|\phi_5|^2-12|\phi_7|^2&=\rho^2~.
\end{align}
For FI parameters satisfying
\begin{align}
\rho^1,\rho^2>0~,
\end{align}
the corresponding excluded set in $\C^7$ is $\{\phi_1=\phi_2=\phi_5=0\}\cup\{\phi_3=\phi_4=\phi_5=0\}$. Taken by itself, the resulting $\C^\ast\times\C^\ast$ quotient does not yet describe $\P^1\times\P^1$. We must also impose the F-term equations. These force $\phi_6=\phi_7=0$ and impose $\phi_5^2 = f(\phi)$, where $f$ is bihomogeneous of degree (2,2) in the two pairs $\{\phi_1,\phi_2\}$ and $\{\phi_3,\phi_4\}$. Crucially, $f$ vanishes whenever $\phi_1=\phi_2=0$ or $\phi_3=\phi_4=0$. Consequently, these loci are excluded by the D-term equations: if, for example, $\phi_1$ and $\phi_2$ vanish simultaneously, then $f=0$ and hence $\phi_5=0$, which is incompatible with the first D-term equation. Thus the excluded set is effectively $\{\phi_1=\phi_2=0\}\cup\{\phi_3=\phi_4=0\}$, and the classical vacuum moduli space is obtained by taking the double covering
\begin{align}
\phi_5^2=f(\phi_1,\dots,\phi_4)~,
\end{align}
of $\P^1\times\P^1$, as claimed. While this classical analysis is independent of the value of $\rho^1-\rho^2$, its sign distinguishes two different large radius phases, with different ambient toric varieties. We will analyze these phases in more detail in section~\ref{s:phases}.

\subsubsection*{Spectator fields}
Before turning to the analysis of the TLSM, we briefly discuss the role of spectator fields. In general, the FI parameters of a GLSM receive one-loop corrections unless the sum of the gauge charges of the chiral multiplets vanishes~\cite{Witten:1993yc}. In a (0,2) model this condition can always be enforced by introducing a pair of spectator fields~\cite{Distler:1993mk}, consisting of one chiral multiplet and one fermi multiplet. For the present model their gauge charges are of the form
\begin{align*}
\arraycolsep=4pt
\begin{array}{cc}
\Phi_{\text{sp}} & \Gamma_{\text{sp}} \\
Q & -Q\\
Q & -Q\\
\end{array}
\end{align*}
where $Q>0$ is defined as minus the total charge of the chiral multiplets under the first $\GU(1)$. Since $\Phi_{\text{sp}}$ and $\Gamma_{\text{sp}}$ carry opposite gauge charges, their contribution to the gauge anomaly vanishes identically. We furthermore introduce the superpotential coupling
\begin{align}
\cW \supset \text{m}\Gamma_{\text{sp}}\Phi_{\text{sp}}~,
\end{align}
where $\text{m}$ is a non-zero mass parameter. The scalar component of the spectator chiral superfield then vanishes in a supersymmetric vacuum, and both spectator multiplets acquire a mass of order $\text{m}$. They therefore decouple from the infrared dynamics---their only role is to render the FI parameters RG invariant. We shall keep their presence implicit in the remainder of the paper.

\section{Phase analysis  of the ``\texorpdfstring{$\Z_6\times\Z_6$}{Z6xZ6}'' TLSM} \label{s:phases}
Having defined a class of TLSMs, we now turn to a detailed analysis of the $k=4$ example.  The analysis of the two remaining models with $k=2,3$ proceeds along similar lines.  Each of these has a LG orbifold (LGO) phase with orbifold group $\Z_{24/k} \times \Z_{24/k}$, and our example has the smallest group $\Z_6 \times \Z_6$.  The TLSM has a $\GU(1)\times\GU(1)$ gauge symmetry, and the matter content carries the charges (the form of the transformations is given above in~(\ref{eq:gaugetransform1}) and (\ref{eq:gaugetransform2})) given by the first two rows of the following table:
\begin{table}[h]
\label{table:charges}
\begin{align*}
\arraycolsep=4pt
\begin{array}{ccccccccccccccccc}
& \Phi_1 & \Phi_2 & \Phi_3 & \Phi_4 & \Phi_5 & \Phi_6 & \Phi_7 & \Phi_8 & \Gamma_1 & \Gamma_2 & \Gamma_3 & \Gamma_4 & \Gamma_5 & \Gamma_6 &   \exp[\Omega] \\
Q^1 & 1 & 1 & 0 & 0 & 1 & 1 & -6 & 0 & 1 & 5 & 0 & 0 & -2 & -2 & 1 \\
Q^2 & 0 & 0 & 1 & 1 & 1 & 1 & 0 & -6 & 0 & 0 & 1 & 5 & -2 & -2 & -1 \\
\hline
q_\sleft^1 & 0 & 0 & 0 & 0 & 0 & 0 & 1 & 0 & -1 & -1 & 0 & 0 & 0 & 0 & 0 \\
q_\sleft^2 & 0 & 0 & 0 & 0 & 0 & 0 & 0 & 1 & 0 & 0 & -1 & -1 & 0 & 0 & 0 \\
q_\sright & 0 & 0 & 0 & 0 & 0 & 0 & 1 & 1 & 0 & 0 & 0 & 0 & 1 & 1 & 0 \\
\end{array}
\end{align*}
\caption{Gauge and global charges of the fields}
\end{table}

The superpotential can be taken to be
\begin{align}
\label{eq:Wgenericenough}
\cW & = \sum_{\alpha=1}^2 (\tau^\alpha - \ff{r^2}{2} w^{\alpha}\Omega )\Upsilon_\alpha + \Gamma_5\left(\Phi_5^2 - f_1\right) + \Gamma_6 \left(\Phi_6^2 -f_2\right)
\nonumber\\
&\qquad + \Phi_7\left( \Gamma_1 \Phi_1^5 + \Gamma_2 \Phi_2\right) + \Phi_8 \left(\Gamma_3 \Phi_3^5 + \Gamma_4\Phi_4\right)~.
\end{align}
The polynomials $f_{1,2}(\Phi_1,\Phi_2,\Phi_3,\Phi_4)$ are of degree $2$ in the pair $\Phi_1,\Phi_2$, as well as in the pair $\Phi_3,\Phi_4$, but otherwise generic.
While this is not the generic form (we will discuss that below), it is generic enough: for generic $f_{1,2}$ the large radius phase geometry is a circle bundle over a smooth K3 surface $M$.  $\cW$ respects a non-anomalous global $\GU(1)^2$ symmetry, as well as a $\GU(1)$ R-symmetry under  which the superspace coordinate $\theta$ and the gauge field strength multiplets $\Upsilon_\alpha$ transform with charge $+1$.  Using c-extremization~\cite{Benini:2012cz} we can choose a basis for these symmetries consistent with the assumption that a linear combination of the symmetry currents flows to the R-current of the SCFT, while the remaining currents flow to holomorphic currents.  
We thus label this symmetry group by $\GUL^1 \times\GUL^2\times\GUR$, and we listed the matter charges in the last three lines of the table above.  The two-point function of the $\GUR$ current is consistent with the SCFT having $c_{\sright} = 15/2 =6+3/2$, where the $6$ is what we expect from the K3 geometry, while the $3/2$ is the central charge associated to the fibered circle.  The gravitational anomaly then determines $c_{\sleft} = 7$, which is again consistent with expectations from the geometry.  For what follows we also recall that the global symmetry and R-charge assignments are ambiguous for gauge-variant fields, since each can be shifted by an arbitrary combination of the gauge symmetry charges.

\subsection{Classical phase diagram}
We now turn to the familiar exercise of linear sigma model analysis:  the construction of the ``phase diagram'' for the FI parameter space.  The classical analysis is straightforward and leads to a diagram with $5$ phases:
\begin{align}
\begin{tikzpicture}[scale=0.75]
\draw[->](-5,0)--(5,0) node[below]{$\rho^1$};
\draw[->](0,-5)--(0,5) node[left]{$\rho^2$};
\draw[thick] (0,0) --(5,0);
\draw[thick] (0,0) --(0,5);
\draw[thick] (0,0)--(-5,0);
\draw[thick] (0,0)--(4,4);
\draw (1,0) node{$\bullet$};
\draw (0,1) node{$\bullet$};
\draw (-1,0) node{$\bullet$};
\draw (0,-1) node{$\bullet$};
\draw (1,1) node{$\bullet$};
\draw (1.75,4) node {LR};
\draw (4,1.75) node {LR'};
\draw (-3,3) node {Hyb};
\draw (3,-3) node {Hyb'};
\draw(-3,-3) node {LGO};
\end{tikzpicture}
\end{align}
There are three qualitatively different limiting behaviors:  large radius (LR),  hybrid (Hyb), and LGO, and these are characterized by choice of excluded sets:
\begin{align}
\text{LGO} & :   \{7\} \cup \{8\}~,&
\text{Hyb}  & : \{ 7\} \cup\{3,4,5,6\}~,&
\text{LR}    & : \{1,2,5,6\} \cup \{3,4,5,6\} \cup\{3,4,7\}~,
\nonumber\\
& & \text{Hyb'} &: \{8\} \cup\{1,2,5,6\}~,&
\text{LR'} &:  \{1,2,5,6\}\cup\{3,4,5,6\}\cup\{1,2,8\}~.
\end{align}
Each list is a shorthand for an  an intersection of hyperplanes in $\C^8$; for example, $\{3,4,7\} $ denotes the subset $\{ \phi_3 = \phi_4 = \phi_7 = 0\} \subset \C^8$, and the ambient toric variety is then the quotient 
\begin{align}
Z = \frac{\C^8 \setminus \cF}{(\C^\ast)^2}~,
\end{align}
where the excluded set $\cF$ is the set determined by the choice of phase.   In the usual linear sigma model philosophy each phase corresponds to a limiting point in the parameter space of the low energy theory, and to describe its physics in our TLSM we must incorporate both the chiral superpotential and the circle fibration encoded in the shift multiplet.  In the remainder of this section we will make a first pass at this description by describing the spaces of classical supersymmetric vacua of the linear sigma model, and in the next section we will refine the analysis by incorporating the redundancies implied by field redefinitions that we discussed in the introduction.  Since the primed phases LR' and Hyb' are equivalent to their unprimed versions by relabeling fields, it suffices to consider just the three unprimed phases.

\subsubsection*{The LGO phase}
The expectation values of $\Phi_7$ and $\Phi_8$ break the gauge symmetry to $\Z_6 \times\Z_6$, and after suitably rescaling the fields and dropping the massive vector multiplets, we find a (0,2) LG model with superpotential
\begin{align}
\cW & = \Gamma_1 \Phi_1^5 + \Gamma_2 \Phi_2+\Gamma_3\Phi_3^5 + \Gamma_4 \Phi_4+ \Gamma_5 (\Phi_5^2-f_1) + \Gamma_6 (\Phi_6^2-f_2)  ~
\end{align}
accompanied by a compact boson with $r^2 =3$.  

A striking feature of this (0,2) LG potential is that it includes a (2,2) locus.  Setting $f_1 = f_2 = 0$ leads to a (0,2) superpotential that can be derived from a (2,2) superpotential
\begin{align}
W & = \tfrac{1}{6} X_1^6 + \tfrac{1}{2} X_2^2 + \tfrac{1}{6} X_3^6 + \tfrac{1}{2} X_4^2 + \tfrac{1}{3} X_5^3 + \tfrac{1}{3} X_6^3~,
\end{align}
where each (2,2) chiral superfield $X_i$ decomposes into a pair of (0,2) chiral superfields $(\Phi_i,\Gamma_i)$, $i=1,\ldots, 6$.    The standard (2,2) computation of the central charge yields $c_{\sright} = 6$, which, when combined with the compact boson yields $c_{\sright} = 15/2$.  An application of c-extremization to the generic (0,2) LG potential $\cW$ reproduces this central charge.  The (2,2) theory is in fact a product of (2,2) minimal models, but the chiral orbifold by $\Z_6\times\Z_6$, which acts on both the LG and compact boson degrees of freedom, leads to a very different SCFT than what would have been obtained by the usual Gepner construction.  We will analyze this in detail in section~\ref{s:lgo}, where we will meet another surprise:  the couplings $f_1$ and $f_2$ are redundant, and the entire LGO phase flows to the same fixed point---the chiral orbifold of the Gepner and circle SCFT.

\subsubsection*{The hybrid phase}
In this phase the expectation value of $\Phi_7$  breaks the first gauge group to $\Z_6$. The superpotential now forces $\phi_1 = \phi_2 = 0$, as well as $\phi_5 = \phi_6 = 0$, and the excluded set $\{3,4,5,6\}$ now implies that $\phi_3$ and $\phi_4$ cannot vanish simultaneously, so that the remaining terms in the superpotential force $\phi_8 = 0$.  Thus, we find a hybrid phase:  the base $B$ is a $\P^1$ parameterized by the superfields $\Phi_3,\Phi_4$, and the remaining degrees of freedom---a  $\Z_6$ orbifold of a compact boson and a (0,2) LG theory---are fibered over it.  The chiral superpotential for the fibered LG theory is
\begin{align}
\cW_{\text{LG fib}} = \Gamma_5\left(\Phi_5^2 - f_1\right) + \Gamma_6 \left(\Phi_6^2 -f_2\right)
+ \left( \Gamma_1 \Phi_1^5 + \Gamma_2 \Phi_2\right) + \Phi_8 \left(\Gamma_3 \Phi_3^5 + \Gamma_4\Phi_4\right)~,
\end{align}
where we should now think of the $\Phi_3$ and $\Phi_4$ as constant background values that determine the coefficients in the superpotential for the superfields $\Phi_1,\Phi_2,\Phi_5,\Phi_6,\Phi_8$ and $\Gamma_1,\Gamma_2,\ldots,\Gamma_6$.  This fits the definition of a ``good hybrid'' given in~\cite{Bertolini:2013xga}, because the R-symmetry does not act on the base geometry.  In this situation we can calculate the central charge $c_{\sright}$ as a sum of three contributions:  the base geometry $B$ yields the usual $3\times \dim_{\C} B$, the compact boson yields $3/2$, and the (0,2) LG model makes the contribution that can be computed by c-extremization.  

The charges relevant for the latter contribution are given by $q_{\sright,\text{Hyb}} = q_{\sright} + \ff{1}{6} Q^1$, where the shift by the gauge charge is made so that the Higgsed multiplet $\Phi_7$ is gauge--neutral.  Making also the same shift for $q_{\sleft}^1$ leads to the following charge assignments for the fiber LG fields:\footnote{We study the generic fiber, in the spirit of~\cite{Bertolini:2013xga,Bertolini:2017lcz}, where the global symmetries corresponding to $Q^1$ and $Q^2$ charge assignments are broken by the expectation values of $\Phi_7$ and $\Phi_3,\Phi_4$.}
\begin{align*}
\arraycolsep=4pt
\begin{array}{ccccccc}
			&\Phi_{1,2,5,6}		&\Phi_8	&\Gamma_1	&\Gamma_2	&\Gamma_{3,4}	&\Gamma_{5,6}\\
q^1			&\ff{1}{6}			&0		&-\ff{5}{6}		&-\ff{1}{6}		&0				&-\ff{2}{6} \\
q^2			&0				&1		&0			&0			&-1				&0 \\
q_{\sright}		&\ff{1}{6}			&1		&\ff{1}{6}		&\ff{5}{6}		&0				&\ff{4}{6}
\end{array}
\end{align*}
The idea of c-extremization is quite simple:  we assume that the superconformal R-symmetry is a linear combination of these three symmetries, which fixes the linear combination of the corresponding conserved currents that flows to the superconformal R-current to be
\begin{align}
J_{\sright}^{\text{ir}} & = J_{\sright} + \alpha J^1 + \beta J^2~,
\end{align}
while any linear combination of $J^1$ and $J^2$ must flow to a holomorphic (possibly trivial) current.  
The undetermined coefficients $\alpha$ and $\beta$ are then fixed by demanding that $J_{\sright}^{\text{ir}}$ has no 't Hooft anomaly with either $J^1$ or $J^2$. Arranging the three currents into $\boldsymbol{J} = (J^1,J^2,J_{\sright})$, we find that the 't Hooft anomaly matrix is 
\begin{align}
\cA(\boldsymbol{J},\boldsymbol{J}) & = 
\begin{pmatrix} \ff{5}{6} & 0 & -\ff{1}{6} \\   0 & 1 & 0 \\ -\ff{1}{6} & 0 & -\ff{7}{6} \end{pmatrix}~,
\end{align}
which fixes the coefficients $\alpha = \ff{1}{5}$ and $\beta = 0$.  Now taking $\boldsymbol{J}' = (J^1, J^2, J_{\sright}^{\text{ir}})$, we find the anomaly matrix
\begin{align}
\cA(\boldsymbol{J}',\boldsymbol{J}') & = \begin{pmatrix} \ff{5}{6} & 0 & 0 \\ 0 & 1 & 0 \\ 0 & 0 & -\ff{1}{3}\cdot \ff{18}{5} \end{pmatrix}~,
\end{align}
from which we read off the LG central charge
\begin{align}
c_{\sright}^{\text{LG}} = \frac{18}{5} = 3 + \ff{3}{5}~.
\end{align}
Combining this with contributions from the circle and the base geometry, we obtain
\begin{align}
c_{\sright}^{\text{Hyb}} = c_{\sright}^{\text{LG}} +3+ \ff{3}{2}  = \ff{15}{2} + \ff{3}{5}~.
\end{align}
This is quite strange: the central charge we obtain in the hybrid limit exceeds our expected value by $3/5$!   We will argue below that the resolution of this discrepancy is that this limiting phase is not connected to the geometric phase by a marginal deformation, but it may be connected to it by a relevant RG flow that will, among other things, decrease the central charge $c_{\sright}$ to the expected $\ff{15}{2}$ value.  However, to make that argument we will first need to discuss the large radius phase, its boundary, as well as TLSM field redefinitions.

\subsubsection*{The large radius phase and its boundary}
We now examine the large radius phase defined by $\rho_2 >\rho_1$ and $\rho_1 >0$, where the excluded set is
\begin{align}
\cF = \{1,2,5,6\} \cup \{3,4,5,6\} \cup \{3,4,7\}~,
\end{align}
while the ideal generated by the chiral superpotential is
\begin{align}
\cI & = \la \phi_5^2-f_1,~ \phi_6^2-f_2,~ \phi_7 \phi_1^5,~ \phi_7 \phi_2,~ \phi_8 \phi_3^5,~ \phi_8 \phi_4\ra~.
\end{align}
Suppose $\phi_7 \neq 0$. Then from $\cI$ we see that we require $\phi_1 = \phi_2 =0$, which in turn implies the vanishing of $f_1$ and $f_2$ and then in turn the vanishing of $\phi_5,\phi_6$.  Thus, any point with $\phi_7 \neq 0$ in the vanishing set of $\cI$ belongs to the excluded set.  The same can be shown for any point with $\phi_8 \neq 0$.  Furthermore, if $\phi_1 = \phi_2 = 0$, then we also see from $\cI$ that $\phi_5 = \phi_6 = 0$, so that any point with $\phi_1 = \phi_2 = 0$ in the vanishing set of $\cI$ belongs to the excluded set.  Putting these observations together, we find that the space of vacua is the vanishing set
\begin{align}
\label{eq:vacuumspace}
M = \{\phi_5^2-f_1 = 0,~\phi_6^2 -f_2 = 0\} &\subset  Z'~, 
&
Z' & = \text{tot}\left( \cO(1,1)^{\oplus 2} \to \P^1 \times \P^1\right)~,
\end{align}
where the pairs $(\phi_1,\phi_2)$ and $(\phi_3,\phi_4)$ are the projective coordinates on the $\P^1\times\P^1$ base, while the $\phi_5,\phi_6$ are the fiber coordinates for the line bundles.  This is exactly the geometry described in section~\ref{ss:abeliancoverings} that motivated our choice of TLSM, and we also find the same trivial bundle structure for the massless left-moving fermions as described above.

Let us now consider a regime deep in the interior of this geometric phase, with $\rho_2 > \rho_1 \gg 0$.
From the supergravity condition reviewed in the introduction we also know that to produce a five-dimensional flux vacuum we need the class $\upomega = \eta_1 -\eta_2$ to be anti-self-dual on $M$.  Using the intersection pairing and the K\"ahler form
\begin{align}
J = \rho_1 \eta_1 + \rho_2 \eta_2~,
\end{align}
we see that
\begin{align}
J . \upomega & = 0~, & \iff && \rho_1 &= \rho_2~.
\end{align}
Thus, supergravity suggests that at least for large $\rho_1$ and $\rho_2$ we need to consider the special locus $\rho_1 = \rho_2$ in the FI parameter space, which turns out to be exactly a phase boundary.   Can we describe this locus in the TLSM parameter space without needing to tackle the gauge theory in its strongly coupled regime?  As we now argue, this is indeed possible.

Setting $\rho_1 = \rho_2 = \rho$, it is convenient to make a change of basis for $(\C^\ast)^2$ action on the $\C^8$ parameterizing the expectation values of the bosonic fields as follows:
\begin{align*}
\arraycolsep=4pt
\begin{array}{ccccccccc}
~		& \phi_1	&\phi_2	&\phi_3	&\phi_4	&\phi_5	&\phi_6	&\phi_7	&\phi_8 \\
Q^1		&1		&1		&0		&0		&1		&1		&-6		&0 \\
Q^1-Q^2	&1		&1		&-1		&-1		&0		&0		&-6		&6	
\end{array}
\end{align*}
This makes it clear that the ambient toric variety has a $\P^1$s worth of singularities, since the locus
\begin{align}
\phi_1 &= \phi_2 = \phi_3 = \phi_4 = \phi_7=\phi_8 = 0~, &  |\phi_5|^2 + |\phi_6|^2 &= \rho
\end{align}
solves the D-terms but is stabilized by the second $(\C)^\ast$ action.  Fortunately, the F-terms save us from having to consider this region in field space: 
the space of TLSM vacua is still described by~(\ref{eq:vacuumspace}), but now each $\P^1$ factor has the same volume determined by $\rho_1 = \rho_2 = \rho >0$.

\subsection{Parameters and field redefinitions}
\label{subsec:TLSMparameters}
The phase analysis left us with two unresolved issues:  how should we interpret the increase in the central charge in the hybrid phase relative to our naive expectations? Is there a way to understand the $\rho_1 = \rho_2$ requirement directly in the TLSM without appealing to a large radius supergravity analysis?  As we will now see, these two puzzles turn out to be intimately related and can be addressed by thinking about the space of TLSM couplings encoded in the holomorphic parameters of $\cW$ modulo the action of field redefinitions.

\subsubsection*{Intuition from supergravity}
Before we examine the TLSM parameter space, we take a moment to review what kind of structure we expect in the large radius phase.\footnote{Our discussion follows~\cite{Melnikov:2012cv}.}  We assume that in that limit the TLSM IR fixed point matches that of a non-linear sigma model for a heterotic flux vacuum and leads to a $5$--dimensional heterotic vacuum with $8$ supercharges.  The supergravity multiplet contains a graviphoton, and the theory also has a universal tensor multiplet that contains the heterotic dilaton.   Leaving out these degrees of freedom, the remaining massless scalars reside in vectors and hypers.  The former is characterized by the unbroken spacetime gauge algebra $\Lg = \Le_8\oplus \Le_8$, with scalars in the adjoint representation of $\Lg$:  these moduli correspond to Wilson line degrees of freedom and parameterize the Coulomb branch of the theory.  The TLSM parameter space does not include these moduli, so we work at the origin of the Coulomb branch.  

However, we do expect the TLSM parameters to describe at least some of the remaining moduli, all of which reside in $19$ hypers.
At a generic point in the moduli space, corresponding to a large smooth $M$, these are all gauge--neutral and correspond to deformations of the metric and $B$-field over $M$.\footnote{Recall that, with the Wilson line set to zero, the circle radius is fixed.}  To understand the dimension, we recall that the supergravity solution requires that the metric on $M$ is conformally hyper-K\"ahler, with the conformal factor determined up to a constant (the value of the dilaton)  by a solution to the Bianchi identity, which reduces to a PDE for the conformal factor over $M$, but the metric must also be chosen so that $\upomega \in H^2(M,\Z)$ is an ASD class, which removes $3$ degrees of freedom from the $57$--dimensional moduli space of fixed volume hyper-K\"ahler metrics on $M$.  In addition, there is a freedom to shift the heterotic B-field by a class in $H^2(M,\R)/H^2(M,\Z)$, but again there is a restriction from the non-trivial fibration:  the B-field modulus proportional to $\upomega$ is in fact lifted through a St\"uckelberg coupling to the $\Lu(1)$ gauge boson associated to the holomorphic current of the compact boson.  Thus, the fibered circle removes exactly one hyper's worth of deformations from the $20$ hypers we usually associate to deformations of K3 moduli in heterotic compactifications.\footnote{As we already noted above, it is not a coincidence that much the same mechanism works for K3 compactifications with line bundles.  This is discussed in more detail in~\cite{Israel:2023tjw}.}

The upshot of this review is that we expect the TLSM parameters, or more precisely motion in the quotient space $\cM_{\text{glsm}}$ defined in~(\ref{eq:Mglsm}), to capture some of these $19$ hyper moduli.  We also expect that rather than describing the full quaternion-K\"ahler space, we will only be able to capture a complex slice through it:  for example, this is already the case for the (2,2) GLSM describing a quartic in $\P^3$, which gives a $1+19$--complex-dimensional slice through the moduli space of the $(4,4)$ SCFT associated to the K3.\footnote{As explained in~\cite{Gomis:2016sab}, this is not limited to the quartic but instead holds for any (2,2) GLSM of a K3 target space.  One way to understand this is simply that the GLSM allows us to describe deformations of an algebraic K3 surface with fixed Picard number, but of course the SCFT has marginal deformations that take the K3 to some more generic non-algebraic surface.}  Similarly, our construction of $M$ presents its geometry as an algebraic variety with $\Pic(M)$ having rank $2$ and a K\"ahler cone generated by the classes $\eta_1,\eta_2$.  Such a K3 surface has an $18$-dimensional space of complex structures~\cite{Aspinwall:1996mn}, and we will see this reflected in the TLSM parameters.

\subsubsection*{The generic superpotential and redefinitions}

As a first step, we write the most general $\cW$ compatible with the gauge and global symmetries described in table~\ref{table:charges}:
\begin{align}
\cW & = \sum_{\alpha=1}^2 (\tau^\alpha - \ff{r^2}{2} w^{\alpha}\Omega )\Upsilon_\alpha + \underbrace{ \Phi_7 \cU_7 + \Phi_8 \cU_8 + \Gamma_5 J_5 + \Gamma_6 J_6}_{=\cW_\Gamma}~,
\end{align}
with
\begin{align}
\cU_7 &= \Gamma_1 F_1^{[5,0]} + \Gamma_2 F_2^{[1,0]}~,  & \#(\cU_7) &= 8~,\nonumber\\
\cU_8 &= \Gamma_3 F_3^{[0,5]} + \Gamma_4 F_4^{[0,1]}~,  & \#(\cU_8) &= 8~,\nonumber\\
J_5 & = F_5^{[0,0]} \Phi_5^2 +F_6^{[0,0]} \Phi_6^2  + F_7^{[0,0]} \Phi_5\Phi_6
+ F_8^{[1,1]} \Phi_5 + F_9^{[1,1]}\Phi_6 + F_{10}^{[2,2]}~, & \#(J_5) & = 20~, \nonumber\\
J_6 & = F_{11}^{[0,0]} \Phi_5^2 +F_{12}^{[0,0]} \Phi_6^2  + F_{13}^{[0,0]} \Phi_5\Phi_6
+ F_{14}^{[1,1]} \Phi_5 + F_{15}^{[1,1]}\Phi_6 + F_{16}^{[2,2]}~, & \#(J_6) & = 20~, \nonumber\\
\end{align}
where the $16$ polynomials $F_s^{[m,n]}$ are homogeneous of degree $m$ in $\Phi_1,\Phi_2$ and of degree $n$ in $\Phi_3,\Phi_4$.  Note that for each term we listed the number of parameters;  $F_s^{[m,n]}$ depends on $\#(F_s^{[m,n]}) =(m+1)(n+1)$ complex parameters.  All in all, the superpotential depends on $2+16+40=58$ complex parameters, where the first two correspond to the $\tau^\alpha$, so that $\cM_{\text{uv}} = \C^{58}$.  There is a locus $\Delta \subset \cM_{\text{uv}}$ where the theory degenerates:  for example, the space of supersymmetric vacua becomes non-compact when we tune the coefficients so that $\cU_7 = \cU_8 = 0$, but every component of $\Delta$ is of at least complex co-dimension $1$, so that a generic point in $\cM_{\text{uv}}$ should correspond to a well-behaved theory with IR limit described by a compact unitary SCFT.

The group of field redefinitions $\cG$ compatible with (0,2) supersymmetry and the symmetries of the TLSM are usefully broken up into two sets:
\begin{align}
\cG_{\text{geo}} : \begin{cases}
\delta \vec{\Phi}_{1,2} & = M_{1} \cdot \vec{\Phi}_{1,2}~ \hspace{2cm} \qquad \# (\delta \vec{\Phi}_{1,2} ) = 4~,\\
\delta \vec{\Phi}_{3,4} & = M_{2} \cdot \vec{\Phi}_{3,4}~ \hspace{2cm}\qquad \# (\delta \vec{\Phi}_{3,4} ) = 4~, \\
\delta \vec{\Phi}_{5,6} & = M_{3} \cdot \vec{\Phi}_{5,6} + \vec{G}^{[1,1]}_1~ \hspace{0.65cm} \qquad \# (\delta \vec{\Phi}_{5,6} ) = 4+2\times 2\times 2= 12~,\\
\delta \vec{\Gamma}_{5,6} & = M_4 \cdot \vec{\Gamma}_{5,6}~\hspace{2cm} \qquad \#(\delta\Gamma_{5,6})  = 4~,
\end{cases}
\end{align}
and
\begin{align}
\cG_{\text{bun}} : 
\begin{cases}
\delta  \Phi_7 & = \sigma_1 \Phi_7~ \hspace{2.5cm}  \#(\delta \Phi_7)  = 1~,\\
\delta  \Phi_8 & = \sigma_2 \Phi_8~ \hspace{2.5cm}  \#(\delta \Phi_8)  = 1~,\\
\delta \Gamma_1 & = \sigma_3 \Gamma_1~  \hspace{2.5cm} \#(\delta\Gamma_1)  = 1~,\\
\delta \Gamma_2 & = \sigma_4 \Gamma_2 + G_2^{[4,0]} \Gamma_1~ \hspace{0.75cm}  \#(\delta\Gamma_2)  = 1+5 = 6~,\\
\delta \Gamma_3 & = \sigma_5 \Gamma_3~ \hspace{2.5cm} \#(\delta\Gamma_3)  = 1~,\\
\delta \Gamma_4 & = \sigma_6 \Gamma_4 + G_3^{[0,4]} \Gamma_3~ \hspace{0.75cm}   \#(\delta\Gamma_4)  = 1+5 = 6~.
\end{cases}
\end{align}
Here the $G$ are homogeneous polynomials in the $\Phi_{1,2,3,4}$, the $M$ are invertible $2\times 2$ matrices, and the $\sigma$ take values in $\C^\ast$.\footnote{We do not consider redefinitions of the linear multiplet because to obtain the desired theory for the heterotic worldsheet we need to decouple its ``radial'' mode, and the remaining field redefinitions are just constant shifts of the periodic scalar that do not induce an additional action on the parameter space beyond that produced by the $\cG$ redefinitions.}

Thus, we have a $58 = 2+56$--dimensional space of parameters---this is our $\cM_{\text{uv}}$---and a $40 = 24 + 16$--dimensional space of infinitesimal field redefinitions.  The latter induce an action on $\cM_{\text{uv}}$, but the action is not effective due to invariance under global and gauge symmetries, as we discuss next.

\subsubsection*{Redefinitions and the FI parameters}

Consider the complexified field redefinition which acts on the $\Phi$ and $\Gamma$ fields as a rescaling:
\begin{align}
\Phi_i &\mapsto e^{-2\pi \ii \mu_i} \Phi_i~,&
\Gamma_A &\mapsto e^{-2\pi \ii \nu_A} \Gamma_A~,
\end{align}
where $\mu_i,\nu_A \in \C$.  Such a redefinition induces, through the Jacobian of the field rescalings in the path integral measure, an action on the $\tau^{\alpha}$ by a shift~\cite{Kreuzer:2010ph}:
\begin{align}
\tau_\alpha \mapsto \tau_\alpha + \sum_{A} Q^\alpha_A \nu_a - \sum_{i} Q^\alpha_i \mu_i~.
\end{align}
Typically we do not consider such an action because it also acts on other couplings in the theory, such as the parameters in $\cW_{\Gamma}$, and when $\cW_{\Gamma}$ is generic enough we can fix most or perhaps all of the freedom in the $\mu_i$ and $\nu_A$ by bringing $\cW_{\Gamma}$ to some fiducial form.  But if $\cW_{\Gamma}$ has a symmetry that acts by phases with
\begin{align}
\mu_i &=  q_i \zeta~, &
\nu_A & = q_A \zeta~, & \zeta &\in \C~,
\end{align}
then this redefinition leaves $\cW_{\Gamma}$ invariant and shifts
\begin{align}
\tau^\alpha \mapsto \tau^\alpha + \left(\sum_A  q_A Q_A^\alpha - \sum_i q_i Q_i^\alpha \right) \zeta~.
\end{align}
Thus, for every anomalous phase symmetry of the classical action the field redefinitions impose an equivalence relation on the complexified FI parameters.   

In the TLSM there is a field redefinition that acts as a constant gauge transformation on the $\Phi_i$ and $\Gamma_A$ with
\begin{align}
\mu_i &=   \sum_\beta Q_i^\beta \xi_\beta~, &
\nu_A& =  \sum_\beta Q_A^\beta \xi_\beta~,
\end{align}
and leaves the remaining fields invariant.   This is a symmetry of $\cW_{\Gamma}$, but it acts on the $\tau^\alpha$ by
\begin{align}
\tau^\alpha \mapsto \tau^\alpha + \sum_{\beta} \cA_{\alpha\beta} \xi_\beta = \tau^\alpha + 
w^\alpha \underbrace{\big(-2 r^2  \sum_\beta w^\beta \xi_\beta\big)}_{=\xi}\, ,
\end{align}
where in the last equality we used~(\ref{eq:anomalycancelation}).   This leads to the equivalence claimed in~(\ref{eq:equivalencecirclebundle}) and has a profound consequence for the phase diagram of the TLSM: there is an invariant parameter $\tau = (\tau_1 + \tau_2)/2$ with imaginary part $\rho$, and two invariant limiting points corresponding to $\rho^1-\rho^2 = \pm \infty$---the two hybrid phases:
\begin{align}
\begin{tikzpicture}[scale=0.6]
\draw[gray,thin, ->](-5,0)--(5,0) node[below]{$\rho^1$};
\draw[gray, thin,->](0,-5)--(0,5) node[left]{$\rho^2$};
\draw[gray,thin] (0,0) --(5,0);
\draw[gray,thin] (0,0) --(0,5);
\draw[gray,thin] (0,0)--(-5,0);
\draw[thick,->] (-4,-4) node[below left] {$\text{LGO}$}--(4,4) node[above right] {$\text{LR}$} node[below] {$\rho$};
\draw (-3,3) node {$\boldsymbol{\ast}$}  node[above left] {Hyb};
\draw (3,-3) node {$\boldsymbol{\ast}$}  node[below right] {Hyb'};
\end{tikzpicture}
\end{align}
This suggests that the TLSM RG flow has three basins of attraction:  the family of SCFTs with a marginal deformation that interpolates between the LGO and LR phases, all with $c_{\sright} = 15/2$ and with the LR phase corresponding to the locus $\rho^1 = \rho^2 \gg 0$, as well as two disconnected hybrid phases with $c_{\sright} = 15/2+3/5$.  Such a hybrid SCFT, if it exists, cannot by itself give rise to a consistent string vacuum, and the jump in the central charge indicates that it is not connected to the LR phase by motion in the SCFT moduli space . However it is plausible that the hybrid phases are connected to the family of flux vacua via RG flow. That is, there may be a relevant deformation of the hybrid CFT that induces an RG flow to the $c_{\sright} = 15/2$ branch.  That would also be consistent with the large radius expectation that a small deformation away from the ASD condition for the circle fiber should induce a non-zero potential in the effective theory. 

We stress that we still expect $\tau$ to interpolate between the LR and LGO phases, and although $\rho = 0$ corresponds to a strong coupling region for the gauge theory, we expect the IR physics to vary smoothly as $\rho$ varies from positive to negative values, at least for generic values of $\vartheta$ and parameters in $\cW_{\Gamma}$.  Unlike in (2,2) GLSMs, where there is typically a non-compact Coulomb branch that opens up at special values of $\tau$, there is no indication of such a pathology in the TLSM.~   This is not to say that every $\tau$ leads to a well-defined SCFT in the IR: it remains an important open problem to find (0,2) GLSM tools that could detect a potential singularity, since it is certainly possible that the heterotic string theory develops an enhanced gauge symmetry when the K3 volume is tuned to some appropriate value.\footnote{We recall that such an enhanced gauge symmetry can occur without the K3 developing an orbifold singularity;  a perfectly smooth K3 can lead to such behavior provided that the bundle, $B$-field, and volume are tuned to special values~\cite{Aspinwall:1996mn}.}

\subsubsection*{Remaining equivalences}
We established that there is a one-parameter family of field redefinitions that leaves $\cW_\Gamma$ invariant and leads to the equivalence $\tau^\alpha \sim\tau^\alpha + w^\alpha \xi$. We also observe that the complexified non-anomalous global symmetries $\GUL^1\times\GUL^2$ leave $\cM_{\text{uv}}$ invariant, as does the diagonal subgroup of the gauge symmetry.
Thus, near a generic point in $\cM_{\text{uv}}$, which we expect to be off the discriminant, the number of non-trivial TLSM deformations is
\begin{align}
\# (\cM_{\text{glsm}}) = 58 - 40 + 3 = 21~.
\end{align}
This value exceeds by $2$ what we naively expect from our discussion of the geometry, but as we will now argue, it is likely that this is associated with the emergence of a rank $2$ trivial bundle realized as a pair of free holomorphic Weyl fermions in the IR.  To do this, we will give an interpretation of the parameters and redefinitions suited to the LR phase.  

As a first step, we consider the $40$ parameters encoded in the $J_5$ and $J_6$ polynomials.  These are acted upon by $\cG_{\text{geo}}$, but the redefinitions that act on $\Phi_{1},\Phi_{2},\ldots,\Phi_{6}$ and $\Gamma_5,\Gamma_6$ as constant gauge transformations leave $J_5$ and $J_6$ invariant.  Thus, near a generic point in $\cM_{\text{uv}}$ the number of ``geometric'' deformations associated to deforming the complete intersection in $\cO(1,1)^{\oplus 2} \to \P^1\times\P^1$ is given by
\begin{align}
\#(\text{geometric}) = \#(J_5) + \#(J_6) - \#(\cG_{\text{geo}}) + 2 = 18~.
\end{align}
This matches the number of complex structure deformations of a K3 surface with Picard rank $2$.   Up to discrete ambiguities we can then fix the deformations of $J_5$ and $J_6$ to some fiducial form, fixing $\cG_{\text{geo}}$ up to the redefinitions that act as global gauge transformations.  

The remaining parameters, which we loosely think of as ``bundle'' deformations, are encoded in the $\cU_7$ and $\cU_8$, and they are acted upon by the remaining redefinitions in $\cG_{\text{bun}}$.  It is not hard to see that a generic $\cU_7$ and $\cU_8$ can be brought to the form
\begin{align}
\cU_7& = \Gamma_1 \Phi_1^5 + \Gamma_2\left(\Phi_2 + \kappa_1 \Phi_1\right)~,&
\cU_8 & = \Gamma_3 \Phi_3^5 + \Gamma_4 \left(\Phi_4 + \kappa_2 \Phi_3\right)~,
\end{align}
which fixes $\cG_{\text{bun}}$ up to the complexified $\GUL^1\times\GUL^2$ transformations.  The remaining parameters $\kappa_1$ and $\kappa_2$ are therefore TLSM parameters that cannot be removed by a field redefinition.  However, we know that these are redundant in the IR, at least in the LR phase because  for any generic value of $\kappa_{1,2}$ we simply find a trivial bundle over $X$. This is a phenomenon already familiar from (2,2) GLSMs: a GLSM can have parameters that cannot be removed by field redefinitions but which nevertheless do not correspond to deformations of the IR SCFT.\footnote{For instance in a (2,2) GLSM for a Calabi-Yau hypersurface in a toric variety a toric divisor class may pull back to a trivial class on the hypersurface.  In that case the corresponding complexified FI parameter decouples from the IR.  In the mirror description this decoupling can be understood via a field redefinition, but not in the original theory~\cite{Kreuzer:2010ph,Melnikov:2019tpl}.}

To summarize, we find that the $21$ deformations in $\cM_{\text{glsm}}$ can be presented as $18$ polynomial deformations of complex structure, $1$ complexified K\"ahler parameter $\tau_1 + \tau_2$, and two redundant ``gauge bundle'' deformations.

\subsection{The special locus \texorpdfstring{$f_1 = f_2 = 0$}{f1=f2=0}}
\label{subsec:speciallocus}
Returning now to our ``generic enough'' superpotential given in~(\ref{eq:Wgenericenough}), we recall that in our discussion of the LGO phase we observed that setting $f_1 = f_2 = 0$ leads to an (2,2) LG model.  Can we make sense of this locus in LR phase as well?  Setting $f_1 = f_2 =0$ does not lead to the appearance of non-compact directions in the space of supersymmetric vacua, and at first sight it seems that the LR phase simply degenerates to a hybrid theory, with base geometry $\P^1\times\P^1$.  However, on second thought this is quite confusing: the base geometry can be made arbitrarily large, and so we would expect it to contribute $c_{\sright} = 6$.  Combining this with the contribution of the compact boson and the LG fiber, it then appears that, just as in the hybrid phases, the central charge exceeds the expected $c_{\sright} = 15/2$ value.

Once again field redefinitions and symmetries shed light on this mystery.  With $f_1 = f_2 = 0$,  $\cW_{\Gamma}$ acquires two new $\GU(1)$ symmetries that act on the $\Phi_{5,6}$ and $\Gamma_{5,6}$ multiplets with charges
\begin{align*}
\arraycolsep=4pt
\begin{array}{ccccccccc}
~		& \Phi_5	&\Phi_6	&\Gamma_5	&\Gamma_6	 \\
~		&1		&0		&-2			&0			 \\
~		&0		&-1		&0			&2		
\end{array}
\end{align*}
and leave the remaining fields invariant.  The diagonal $\GU(1)$ subgroup is non-anomalous, but the anomaly in the first $\GU(1)$ means there is an additional equivalence on the FI parameters:
\begin{align}
\tau = \tau_1 + \tau_2 \sim \tau +  \zeta~.
\end{align}
Thus, on the $f_1 = f_2 = 0$ locus $\tau$ ceases to be a meaningful parameter!  Since the LGO fixed point is expected to be at finite distance in the SCFT moduli space, a plausible explanation is that when $f_1 = f_2 = 0$ the RG flow drives the TLSM to the (2,2) LGO fixed point.  The additional non-anomalous $\GU(1)$ symmetry means there is a symmetry enhancement beyond $\Le_8\oplus\Le_8$, and we will show below that in fact the spacetime gauge group at the (2,2) LGO point is $\GE_8\times\GE_8\times\SO(4)\times\GU(1)$, with all of the massless hypers charged with respect to the $\SO(4)\times\GU(1)$ factor.  The ``disappearance'' of $\tau$ on the $f_1 = f_2 = 0$ locus then has a spacetime interpretation: while we might be able to map the TLSM parameters to expectation values of the massless hypers, there is no reason for each UV parameter to map to a flat direction of the $\cN=2$ spacetime potential for the charged hypers.\footnote{This is a phenomenon that is familiar from studies of (0,2)--preserving deformations of (2,2) GLSMs, such as the GLSM for the quintic~\cite{McOrist:2011bn}.}   We will discuss this in more detail once we analyze the massless spectrum in the LGO phase, and we will argue that setting $f_1 = f_2 = 0$ or scaling $\rho\to-\infty$ while keeping the other couplings in $\cW$ fixed lead to the same IR fixed point corresponding to the vacuum with the $\GE_8\times\GE_8\times\SO(4)\times\GU(1)$ symmetry realized as a (0,2) rational SCFT.

\section{Dressed elliptic genus} \label{s:dressedellgen}
For any (0,2) SCFT in two dimensions with a non-anomalous $\GU(1)_\sleft$ symmetry, the elliptic genus is defined by~\cite{Witten:1986bf}
\begin{equation}
\label{eq:def_ell}
\mathbb{E}(\tau|z) = \text{Tr}_{\textsc{rr}}\left( (-1)^F \ee^{2\ii\pi z J_{\sleft0}} \tq^{L_0-c_\sleft/24} \bar{\tq}^{\bar{L}_0-c_\sright/24} 
\right)~,
\end{equation}
where $J_{\sleft0}$ is the zero-mode of the $\GU(1)_\sleft$ holomorphic current and $(-1)^F$ the fermion number parity operator. Being protected by supersymmetry, it is constant across the moduli space of the SCFT. Elliptic genera of (0,2) Landau--Ginzburg orbifolds have been computed long ago~\cite{Kawai:1993jk}. For a (0,2) SCFT associated with a non-linear sigma model on a Calabi--Yau manifold  endowed with a holomorphic vector bundle, the elliptic genus can be given a geometrical interpretation as the holomorphic Euler characteristic of some formal bundle~\cite{Kawai:1994qy}. More recently, the elliptic genera of (0,2) GLSMs were obtained using supersymmetric localization~\cite{Benini:2013xpa}, thereby providing further evidence that GLSMs connect LGO and large-volume limits of  heterotic Calabi--Yau compactifications.  

Most of this story carries over to heterotic flux vacua obtained as $T^2$ bundles over K3 and the associated TLSMs, with some modifications. As the right-moving fermions associated with the torus coordinates are free, the corresponding zero-modes in the functional measure cannot be saturated, and the ordinary elliptic genus~\eqref{eq:def_ell} vanishes. As explained in~\cite{Israel:2015aea,Israel:2016xfu}, one can define a {\it dressed elliptic genus}, which retains most of the properties of the ordinary elliptic genus at the expense of holomorphicity, as follows:
\begin{equation}
\label{eq:def_ell_dr}
\mathbb{E}(\tau,\bar{\tau}|z) = \frac{1}{\bar{\eta}(\bar \tau)^2} 
\text{Tr}_{\textsc{rr}}\left( J_{\sright0} (-1)^F \ee^{2\ii\pi z J_{\sleft0}} \tq^{L_0-c_\sleft/24} \bar{\tq}^{\bar{L}_0-c_\sright/24}
\right)~.
\end{equation}
The insertion of the right-moving R-current provides the necessary fermionic zero-modes. While this index is independent of the K3 moduli, it does depend on the (quantized) moduli of the two-torus. 

\subsection{Geometric formula}
The dressed elliptic genus can be expressed purely in terms of the geometrical data of the compactification, namely the principal bundle over the K3 surface $M$, the tangent bundle $T_M$ of the base and the vector bundle $\cE$ over it. One first defines the formal bundle
\begin{equation}\label{eq:formalbundle}
\mathbb{V}_{\tq,\ty} = \bigotimes_{n=0}^\infty \bigwedge\nolimits_{-\tq^n\ty } \cE^\ast \otimes 
\bigotimes_{n=1}^\infty \bigwedge\nolimits_{-\tq^n/\ty} \cE \otimes 
\bigotimes_{n=1}^\infty \Sym_{\tq^n} T_M^\ast \otimes 
\bigotimes_{n=1}^\infty \Sym_{\tq^n} T_M~,
\end{equation}
where 
\begin{align}
\bigwedge\nolimits_{\text{t}} \cE &= 1+ \text{t}\, \cE + \text{t}^2\wedge^2 \cE + \cdots~,&
\Sym_{\text{t}} T_M &= 1+ \text{t}\, T_M + \text{t}^2 \, \Sym^2T_M + \cdots~,
\end{align}
with $\wedge^\ell$ and $\Sym^\ell$ denoting respectively the $\ell$-th exterior product and the $\ell$-th symmetric product. For a six-dimensional target geometry of the factorized form $X\times S^1$, where $X$ is the total space of a principal $\GU(1)$ bundle over the surface $M$, this is supplemented by the circle bundle data: the quantized radius $r=\sqrt{k}$ and the ASD class $\upomega\in\operatorname{Pic}(M)$ specifying the fibration. In addition, the free circle is characterized by a second, unconstrained radius parametrizing the corresponding Narain CFT, whose modular invariant partition function we denote by $Z_\Gamma(\tau,\bar{\tau})$. The dressed elliptic genus~\eqref{eq:def_ell_dr} for the (0,2) NLSM associated with this geometrical data is then given by 
\begin{equation}
\label{eq:gen_index}
\mathbb{E}(\tau,\bar{\tau}|z) = 
\tq^{\frac{\text{rk}(\cE)-2}{12}} \ty^{-\text{rk}(\cE)/2} \int_M \text{ch}\, (\mathbb{V}_{\tq,\ty})\, \text{td}\, (T_M) 
\sum_{m\in\Z_{2k}}
\frac{\Theta_{m,k}\left(\tau\left|\frac{\upomega}{\ii\pi}\right.\right)}{\eta(\tau)}
\frac{\bar{\Theta}_{m,k}(\bar{\tau}|0)}{\bar{\eta}(\bar{\tau})} Z_\Gamma(\tau,\bar{\tau})~,
\end{equation}
where $\ty=\ee^{2\pi\ii z}$.

In the present case, the background is purely geometric, with a trivial rank-two bundle, as explained in section~\ref{s:tlsms}. The fibered circle has radius $r^2=3$, corresponding in the CFT to a $\Lu(1)$ affine algebra at level $3$. Factorizing the total Chern class of the K3 tangent bundle as $c(T_M) = (1+\upupsilon)(1-\upupsilon)$, the general formula~(\ref{eq:gen_index}) reduces to
\begin{equation}
\mathbb{E}(\tau,\bar \tau|z) = Z_\Gamma(\tau,\bar{\tau}) \left(\frac{\ii\vartheta_1(\tau|z)}{\eta (\tau)}\right)^2\sum_{m \in \mathbb{Z}_6} 
\frac{\bar{\Theta}_{m,3} (\bar \tau| 0)}{\bar \eta (\bar \tau)}
\int_M \upupsilon \wedge \upupsilon \left(\frac{-\ii\eta (\tau)}{\vartheta_1(\tau|\frac{\upupsilon}{2\ii\pi})}\right)^2
\frac{\Theta_{m,3} (\tau|\frac{\upomega}{\ii\pi} )}{\eta (\tau)}~.
\end{equation}
Using the same type of expansion as in~\cite{Israel:2016xfu}, which simplifies here because the gauge bundle is trivial, the integral over the surface $M$ can be carried out explicitly. We begin with the Taylor expansion
\begin{equation}
\vartheta_1(\tau|z) = 2\pi\eta(\tau)^3 z \left( 1-\frac{\pi^2}{6}E_2(\tau)z^2 +O(z^4)\right)~, 
\end{equation}
where $E_2$ is the second Eisenstein series, which is a quasi-modular form. Substituting this expansion into the integrand and using the identity~\eqref{eq:heateq}, we obtain
\begin{multline}
\mathbb{E}(\tau,\bar \tau|z) = Z_\Gamma(\tau,\bar{\tau}) \frac{\vartheta_1(\tau|z)^2}{\eta(\tau)^6} \\ 
\times \sum_{m \in \mathbb{Z}_6} \frac{\bar{\Theta}_{m,3}(\bar \tau|0)}{\bar{\eta}(\bar \tau)} 
\frac{1}{\eta(\tau)} \int_M \left(\frac{1}{12} E_2 (\tau) \,\upupsilon\wedge\upupsilon-6\,\upomega\wedge\upomega \frac{1}{2\ii\pi} \frac{\p}{\p\tau}
\right) \Theta_{m,3} (\tau|0)~.
\end{multline}
Using $c_2(T_M)=-\upupsilon.\upupsilon=24$ and $\upomega.\upomega=-8$, the integral can be evaluated to give
\begin{equation}
\label{eq:index_geom_final}
\mathbb{E}(\tau,\bar \tau|z) = 48\,
 Z_\Gamma(\tau,\bar{\tau}) \frac{\vartheta_1(\tau|z)^2}{\eta(\tau)^6} 
\sum_{m \in \mathbb{Z}_6} \frac{\bar{\Theta}_{m,3}(\bar \tau|0)}{\bar{\eta}(\bar \tau)} 
\frac{1}{\eta(\tau)} \Big(\frac{1}{2\ii\pi} \frac{\partial}{\partial \tau} - \frac{1}{24} E_2 (\tau) 
\Big) \Theta_{m,3} (\tau|0)~.
\end{equation}
In this expression we recognize the action of the Serre derivative on a modular form $f$ of weight $w$,
\begin{equation}
\text{D}(f) (\tau)\, \stackrel{\textrm{def}}{=}\, \left(\frac{1}{2\ii\pi}\frac{\partial}{\partial \tau} - \frac{w}{12} E_2 (\tau) \right) f (\tau)~,
\end{equation}
which yields a modular form $\text{D}(f)$ of weight $w+2$. Here $w=\frac{1}{2}$, consistently with the modular weight of the theta function $\Theta_{m,k}$.

\subsection{Result from supersymmetric localization}
As explained in~\cite{Israel:2015aea}, the dressed elliptic genus of a TLSM can be computed using supersymmetric localization, taking into account the cancellation between the variation of the measure and that of the field-dependent FI coupling under the action of the localization supercharge. Consider a TLSM with a rank-$r$ abelian gauge group and a $\GU(1)_\sleft^1\times\GU(1)_\sleft^2$ non-anomalous global symmetry, with chemical potentials $z_{s=1,2}$ turned on. We take the following generic field content:
\begin{itemize}
\item[-] a set of (0,2) vector mutiplets labeled by $\alpha = 1,\ldots, r$;
\item[-] a set of (0,2) chiral multiplets $\Phi_i$ of gauge and global charges $(Q^\alpha_i,q_{\sleft i}^s)$;
\item[-] a set of fermi multiplets $\Gamma_A$ of charges $(Q^\alpha_A,q_{\sleft A}^s)$;
\item[-] a shift multiplet $\Omega$ of radius $r=\sqrt{k}$, with shift charges $w^\alpha$.
\end{itemize}
In addition, the theory contains the compact boson associated with the free circle of $X\times S^1$, together with its right-moving superpartner. The dressed elliptic genus is then obtained as a Jeffrey--Kirwan (JK) residue,
\begin{align}
\mathbb{E}(\tau,\bar{\tau}|z_s) = \sum_{u^{\ast}\in\mathcal{M}_{\text{sing}}}\underset{u=u^{\ast}}{\text{JK-Res}}\left(\text{Q}(u^{\ast}),\xi\right)\,\Delta(u_1,\dots,u_r) \mathop{\wedge}_{\alpha=1}^{r}{\rm d}u_{\alpha }~,
\label{eq:local_genus}
\end{align}
where
\begin{multline}
\Delta(u_1,\dots,u_r) =
\left(-2\ii\pi\eta(\tau)^{2}\right)^r 
\prod_{\Phi_{i}}\frac{\ii\eta(\tau)}{\vartheta_1(\tau|Q^\alpha _{i}u_\alpha +q_{\sleft i}^s z_s)}
\prod_{\Gamma_{A}}\frac{\ii\vartheta_1(\tau|Q_A^\alpha u_\alpha +q_{\sleft A}^s z_s)}{\eta(\tau)} \\
\qquad\times \sum_{m\in\Z_{2k}} \frac{\Theta_{m,k} (\tau|2 w^\alpha u_\alpha)}{\eta (\tau)} 
\frac{\bar{\Theta}_{m,k} (\bar \tau| 0)}{\bar \eta (\bar \tau)} Z_\Gamma(\tau,\bar{\tau})
\end{multline}
is the one-loop determinant in the space $\mathfrak{M}$ of gauge holonomies; see~\cite{Benini:2013xpa,Israel:2015aea} for details. More explicitly, the determinants of the chiral multiplets have poles in $\mathfrak{M}$ along the hyperplanes defined by $H_i = \{ Q_i^\alpha u_\alpha + q_{\sleft i}^s z_s = 0 \mod \mathbb{Z} + \tau \mathbb{Z} \}$. We denote then by $\mathfrak{M}_{\text{sing}}^\ast$ the set of isolated points in $\mathfrak{M}$ where at least $r$ such hyperplanes meet, and by $\text{Q}(u^\ast)$ the set of charges of chiral multiplets giving poles at $u^\ast$. Finally, $\xi$ specifies a choice of chamber that will provide expressions of the (dressed) elliptic genus suitable for interpretation in various phases, geometric, Landau--Ginzburg or hybrid if any.\footnote{In abelian GLSMs, we may be tempted to identify the parameters $\xi^\alpha$ specifiying the chamber with the FI couplings $\tau^\alpha$, given that the chambers are in one-to-one correspondance to the various phases of the GLSM. However there is no justification for such identification, since the FI couplings in the GLSM action are exact with respect to the localization supercharge~\cite{Benini:2013xpa}.}

In the present case, we obtain from the charges in table~\ref{table:charges} the following one-loop determinant, as a function of the two gauge holonomies $(u_1,u_2)$:
\begin{multline}
\Delta (u_1,u_2) = \left(-2\ii\pi\eta(\tau)^{2}\right)^{2} Z_\Gamma(\tau,\bar{\tau}) \notag \\ \times \! 
\left(\frac{\ii\eta(\tau)}{\vartheta_1(\tau|u_1)}\right)^2\left(\frac{\ii\eta(\tau)}{\vartheta_1(\tau|u_2)}\right)^2
\left(\frac{\ii\eta(\tau)}{\vartheta_1(\tau|u_1 + u_2)}\right)^2
\frac{\ii\eta(\tau)}{\vartheta_1(\tau|-6u_1+z_1)}
\frac{\ii\eta(\tau)}{\vartheta_1(\tau|-6u_2+z_2)} \notag \\ \times \!
\frac{\ii\vartheta_1(\tau|u_1-z_1)}{\eta(\tau)} \frac{\ii\vartheta_1(\tau|5 u_1-z_1)}{\eta(\tau)}
\frac{\ii\vartheta_1(\tau|u_2-z_2)}{\eta(\tau)} \frac{
\ii\vartheta_1(\tau|5 u_2-z_2)}{\eta(\tau)}
\left( \frac{\ii\vartheta_1(\tau|-2u_1-2u_2)}{\eta(\tau)}\right)^2\notag \\ \times \, 
\sum_{m \in \mathbb{Z}_6} 
\frac{\Theta_{m,3} (\tau|2 u_1-2u_2)}{\eta (\tau)} 
\frac{\bar{\Theta}_{m,3} (\bar \tau| 0)}{\bar \eta (\bar \tau)}~.
\end{multline}

\subsubsection*{JK residue in the LR chamber}
In order to express the dressed elliptic genus in a form adapted to the geometric LR phase of the TLSM, we compute the JK residue at the degenerate intersection of the hyperplanes
\begin{align}
H_{1,2} \; &: \; \{ u_1 = 0 \mod \mathbb{Z} + \tau \mathbb{Z} \}~,\nonumber\\
H_{3,4} \; &: \; \{ u_2 = 0 \mod \mathbb{Z} + \tau \mathbb{Z} \}~,\nonumber\\
H_{5,6} \; &: \; \{ u_1+u_2 = 0 \mod \mathbb{Z} + \tau \mathbb{Z} \}~,
\end{align}
which reduces to the single point $u_1=u_2=0$. The computation of the JK residue at such a degenerate intersection requires some care; see~\cite{Benini:2013xpa}. Following the notation of that reference, we consider the flags 
\begin{align}
F &= \{ 0 \subset \text{Span}(Q_1) \subset \mathbb{R}^2 \} ~,&
F' &= \{ 0 \subset \text{Span}(Q_3) \subset \mathbb{R}^2 \} ~,&
F'' &= \{ 0 \subset \text{Span}(Q_5) \subset \mathbb{R}^2 \}~,
\end{align}
and their associated ``kappa-vectors''
\begin{align}
\kappa &= \{Q_1,Q_1+Q_2+Q_3 \}&
\kappa' &= \{Q_3,Q_1+Q_2+Q_3 \}&
\kappa'' &= \{Q_5, Q_1+Q_2+Q_3 \} 
\nonumber\\
&= \{ (1,0), \, (2,2) \}~,&
&= \{ (0,1), \, (2,2) \}~,&
& = \{ (1,1), \, (2,2) \}~, 
\end{align}
with
\begin{align}
\nu (F) &= 1,&
\nu (F') &= -1 ~,&
\nu (F'') &= 0~.
\end{align}
This determines the JK prescription in the chamber $\xi^1,\xi^2 >0$. In the sub-chamber 
\begin{equation}
s_+ (F) = \text{Cone} \big((1,0), \, (2,2) \big) = \{ \xi^1 >\xi^2 >0 \}~,
\end{equation} 
the JK residue is given by the iterated residue
\begin{equation}
\label{eq:it_res}
\mathbb{E}(\tau,\bar{\tau}|z_1,z_2) = + \operatorname*{Res}_{u_2=0} \left(\operatorname*{Res}_{u_1=0}\, \Delta (u_1,u_2) \right)~,
\end{equation}
while in the sub-chamber 
\begin{equation}
s_+ (F') = \text{Cone} \big((0,1), \, (2,2) \big) = \{ \xi^2 > \xi^1 >0 \}~,
\end{equation} 
we instead have
\begin{equation}
\mathbb{E}(\tau,\bar{\tau}|z_1,z_2) = - \operatorname*{Res}_{u_1=0} \Big( \operatorname*{Res}_{u_2=0} \,\Delta (u_1,u_2) \Big) ~.
\end{equation}
The prescription does not apply at $\xi^1=\xi^2$ (i.e. at $s_+ (F'')$). However, the JK residue does not jump across boundaries between sub-chambers so the result can be obtained by evaluating it in either the $s_+(F)$ or the $s_+(F')$ sub-chamber.

Let us perform the computation in the sub-chamber $\xi^1>\xi^2$, for which the JK prescription reduces to the iterated residue~\eqref{eq:it_res}. To simplify the notation, we factor out the non-holomorphic contributions to the one-loop determinant and write
\begin{align}
\Delta (u_1,u_2) = Z_\Gamma(\tau,\bar{\tau})\sum_{m\in\Z_6} \frac{\bar{\Theta}_{m,3}(\bar{\tau}|0)}{\bar{\eta}(\bar{\tau})}\Delta_m(u_1,u_2)~.
\end{align}
We will leave the dependence on $\tau$ implicit in the intermediate expressions. The first step is to evaluate the residue in the $u_1$-plane. The integrand $\Delta_m(u_1,u_2)$ has a double pole at $u_1=0$, originating from the factor $1/\vartheta_1(u_1)^2$. One finds
\begin{align}
\mathop{\operatorname{Res}}_{u_1=0}\Delta_m (u_1,u_2)  &= \frac{1}{\eta} 
\frac{\vartheta_1(u_2-z_2)\vartheta_1(5 u_2-z_2)}{\vartheta_1(u_2)^2\vartheta_1(-6u_2+z_2)}
\nonumber\\ &\quad \times 
\frac{\p}{\p u_1} \left[ \frac{\vartheta_1(u_1-z_1)\vartheta_1(5 u_1-z_1)}{\vartheta_1(-6u_1+z_1)} \frac{\vartheta_1(-2u_1-2u_2)^2}{\vartheta_1(u_1+u_2)^2}\Theta_{m,3} (2u_1-2u_2)\right]_{u_1=0}~.
\end{align}
The dependence on $z_1$ is entirely contained in the first three Jacobi theta functions in the bracket. These satisfy the identity
\begin{align}
&\frac{\p}{\p u_1}\left[\frac{\vartheta_1(u_1-z_1)\vartheta_1(5 u_1-z_1)}{\vartheta_1(-6u_1+z_1)}\right]_{u_1=0}= 0~,
\label{eq:Jacobifactorization}
\end{align}
so that the resulting contribution is an overall multiplicative factor $\vartheta_1(z_1)$. For the remaining terms, one obtains
\begin{align}
\frac{\p}{\p u_1} \left[\frac{\vartheta_1(-2u_1-2u_2)^2}{\vartheta_1(u_1+u_2)^2}\right]_{u_1=0} &= 2\frac{\vartheta_1(2u_2)^2}{\vartheta_1(u_2)^2} \lambda(u_2)~,
\end{align}
where we introduced the function
\begin{align}
\lambda(u)= 2\frac{\vartheta_1'(2u)}{\vartheta_1(2u)}-\frac{\vartheta_1'(u)}{\vartheta_1(u)}~.
\end{align}
The residue at $u_1=0$ then takes the form
\begin{align}
&\mathop{\operatorname{Res}}_{u_1=0}\Delta_m (u_1,u_2)  = \nonumber\\
&\qquad\frac{2\vartheta_1(z_1)}{\eta}
\frac{\vartheta_1(u_2-z_2)\vartheta_1(5 u_2-z_2)\vartheta_1(2u_2)^2}{\vartheta_1(-6u_2+z_2)\vartheta_1(u_2)^4}
\left(\Theta_{m,3}' (-2u_2)+\lambda(u_2)\Theta_{m,3}(-2u_2)\right)~.
\end{align}
It remains to evaluate the residue at $u_2=0$. The function $\lambda(u)$ is regular at the origin, with expansion
\begin{align}
\lambda(u) &=-\pi^2 E_2\, u+O(u^3)~,
\end{align}
so the integrand has a double pole at $u_2=0$, due to the factor $\vartheta_1(2u_2)^2/\vartheta_1(u_2)^4$. Evaluating the residue gives
\begin{align}
\mathop{\operatorname{Res}}_{u_2=0}&\mathop{\operatorname{Res}}_{u_1=0}\Delta_m(u_1,u_2)  = 
\nonumber\\
& \frac{2\vartheta_1(z_1)}{\pi^2\eta^7}\frac{\p}{\p u_2}\left[
\frac{\vartheta_1(u_2-z_2)\vartheta_1(5 u_2-z_2)}{\vartheta_1(-6u_2+z_2)} 
 \left( \Theta_{m,3}' (-2u_2)+\lambda(u_2)\Theta_{m,3} (-2u_2)\right) \right]_{u_2=0}~.
\end{align}
Using again the identity~\eqref{eq:Jacobifactorization}, the $z_2$ dependence factorizes into an overall factor $\vartheta_1(z_2)$, yielding
\begin{align}
\mathop{\operatorname{Res}}_{u_2=0}\mathop{\operatorname{Res}}_{u_1=0}\Delta (u_1,u_2)  &= \frac{2\,\vartheta_1(z_1)\vartheta_1(z_2)}{\pi^2\eta^7}\left(
-2\Theta_{m,3}'' (0) -\pi^2 E_2\,\Theta_{m,3} (0)\right)~.
\end{align}
Finally, using the heat equation~\eqref{eq:heateq} obeyed by the theta functions $\Theta_{m,3}$, the result can be rewritten as
\begin{align}
\mathop{\operatorname{Res}}_{u_2=0}\mathop{\operatorname{Res}}_{u_1=0}\Delta_m (u_1,u_2)  &= \frac{48}{\eta^7} \vartheta_1(z_1)\vartheta_1(z_2) \left(
\frac{1}{2\pi\ii}\frac{\p}{\p\tau}-\frac{E_2}{24}  \right)\Theta_{m,3} (0)~.
\end{align}
Comparing this expression with the geometric formula~\eqref{eq:index_geom_final}, we find exact agreement upon identifying $z=z_1=z_2$. The localization computation of the dressed elliptic genus in the LR phase therefore reproduces the geometric result. It provides evidence that the interpretation of the LR phase (an $S^1$ fibered over K3 with a trivial gauge bundle) is correct. 

\subsubsection*{JK residue in the LGO chamber}
To obtain a representation of the dressed elliptic genus adapted to the LGO phase of the TLSM, we choose the JK residue associated with the intersection of the two hyperplanes
\begin{align}
H_7 \, &: \ \{ -6u_1 + z_1 = 0 \mod \mathbb{Z} + \tau \mathbb{Z} \}~,&
H_8 \, &: \ \{ -6u_2 + z_2 = 0 \mod \mathbb{Z} + \tau \mathbb{Z} \}~.
\end{align}
We parameterize their intersections as
\begin{align}
-6u_1 + z_1 & = - a_1\tau - b_1~,&
-6u_2 + z_2 & = - a_2\tau - b_2~,
\end{align}
with $a_s,b_s \in \Z_6$. The dressed elliptic genus is then given by
\begin{equation}
\mathbb{E} (\tau,\bar \tau|z_1,z_2) = \frac{1}{(2\ii\pi)^2} \sum_{a_1,b_1,a_2,b_2 \in \mathbb{Z}_6} 
\oint_{\frac{1}{6}(z_1+a_1\tau + b_1)}\!\!\!\!\!\!  \dd u_1 \oint_{\frac{1}{6}(z_2+a_2\tau + b_2)} \!\!\!\!\!\!   \dd u_2  \ 
\Delta (u_1,u_2) ~.
\end{equation}
Evaluating the residues using~\eqref{eq:theta_shift} and~\eqref{eq:theta_resid}, one obtains
\begin{align}
\mathbb{E}&(\tau,\bar{\tau}|z_1,z_2) = \frac{1}{6} \sum_{a_1,b_1\in \mathbb{Z}_6} \frac{1}{6} \sum_{a_2,b_2 \in \mathbb{Z}_6}
(-1)^{a_1+b_1+a_2+b_2} \tq^{\frac{1}{2}(a_1^2+a_2^2)}
\nonumber\\
&\times \frac{ \vartheta_1(\tau|\frac{z_1+a_1\tau + b_1}{6}-z_1) \vartheta_1(\tau|5\frac{z_1+a_1\tau + b_1}{6}-z_1) \vartheta_1(\tau|\frac{z_2+a_2\tau + b_2}{6}-z_2) \vartheta_1(\tau|5\frac{z_2+a_2\tau +b_2}{6}-z_2)}{\vartheta_1(\tau|\frac{z_1+a_1\tau + b_1}{6})^2\,\vartheta_1(\tau|\frac{z_2+a_2\tau + b_2}{6})^2} \nonumber\\
&\times \frac{ \vartheta_1(\tau|-\frac{z_1+z_2+(a_1+a_2)\tau + (b_1+b_2)}{3})^2 }{\vartheta_1(\tau|\frac{z_1+z_2+(a_1+a_2)\tau + (b_1+b_2)}{6})^2} \sum_{m\in \Z_6} \frac{\Theta_{m,3} (\tau|\frac{z_1-z_2+(a_1-a_2)\tau + (b_1-b_2)}{3})}{\eta(\tau)} 
\frac{\bar{\Theta}_{m,3}(\bar{\tau}|0)}{\bar{\eta}(\bar{\tau})}~.
\end{align}
This expression can be simplified using the identity~(\ref{eq:theta_shift}) and performing the change of basis
\begin{align}
a&=a_1\mod6~,&
c&=a_2-a_1\mod6~,\nonumber\\
b&=b_1\mod6~,&
d&=b_2-b_1\mod6~.
\end{align}
After applying the corresponding shifts, the dependence associated with $(\Phi_2,\Gamma_2)$ and $(\Phi_4,\Gamma_4)$ cancels, and the elliptic genus takes the form
\begin{multline}
\label{eq:ellip_LGO}
\mathbb{E} (\tau,\bar \tau| z_1,z_2) = 
 \frac{1}{6} \sum_{a,b\in \mathbb{Z}_6} 
 \ee^{-\ii\pi a (z_1+z_2)}\ 
 \textcolor{purple}{\frac{1}{6} \sum_{c,d \in \mathbb{Z}_6}}
\textcolor{purple}{
 \ee^{-\ii\pi a (c\tau +d)}
\ee^{\frac{2\ii\pi c}{6} (z_2+c\tau +d+a\tau +b ) }} \\
\underbrace{\left(\frac{\vartheta_1(\tau|\frac{z_1}{6}-z_1+\frac{a\tau + b}{6})}{\vartheta_1(\tau|\frac{z_1}{6}+\frac{a\tau + b}{6})}\right)}_{(\Phi_1,\Gamma_1)}
\underbrace{\left(\frac{\vartheta_1(\tau|\frac{z_2}{6}-z_2 +\frac{a\tau + b}{6} \textcolor{purple}{+\frac{c\tau + d}{6}})}{
\vartheta_1(\tau|\frac{z_2}{6}+\frac{a\tau + b}{6} \textcolor{purple}{+\frac{c\tau + d}{6}})}\right)}_{(\Phi_3,\Gamma_3)}\\
\underbrace{\left(\frac{\vartheta_1(\tau|\frac{z_1+z_2}{6} -\frac{z_1+z_2}{2} + \frac{a\tau + b}{3} \textcolor{purple}{-\frac{c\tau + d}{3}} )}{
\vartheta_1(\tau|\frac{z_1+z_2}{6}+ \frac{a\tau + b}{3}  \textcolor{purple}{+\frac{c\tau + d}{6})}}\right)^2}_{
(\Phi_5,\Gamma_5)\ \text{and}\ (\Phi_6,\Gamma_6)}
\underbrace{\sum_{m \in \mathbb{Z}_6} 
\frac{\Theta_{m,3} \big(\tau|\frac{z_1-z_2}{3}\textcolor{purple}{-\frac{c\tau +d}{3}}\big)}{\eta (\tau)} 
\frac{\bar{\Theta}_{m,3} (\bar \tau| 0)}{\bar \eta (\bar \tau)}}_{\text{circle}}~.
\end{multline}
To understand the content of this expression, let us first consider the $(c,d)=(0,0)$ sector of the second $\mathbb{Z}_6$ orbifold. We only keep track of the diagonal $\GU(1)_\sleft$ global symmetry by setting the fugacities to $z=z_1=z_2$. In this sector, we find
\begin{align}
\mathbb{E}_{(0,0)}(\tau,\bar{\tau}|z) &= \sum_{m \in \Z_6} \frac{\Theta_{m,3}(\tau|0)}{\eta(\tau)} \frac{\bar{\Theta}_{m,3}(\bar{\tau}|0)}{\bar{\eta}(\bar{\tau})} \nonumber\\
&\quad \times\frac{1}{6} \sum_{a,b\in\Z_6} \ee^{-2\ii\pi a z}
\underbrace{ \left(\frac{\vartheta_1(\tau|\frac{z}{6}-z+\frac{a\tau + b}{6})}{\vartheta_1(\tau|\frac{z}{6}+\frac{a\tau + b}{6})}\right)^2 }_{(\Phi_1,\Gamma_1)\text{ and }(\Phi_3,\Gamma_3)}
\underbrace{ \left(\frac{\vartheta_1(\tau|\frac{z}{3}-z+\frac{a\tau + b}{3})}{\vartheta_1(\tau|\frac{z}{3}+\frac{a\tau + b}{3})}\right)^2}_{(\Phi_5,\Gamma_5)\text{ and }(\Phi_6,\Gamma_6)}~.
\end{align}
The result factorizes into the partition function of a free compact boson (first line) and the elliptic genus of a (2,2) K3 SCFT (second line). The latter appears in the same presentation as one would obtain from the Gepner model corresponding to the diagonal $\mathbb{Z}_6$ orbifold of: 
\begin{enumerate}
\item a pair of minimal models with $k=6$, corresponding to $(\Phi_1,\Gamma_1)$ and $(\Phi_3,\Gamma_3)$;
\item a pair of minimal models with $k=3$, corresponding to $(\Phi_5,\Gamma_5)$ and $(\Phi_6,\Gamma_6)$.
\end{enumerate}
On top of this, the second $\mathbb{Z}_6$ orbifold, highlighted in color in~\eqref{eq:ellip_LGO}, acts in three distinct ways:
\begin{enumerate}
\item a $\mathbb{Z}_6$ action on $(\Phi_3,\Gamma_3)$ preserving $(2,2)$ supersymmetry;
\item a $\mathbb{Z}_6$ action on $(\Phi_5,\Gamma_5)$ and $(\Phi_6,\Gamma_6)$ breaking $(2,2)$ worldsheet supersymmetry to $(0,2)$, 
as the fermi and chiral multiplets carry distinct orbifold charge (respectively $-2$ and $1$);
\item a $\mathbb{Z}_6$ shift in the left-moving lattice of the compact free boson.
\end{enumerate}
This interpretation is consistent with the phase analysis made in section~\ref{s:tlsms} and will be further confirmed in the next section by an explicit analysis of the LGO phase of the linear model.

\section{The LG orbifold phase} \label{s:lgo}
We now turn to the LG phase of the linear model. As discussed in the phase analysis above, for $\rho\ll0$ the low energy degrees of freedom are described by the product of a LG model and a circle CFT, orbifolded by a discrete group. For a particular choice of superpotential couplings, the LG theory becomes exactly solvable and admits a description as a Gepner SCFT, with a non-standard orbifold symmetry group that includes an action on the quantized circle. The Gepner orbifold leads to a dramatic enhancement of the spacetime gauge symmetry and, quite remarkably, to the appearance of free left-moving fermions in the SCFT spectrum, as anticipated from the non-linear sigma model description in the large radius phase. This solvable description will allow us to determine explicitly the (0,2) marginal operators and identify them with gauge-invariant operators in the linear model. We will further comment on the deformations that take the theory away from the LG phase.

\subsection{The \texorpdfstring{$\Z_6\times\Z_6$}{Z6xZ6} orbifold}
In the LG phase the chiral superfields $\Phi_7$ and $\Phi_8$ acquire expectation values, which may be set to one by a super-gauge transformation. The $\GU(1)\times\GU(1)$ gauge symmetry of the linear model is thereby broken to the $\Z_6\times\Z_6$ subgroup that leaves these expectation values invariant. The low energy degrees of freedom consist of an equal number of chiral and fermi multiplets, with (0,2) superpotential
\begin{align}
\cW_{\text{LG}} = \Gamma_1\Phi_1^5+\Gamma_2\Phi_2+\Gamma_3\Phi_3^5+\Gamma_4\Phi_4+\Gamma_5\Phi_5^2+\Gamma_6\Phi_6^2 -\Gamma_5f_1(\Phi)-\Gamma_6f_2(\Phi)~,
\label{eq:LGsuperpotential}
\end{align}
together with the degrees of freedom associated with the shift multiplet. These comprise a compact scalar of radius $r=\sqrt{3}$, a free non-compact scalar, and two free right-moving Majorana--Weyl fermions. Since the orbifold acts only on the compact scalar, the non-compact scalar and the free right-moving fermions are decoupled---we will therefore leave them implicit below. The non-trivial part of the SCFT is thus of the form
\begin{align}
(\cC_{\text{LG}}\times S^1)/(\Z_6\times\Z_6)~,
\end{align}
where $S^1$ denotes the Narain CFT of the circle at the quantized radius $r$. The orbifold action is chiral. Denoting its generators by $g_1$ and $g_2$, their action on the superfields is
\begin{align*}
\arraycolsep=4pt
\begin{array}{cccccccccccccccccc}
& \Phi_1 & \Phi_2 & \Phi_3 & \Phi_4 & \Phi_5 & \Phi_6 & \Gamma_1 & \Gamma_2 & \Gamma_3 & \Gamma_4 & \Gamma_5 & \Gamma_6 & S^1 \\
g_1 & \zeta & \zeta & 1 & 1 & \zeta & \zeta & \zeta & \zeta^5 & 1 & 1 & \zeta^4 & \zeta^4 & g_\cir \\
g_2 & 1 & 1 & \zeta & \zeta & \zeta & \zeta & 1 & 1 & \zeta & \zeta^5 & \zeta^4 & \zeta^4 & g_\cir^{-1} \\
\end{array}
\end{align*}
where $\zeta =\ee^{\frac{2\pi\ii}{6}}$. It is useful to regard each pair $(\Phi_i,\Gamma_i)$ as a (2,2) chiral multiplet. In this language, the LG superpotential~\eqref{eq:LGsuperpotential} describes a (0,2) deformation of the (2,2) locus defined by $f_1=f_2=0$. Even on this locus, however, as a consequence of the asymmetric action of the orbifold on left- and right-movers, the theory preserves only (0,2) supersymmetry.

The $\Z_6\times\Z_6$ action on the circle CFT is an asymmetric shift orbifold, generated by
\begin{align}
g_\cir = \exp(2\pi\ii J^\cir_{\sleft 0})~,
\label{eq:circleaction}
\end{align}
where $J^\cir_\sleft$ is the left-moving current of the circle CFT. We normalize this current so that its charges satisfy $q^\cir_\sleft\in\frac{1}{6}\Z$, with $6q^\cir_\sleft$ taking primitive integer values. Thus $g_\cir$ generates an effective $\Z_6$ symmetry. There are several ways to understand the form of this action. First, the coupling of the shift multiplet to the (0,2) vector multiplets in the TLSM is asymmetric: while the kinetic term is canonical, the axial coupling of $\Omega$ to the super-gauge curvatures distinguishes the left- and right-moving components of the abelian gauge fields. More directly, an orbifold acting only on the LG degrees of freedom would be inconsistent: the $\Z_6\times\Z_6$ symmetry is the remnant of the gauge symmetry of the linear model, and its action on the LG theory is thus anomalous. The anomaly must therefore be canceled by the action on the circle: this requires an asymmetric shift, involving the winding modes. Indeed, if $\cV_{\text{n},\text{w}}$ is a circle vertex operator with momentum and winding numbers $\text{n},\text{w}\in\Z$, then
\begin{align}
g_\cir \circ\cV_{\text{n},\text{w}} = \zeta^{\text{n}+3\text{w}}\,\cV_{\text{n},\text{w}}~.
\end{align}
The winding contribution is precisely what is required to compensate the anomaly of the LG orbifold. The same action is confirmed by the dressed elliptic genus computed in the previous section: in the LG expression for the dressed elliptic genus~\eqref{eq:ellip_LGO}, the circle contribution supplies exactly the phase associated with \eqref{eq:circleaction}.

The orbifold action on the circle leads to a non-standard realization of the GLSM global symmetries in the LG phase. Indeed, in the basis of $\GU(1)_\sleft^1\times\GU(1)_\sleft^2\times\GU(1)_\sright$ charges introduced in the previous section, the superfields $\Phi_7$ and $\Phi_8$ carry non-zero charges. Since the GLSM global symmetries are defined only up to global gauge transformations, we may redefine the charges as
\begin{align}
q_\sleft^1 &\to q_\sleft^1+\tfrac{1}{6}Q^1~,&
q_\sleft^2 &\to q_\sleft^2+\tfrac{1}{6}Q^2~,&
q_\sright &\to q_\sright+\tfrac{1}{6}(Q^1+Q^2)~,
\end{align}
where $Q^1,Q^2$ denote the gauge charges. In this basis $\Phi_7$ and $\Phi_8$ are neutral and may consistently be set to one. The resulting charge assignment is given in the following table.
\begin{align*}
\arraycolsep=4pt
\begin{array}{ccccccccccccccc}
& \Phi_1 & \Phi_2 & \Phi_3 & \Phi_4 & \Phi_5 & \Phi_6 & \Gamma_1 & \Gamma_2 & \Gamma_3 & \Gamma_4 & \Gamma_5 & \Gamma_6 & \theta & S^1 \\
q_\sleft^1 & \frac{1}{6} & \frac{1}{6} & 0 & 0 & \frac{1}{6} & \frac{1}{6} & -\frac{5}{6} & -\frac{1}{6} & 0 & 0 & -\frac{1}{3} & -\frac{1}{3} & 0 & +q^\cir_\sleft \\
q_\sleft^2 & 0 & 0 & \frac{1}{6} & \frac{1}{6} & \frac{1}{6} & \frac{1}{6} & 0 & 0 & -\frac{5}{6} & -\frac{1}{6} & -\frac{1}{3} & -\frac{1}{3} & 0 & -q^\cir_\sleft \\
q_\sright & \frac{1}{6} & \frac{1}{6} & \frac{1}{6} & \frac{1}{6} & \frac{1}{3} & \frac{1}{3} & \frac{1}{6} & \frac{5}{6} & \frac{1}{6} & \frac{5}{6} & \frac{1}{3} & \frac{1}{3} & 1 & 0 \\
\end{array}
\end{align*}
The redefinition makes the $\GU(1)_\sleft^1\times\GU(1)_\sleft^2$ symmetry act non-trivially in the circle CFT. This description also makes the orbifold action transparent. At low energies the global symmetries are realized as a $\Lu(1)_\sleft^1\oplus\Lu(1)_\sleft^2\oplus\Lu(1)_\sright$ current algebra. Here, $\Lu(1)_\sright$ is the R-symmetry of the right-moving N=2 superconformal algebra (SCA) associated to the LG model,\footnote{The corresponding N=2 SCA has central charge $c_\sright=6$ and is enhanced to an N=4 superconformal symmetry with R-symmetry $\su(2)_\sright\supset\Lu(1)_\sright$. This should be distinguished from the N=2 SCA of the SCFT associated with the full target space $X_5\times\R$, which has central charge $c_\sright=9$. This splitting of the superconformal symmetry is consistent with the non-linear sigma model analysis of~\cite{Melnikov:2010pq} and with general expectations for extended spacetime supersymmetry in four dimensions~\cite{Banks:1988yz}: the right-moving algebra must decompose into a $c_\sright=6$ N=4 SCA together with a free $c_\sright=3$ N=2 SCA.} while $\Lu(1)_\sleft^1\oplus\Lu(1)_\sleft^2$ is a left-moving current algebra. Denoting the corresponding left-moving currents by $J_\sleft^1$ and $J_\sleft^2$, the orbifold generators can be written as
\begin{align}
g_1 &= \exp(2\pi\ii J^1_{\sleft0})~,&
g_2 &= \exp(2\pi\ii J^2_{\sleft0})~.
\end{align}

As a final general observation, the LG superpotential~\eqref{eq:LGsuperpotential} contains mass terms for the multiplets $(\Phi_2,\Gamma_2)$ and $(\Phi_4,\Gamma_4)$. Their masses are proportional to the expectation values of $\Phi_7$ and $\Phi_8$, respectively, which we have set to one in~\eqref{eq:LGsuperpotential} for notational convenience. Deep in the LGO phase these expectation values become arbitrarily large, and the massive multiplets can be integrated out. We will therefore omit them in what follows. The $\Z_6\times\Z_6$ action on the remaining superfields is still chiral, but the two combinations
\begin{align}
g &= g_1 g_2~,&
g' &= g_1 g_2^{-1}~,
\end{align}
act symmetrically on the left- and right-moving massless fermions. The diagonal combination $g$ acts only on the LG degrees of freedom and implements the projection onto operators of integral R-charge. The anti-diagonal combination $g'$ combines a non-chiral action on the (2,2) multiplets $(\Phi_i,\Gamma_i)$ with a shift on the circle. For later convenience, we also introduce the corresponding left-moving currents
\begin{align}
J_\sleft &= J_\sleft^1+J_\sleft^2~,&
J_\sleft' &= J_\sleft^1-J_\sleft^2~.
\end{align}

\subsection{Gepner locus}
At a locus in the GLSM parameter space $\cM_{\text{uv}}$, the LG phase admits a description as a Gepner orbifold. Indeed, starting from the superpotential~\eqref{eq:LGsuperpotential}, we may deform all coefficients in the polynomials $f_1$ and $f_2$ to zero. The superpotential then takes the form
\begin{align}
\cW_{\text{LG}} = \Gamma_1\Phi_1^5+\Gamma_3\Phi_3^5+\Gamma_5\Phi_5^2+\Gamma_6\Phi_6^2~,
\label{eq:Gepnersuperpotential}
\end{align}
and is in fact (2,2) supersymmetric: it is the Fermat superpotential for the $(2,2)$ chiral multiplets $(\Phi_i,\Gamma_i)$. In the IR the LG theory flows to a product of N=2 minimal models,
\begin{align}
\cC_{\text{LG}} = \cA_6\times\cA_6\times\cA_3\times\cA_3~,
\label{eq:Gepnermodel}
\end{align}
where $\cA_k$ denotes the minimal model at level $k$, with total central charge
\begin{align}
c_\sleft(\cC_{\text{LG}})=c_\sright(\cC_{\text{LG}})=6~.
\end{align}

At the Gepner locus the IR SCFT has an enhanced current algebra. The corresponding symmetries are already visible in the GLSM: when the polynomials $f_1$ and $f_2$ are set to zero, the GLSM acquires additional global symmetries. Their action on the IR theory is transparent from the explicit description~\eqref{eq:Gepnermodel}. Each non-trivial minimal model factor contributes a left- and a right-moving $\Lu(1)$ current, which we denote by $J_\sleft^{\mm,i}$ and $J_\sright^{\mm,i}$, respectively. The circle CFT contributes one additional current of each chirality, denoted by $J_\sleft^\cir$ and $J_\sright^\cir$. Thus $\cC_{\text{LG}}\times S^1$ contains a $\Lu(1)^{\oplus 5}$ current algebra in both the left- and right-moving sectors. We normalize the currents so that their non-vanishing two-point functions are
\begin{align}
z_{12}^2\la J^{\text{min},i}_\sleft(z_1) J^{\text{min},i}_\sleft(z_2) \ra &= \frac{k_i-2}{k_i}~,&
z_{12}^2\la J^\cir_\sleft(z_1) J^\cir_\sleft(z_2)\ra &= \frac{1}{6}~,
\end{align}
with analogous expressions for the right-moving currents. The $\GU(1)_\sleft^1\times\GU(1)_\sleft^2$ global symmetry of the GLSM is realized in the IR by the linear combinations
\begin{align}
J_\sleft^1 &= J_\sleft^{\text{min},1}+\tfrac{1}{2}J_\sleft^{\text{min},5}+\tfrac{1}{2}J_\sleft^{\text{min},6}+J_\sleft^\cir~,
\nonumber\\
J_\sleft^2 &= J_\sleft^{\text{min},3}+\tfrac{1}{2}J_\sleft^{\text{min},5}+\tfrac{1}{2}J_\sleft^{\text{min},6}-J_\sleft^\cir~.
\end{align}
Using the two-point functions above, one finds
\begin{align}
J_\sleft^1(z_1)J_\sleft^1(z_2)&\sim \frac{1}{z_{12}^2}~,&
J_\sleft^2(z_1)J_\sleft^2(z_2)&\sim \frac{1}{z_{12}^2}~,&
J_\sleft^1(z_1)J_\sleft^2(z_2)&\sim 0~.
\end{align}
The levels agree with the 't Hooft anomalies of the UV linear model.

\subsubsection*{Partition function of the Gepner orbifold}
The IR SCFT is rational at the Gepner locus. Its left-moving representations in the NS sector can be labeled by a finite set of indices
\begin{align}
\boldsymbol{l} &= (l_1,l_3,l_5,l_6)~\in\{0,\dots,4\}^2\times\{0,1\}^2~,\nonumber\\
\boldsymbol{m} &= (m_1,m_3,m_5,m_6)~\in(\Z_{12})^2\times(\Z_6)^2~,
\end{align}
subject to the usual field identifications, reviewed in appendix~\ref{app:minimalmodels}. We have omitted the labels associated with the two trivial factors, since they are fixed to $(l_2,m_2)=(l_4,m_4)=(0,0)$. We denote the corresponding characters by $\chi^{\boldsymbol{l}}_{\boldsymbol{m}}(\tau)$. The circle CFT is rational as well at the special radius $r^2=3$, with six primary representations of the extended chiral algebra, labeled by $m\in\Z_6$. In the supergravity description, this rationality follows necessarily from quantization of the $H$-flux~\cite{Melnikov:2012cv}. As we will see, it also plays a central role in the LGO phase.

The $\Z_6\times\Z_6$ action is chiral: it acts asymmetrically on the (0,2) components of the multiplets $(\Phi_5,\Gamma_5)$ and $(\Phi_6,\Gamma_6)$. In terms of the charges of the minimal model and circle factors, the two generators act by phases
\begin{align}
\ee^{2\pi\ii q_\sleft^1} &= \ee^{2\pi\ii (q_\sleft^{\mm,1}+ q_\sleft^\cir)+\pi\ii( q_\sleft^{\mm,5}+q_\sleft^{\mm,6})}~,&
\ee^{2\pi\ii q_\sleft^2} &= \ee^{2\pi\ii (q_\sleft^{\mm,3}-q_\sleft^\cir)+\pi\ii( q_\sleft^{\mm,5}+q_\sleft^{\mm,6})}~.
\end{align}
Consider, for example, the minimal model associated with $(\Phi_5,\Gamma_5)$. The two components of a supersymmetry multiplet do not have the same value of $\exp(\pi\ii q_\sleft^{\mm,5})$, since the supercurrents $G_{\mm,5}^{\pm}$ of the minimal model carry charges $q_\sleft^{\mm,5}=\pm1$. The same applies to the minimal model associated with $(\Phi_6,\Gamma_6)$. It is therefore useful to resolve the corresponding N=2 representations into their individual components. Accordingly, we write the characters of the LG theory as
\begin{align}
\chi^{\boldsymbol{l}}_{\boldsymbol{m}}{}^{(\boldsymbol{s})} &= \chi^{l_1}_{m_1} \chi^{l_3}_{m_3} \chi^{l_5}_{m_5}{}^{(s_5)}\chi^{l_6}_{m_6}{}^{(s_6)}~,&
\overline{\chi}^{\boldsymbol{l}}_{\boldsymbol{m}} &= \overline{\chi}^{l_1}_{m_1}\overline{\chi}^{l_3}_{m_3} \overline{\chi}^{l_5}_{m_5}\overline{\chi}^{l_6}_{m_6}~,
\end{align}
where $\boldsymbol{s}=(s_5,s_6)\in(\Z_4)^2$ labels the individual components of the supersymmetry multiplets. The partition function of $\cC_{\text{LG}}\times S^1$ factorizes as $Z_{\text{LG}}\times Z_\cir$ where, in the (NS,NS) sector,
\begin{align}
Z_{\text{LG}} &=\frac{1}{2^4}\sum_{\boldsymbol{l},\boldsymbol{m},\boldsymbol{\varepsilon}} 
\chi^{\boldsymbol{l}}_{\boldsymbol{m}}{}^{(2\boldsymbol{\varepsilon})} \overline{\chi}^{\boldsymbol{l}}_{\boldsymbol{m}}~,&
Z_\cir & = \sum_{m} \chi^{\Lu(1)}_m\overline{\chi}^{\Lu(1)}_m~.
\end{align}
Here $\boldsymbol{\varepsilon}\in(\Z_2)^2$ labels the components of the two supersymmetry multiplets discussed above. Summing over $\boldsymbol{\varepsilon}$ reassembles these components into supersymmetry multiplets. The overall factor in $Z_{\text{LG}}$ accounts for the field identifications.

We now turn to the orbifold theory $(\cC_{\text{LG}}\times S^1)/(\Z_6\times\Z_6)$. Its (NS,NS) partition function may be written as
\begin{align}
Z= \frac{1}{6^2}\sum_{a\in(\Z_6)^2}\sum_{b\in(\Z_6)^2} Z^{(a,b)}~,
\end{align}
where $Z^{(a,b)}$ denotes the contribution from the sector twisted by $g_1^{a_1}g_2^{a_2}$, with an insertion of the symmetry $g_1^{b_1}g_2^{b_2}$. The twisted Hilbert space is the tensor product of the corresponding Hilbert spaces of the LG model and the circle CFT, with $Z^{(a,b)}=Z_{\text{LG}}^{(a,b)}\times Z_\cir^{(a,b)}$. The spectrum of the unprojected twisted sectors can be determined explicitly: the corresponding twisted partition functions are
\begin{align}
Z_{\text{LG}}^{(a,0)} &=\frac{1}{2^4}\sum_{\boldsymbol{l},\boldsymbol{m},\boldsymbol{\varepsilon}} 
\chi^{\boldsymbol{l}}_{\boldsymbol{m}+\boldsymbol{\mu}_a}{}^{(2\boldsymbol{\varepsilon}+\boldsymbol{\nu}_a)} \overline{\chi}^{\boldsymbol{l}}_{\boldsymbol{m}}~,&
Z_\cir^{(a,0)} & = \sum_{m} \chi^{\Lu(1)}_{m+a_1-a_2}\overline{\chi}^{\Lu(1)}_{m}~,
\label{eq:LGOpartitionfunction}
\end{align}
where
\begin{align}
\boldsymbol{\mu}_a &=(2a_1,2a_2,a_1+a_2,a_1+a_2)~,&
\boldsymbol{\nu}_a &=(a_1+a_2,a_1+a_2)~.
\end{align}
The orbifold partition function is obtained by projecting each twisted sector onto $\Z_6\times\Z_6$-invariant states and summing over the twisted sectors. As shown in appendix~\ref{app:modularinvariance}, the orbifold projection removes all states of fractional spin other than half-integer spin, so that the resulting (NS,NS) spectrum satisfies
\begin{align}
h_\sleft-h_\sright\in\tfrac{1}{2}\Z~.
\end{align}
The modular properties can be determined explicitly. The full partition function obeys
\begin{align}
Z(\tau+2,\bar{\tau}+2) &= Z(\tau,\bar{\tau})~,&
Z(-1/\tau,-1/\bar{\tau}) &= Z(\tau,\bar{\tau})~.
\end{align}

\subsubsection*{Remark on the anomalous shift}
As emphasized above, the orbifold is well-defined only when acting on the full theory $\cC_{\text{LG}}\times S^1$. The restriction of the orbifold action to either factor separately defines an anomalous $\Z_6\times\Z_6$ symmetry. This anomaly is already visible in the modular properties of the corresponding twisted partition functions. Indeed, although the twisted Hilbert spaces $\cH^a$ factorize as $\cH^a=\cH^a_{\text{LG}}\otimes\cH^a_\cir$ and the corresponding twisted partition functions are individually well-defined, they transform under $\SL(2,\Z)$ with non-trivial phases:
\begin{align}
Z_{\text{LG}}^{(a,b)}(-1/\tau,-1/\bar{\tau}) &= \zeta^{(a_2-a_1)(b_1-b_2)}Z_{\text{LG}}^{(b,-a)}(\tau,\bar{\tau})~,
\nonumber\\
Z_\cir^{(a,b)}(-1/\tau,-1/\bar{\tau}) &= \zeta^{(a_1-a_2)(b_1-b_2)}Z_\cir^{(b,-a)}(\tau,\bar{\tau})~,
\end{align}
where $\zeta=\ee^{\frac{2\pi\ii}{6}}$. These phases are precisely the remnant of the GLSM gauge anomalies.

The anomalous transformation of the circle partition function has a simple interpretation, because the $\Z_6\times\Z_6$ action on the circle CFT is a shift orbifold. The consistency conditions for modular invariance of asymmetric orbifolds were understood long ago~\cite{Narain:1986qm}. In the present setup, only a $\Z_6$ subgroup of the $\Z_6\times\Z_6$ symmetry acts non-trivially on the circle CFT.~ In the notation of~\cite{Israel:2025ouq}, the corresponding shift orbifold is specified by the lattice vector $\boldsymbol{\sigma}=\boldsymbol{e}+r^2\boldsymbol{e}^\ast$, with $r^2=3$, for which $\boldsymbol{\sigma}\cdot\boldsymbol{\sigma}=6$. This fails the consistency condition for the orbifold: spin integrality in the $S^1/\Z_6$ theory requires $\boldsymbol{\sigma}\cdot\boldsymbol{\sigma}\in12\Z$. The anomalous modular transformation above is the manifestation of this inconsistency.

\subsection{Symmetry enhancement and fermion decoupling}
The partition function reveals a substantial enhancement of the left-moving symmetry of the IR CFT to
\begin{align}
\mathfrak{g}_\sleft =\su(2)_\sleft\oplus\su(2)_\sleft'\oplus\Lu(1)_\sleft''\oplus\so(4)~.
\label{eq:leftmovingKM}
\end{align}
Part of this enhancement can be anticipated from the GLSM analysis. For instance, the $\Lu(1)_\sleft''$ symmetry corresponds to one of the extra non-anomalous global symmetry that emerges at the locus $f_1=f_2=0$, identified in section~\ref{subsec:speciallocus}. The $\su(2)_\sleft$ factor also has a natural explanation: the theory admits the presentation $(\cC_{\text{LG}}/\Z_6\times S^1)/\Z_6$, where only the second $\Z_6$ acts chirally. In this description, $\su(2)_\sleft$ is identified with the left-moving R-symmetry of the K3 Gepner model $\cC_{\text{LG}}/\Z_6$, which is preserved by the chiral orbifold. By contrast, the additional $\so(4)$ symmetry is intrinsically a quantum symmetry that emerges only deep in the LGO phase.

To determine the enhanced chiral algebra~\eqref{eq:leftmovingKM}, we first note that there are $13$ holomorphic (NS,NS) states with $h_\sleft=1$, distributed among the $\Z_6\times\Z_6$ twisted sectors as follows.
\begin{align}
\begin{array}{c|cccccc}
&\,0\,&\,1\,&\,2\,&\,3\,&\,4\,&\,5\, \\ \hline
0&5&1&&&&1 \\
1&1&1&&&&1 \\
2&&&&&& \\
3&&&&&& \\
4&&&&&& \\
5&1&1&&&&1 \\
\end{array}
\end{align}
The five currents in the untwisted sector generate the manifest $\Lu(1)^{\oplus5}$ symmetry of the minimal models and the circle CFT. A useful basis for these currents is
\begin{align}
J_\sleft &= J_\sleft^{\min,1}+J_\sleft^{\min,3}+J_\sleft^{\min,5}+J_\sleft^{\min,6}~,
\nonumber\\
J_\sleft' &= J_\sleft^{\min,1}-J_\sleft^{\min,3}+2 J_\sleft^\cir~,
\nonumber\\
J_\sleft'' &= 3(J_\sleft^{\min,5}-J_\sleft^{\min,6})~,
\nonumber\\
K_\sleft &= 2(J_\sleft^{\min,1}-J_\sleft^{\min,5}-J_\sleft^{\min,6}-2J_\sleft^\cir)~,
\nonumber\\
K_\sleft' &=2(J_\sleft^{\min,3}-J_\sleft^{\min,5}-J_\sleft^{\min,6}+2J_\sleft^\cir)~.
\label{eq:enhancedsymmetrycurrents}
\end{align}
The first two are the $\Lu(1)_\sleft\oplus\Lu(1)'_\sleft$ currents associated with the GLSM global symmetries introduced above. The remaining currents obey the OPE
\begin{align}
J_\sleft'' (z_1)J_\sleft''(z_2)&\sim \frac{6}{z_{12}^2}~,&
K_\sleft(z_1)K_\sleft(z_2)&\sim \frac{8}{z_{12}^2}~,&
K_\sleft'(z_1)K_\sleft'(z_2)&\sim \frac{8}{z_{12}^2}~.
\label{eq:enhancedcurrentsOPE}
\end{align}
The orbifold projection ensures that the charges of all states are integral with respect to these five currents. The $h_\sleft=1$ holomorphic operators in the twisted sectors enhance the $\Lu(1)^{\oplus 4}$ symmetry generated by $J_\sleft$, $J_\sleft'$, $K_\sleft$ and $K_\sleft'$ to $\su(2)^{\oplus 4}$. Note that the $\su(2)$ current algebras associated to $K_\sleft$ and $K_\sleft'$ have level $4$, as follows from~\eqref{eq:enhancedcurrentsOPE}. The global symmetry of the LG theory is thus enhanced to~\eqref{eq:leftmovingKM}. The Sugawara central charge of the left-moving current algebra is equal to
\begin{align}
c_\sleft(\Lg_\sleft)=7~,
\end{align}
saturating the entire left-moving central charge of the CFT.

There is a further subtlety concerning the global form of the $\so(4)$ factor. The $\Z_2$ center generated by
\begin{align}
\exp(\pi\ii(K_{\sleft 0}-K_{\sleft0}'))
\end{align}
acts trivially on the CFT, so that only real representations of the $\so(4)=\su(2)\oplus\su(2)$ symmetry associated to $K_\sleft$ and $K'_\sleft$ occur. A related property is the presence, in the purely left-moving spectrum, of states with integer wight $h_\sleft=2$ transforming in the primary representation $(\rep{5},\rep{5})$. This representation is a $\Z_2$ simple current of the $\su(2)\oplus\su(2)$ current algebra at level $4$, and its appearance as a chiral operator signals the non-simply connected global form $\SO(4)=(\SU(2)\times\SU(2))/\Z_2$.\footnote{The relation between simple currents of Kac--Moody algebras and the corresponding center symmetries is well known~\cite{Fuchs:1990wb}. In the heterotic context, it implies that a non-trivial topology of the spacetime gauge group is reflected on the worldsheet by an extended left-moving current algebra and the presence of purely left-moving Kac--Moody primaries of integral spin in the heterotic CFT spectrum, as explained recently in~\cite{Lockhart:2026gvb}.}

\subsubsection*{Free fermions}
The presence of the enhanced $\SU(2)\times\SU(2)'$ symmetry is natural from the GLSM perspective. In the large radius phase, the left-moving fermions are sections of a trivial rank-two bundle. We therefore expect their coupling to the NLSM scalar fields---specified by a connection on this bundle---to flow in the IR to the flat connection on $\cO\oplus\cO$, so that the fermions decouple and their bilinears generate an $\su(2)_\sleft\oplus\su(2)_\sleft'$ current algebra at level one. While this decoupling is only conjectural in the large radius phase of the linear model, it can be verified directly at the Gepner point. Indeed, the (NS,NS) spectrum of the Gepner SCFT contains four purely left-moving states with weight $h_\sleft=\frac{1}{2}$, corresponding to two free left-moving Weyl fermions. They arise in the twisted sectors $(a_1,a_2)=(1,0),(0,1),(5,0),(0,5)$ and transform in the representation $(\rep{2},\rep{2},\rep{1})_0$ of the current algebra~\eqref{eq:leftmovingKM}. This strongly suggests a factorization
\begin{align}
(\cC_{\text{LG}}\times S^1)/(\Z_6\times\Z_6) = \cC_{\text{free}} \times \cC_{\text{int}}~,
\label{eq:CFTfactorization}
\end{align}
where $\cC_{\text{free}}$ is the theory of two free left-moving Weyl fermions, while $\cC_{\text{int}}$ is an interacting theory with $c_\sleft(\cC_{\text{int}})=5$ and $c_\sright(\cC_{\text{int}})=7$. In the large radius description, these central charges are accounted for by the five real scalars describing the geometry $X\to M$ and the two right-moving Weyl fermions valued in $T_M$. The full SCFT contains one additional free right-moving fermion, the superpartner of the compact scalar, giving $c_\sright=\frac{15}{2}$.

If the factorization~\eqref{eq:CFTfactorization} holds, the (NS,NS) partition function should split as
\begin{align}
Z = Z_{\text{free}} \times Z_{\text{int}}~.
\end{align}
The free fermion contribution can in turn be decomposed in terms of characters of the left-moving $\su(2)_\sleft\oplus\su(2)_\sleft'$ current algebra as
\begin{align}
Z_{\text{free}} =\chi^{\su(2)_\sleft}_{\rep{1}} \chi^{\su(2)'_\sleft}_{\rep{1}}+\chi^{\su(2)_\sleft}_{\rep{2}}\chi^{\su(2)'_\sleft}_{\rep{2}}~.
\end{align}
We tested this factorization by refining the orbifold partition function $Z$ with fugacities $z$ and $z'$ keeping track of the $\Lu(1)_\sleft\oplus\Lu(1)_\sleft'$ charges. The resulting partition function admits an expansion in the modular parameters $\text{q}=\ee^{2\pi\ii\tau}$ and $\bar{\text{q}}$, with coefficients that are polynomials in $\text{y}=\ee^{2\pi\ii z}$ and $\text{y}'=\ee^{2\pi\ii z'}$. We verified, through the first few hundred terms, that the quotient $Z/Z_{\text{free}}$ is independent of $z$ and $z'$ and has an expansion in $\text{q}$ and $\bar{\text{q}}$ with non-negative coefficients, as expected for a unitary CFT. We take this as strong evidence that the factorization holds to all orders.

Although the presence of free fermions was anticipated from our GLSM analysis, their emergence in the LGO phase is nonetheless remarkable. Indeed, viewing the LGO phase as a $\Z_6$ orbifold of a (4,4) K3 SCFT combined with a shift action on a circle, it is striking that this $\Z_6$ orbifold suffices to detach and decouple the two left-moving Weyl fermions of the K3 theory. Alternatively, the LG theory admits a free-field description in which each minimal model is represented by a compact boson and a collection of parafermions~\cite{Fateev:1985mm,Qiu:1986zf,Gepner:1987qi}. From this perspective, the free fermions we identify should emerge as composite operators built from the boson--parafermion system and twisted by the $\Z_6\times\Z_6$ orbifold action. Identifying these composite operators would provide an additional check that the LGO phase indeed describes a stringy limit of the large-radius flux compactification.

It remains to describe the Ramond sectors of the worldsheet theory. Starting from the (NS,NS) sector, the (R,NS), (NS,R), and (R,R) sectors can be obtained by chiral spectral flow. The left-moving spectral flow operator, constructed by bosonizing the $\Lu(1)_\sleft$ current, acts trivially on the interacting factor $\cC_{\text{int}}$, since all its states are neutral under $\Lu(1)_\sleft$. It therefore acts only on the left-moving fermions in $\cC_{\text{free}}$, giving the Ramond partition function
\begin{align}
Z^{\text{R}}_{\text{free}} = \chi^{\su(2)_\sleft}_{\rep{2}} \chi^{\su(2)'_\sleft}_{\rep{1}}+\chi^{\su(2)_\sleft}_{\rep{1}}\chi^{\su(2)'_\sleft}_{\rep{2}}~.
\end{align}
This decomposition also makes the left-moving fermion number manifest, since the corresponding operator is represented by $\ee^{\pi\ii J_{\sleft0}}$. In particular, the left-moving degrees of freedom of $\cC_{\text{int}}$ are purely bosonic, in agreement with the large radius description, where they arise from five compact scalars. The right-moving spectral flow operator is constructed from the right-moving $\Lu(1)_\sright\subset\su(2)_\sright$ R-symmetry of the $c_\sright=6$ Gepner model and acts non-trivially only on the degrees of freedom in $\cC_{\text{int}}$. As before, the resulting partition function does not include the right-moving fermion associated with the circle. The full SCFT is therefore obtained by combining the NS and R sectors of $\cC_{\text{int}}$ with a free right-moving Majorana--Weyl fermion, with the latter taken in the corresponding NS or R sector.

When using the LGO theory as a background for the heterotic string, we must supplement it with additional worldsheet degrees of freedom:  non-compact scalars describing the Minkowski $\R^{1,4}$ factor, their right-moving superpartners, the heterotic ghost system, and the remaining $\Le_8\oplus\Le_8$ degrees of freedom. We retain one $\Le_8$ factor as a ``hidden'' level-one current algebra. For the other, we use a fermionic realization in terms of eight left-moving Weyl fermions with common boundary conditions, two of which are included in the LGO sector. The fermion boundary condition is imposed by a chiral GSO projection, generated by the fermion number operator $\ee^{\pi\ii F_\sleft}$. The six free fermions generate an $\so(12)$ current algebra at level one through their bilinears. Their fermion number realizes a $\Z_2$ center symmetry of $\so(12)$: it acts trivially on the spinor representation $\rep{32}$ and with a minus sign on the vector and conjugate-spinor representations $\rep{12}$ and $\rep{32'}$. Here $\rep{32}\oplus\rep{32}'$ are the Ramond ground states. Combining the LGO theory with the six free fermions enlarges the $\so(12)\oplus\su(2)_\sleft\oplus\su(2)_\sleft'$ symmetry: the GSO projection retains additional currents transforming as $(\rep{12},\rep{2},\rep{2})$ in the (NS,NS) sector, and as $(\rep{32},\rep{1},\rep{2})\oplus(\rep{32'},\rep{2},\rep{1})$ in the (R,NS) sector. 
These currents precisely reconstruct the second $\Le_8$ current algebra. 

Thus, including the enhanced currents from $\cC_{\text{int}}$, the spacetime gauge symmetry at the Gepner point is
\begin{align}
G=\GE_8\times\GE_8\times\SO(4)\times\GU(1)~.
\end{align}
Moving away from the Gepner point will break part of this enhanced symmetry. Marginal deformations arising from $\cC_{\text{int}}$, however, preserve the $\GE_8\times\GE_8$ subgroup. In particular, this remains true upon moving to the large radius phase, since the deformation does not affect the left-moving free fermions. In the next section, we make these ideas more precise by analyzing the marginal (0,2) deformations of the SCFT.

\subsection{Marginal deformations}
\label{subsec:marginaldeformations}
At the Gepner point, we can obtain a detailed description of the marginal deformations of the low-energy SCFT and compare them with the expectations from the GLSM. As discussed above, the internal sector of the heterotic CFT can be viewed as consisting of two factors: an $\Le_8\oplus\Le_8$ current algebra at level one, with $c_\sleft=16$, and a ``geometric'' factor $\cC_{\text{geom}}$ associated with the (0,2) non-linear sigma model on $X\times\R$, with $c_\sleft=6$ and $c_\sright=9$. The latter further decomposes into the interacting theory $\cC_{\text{int}}$ and a free sector containing a non-compact real scalar and a right-moving Weyl fermion.

The (0,2) marginal supersymmetric deformations are generated by right-chiral primary operators with $h_\sleft=1$ and $\Lu(1)_\sright$ charge $q_\sright=1$. They fall into two classes. The first consists of deformations obtained by pairing a current of $\Lg_\sleft$ with a purely right-moving operator of weight $h_\sright=\frac{1}{2}$, namely the free right-moving fermion. The second consists of (0,2) marginal deformations of $\cC_{\text{geom}}$ itself. At the Gepner point, we can characterize the latter by analyzing the marginal operators of the $(\cC_{\text{LG}}\times S^1)/(\Z_6\times\Z_6)$ orbifold theory. There are $104$ states with $h_\sleft=1$ and $h_\sright=\frac{1}{2}$, distributed among the $\Z_6\times\Z_6$ twisted sectors as follows.
\begin{align}
\begin{array}{c|cccccc}
\multicolumn{6}{c}{q_\sright=+1} \\[4pt]
&\,0\,&\,1\,&\,2\,&\,3\,&\,4\,&\,5\, \\ \hline
0&4&&&&2&4 \\
1&&4&4&2&& \\
2&&4&4&2&& \\
3&&2&2&2&& \\
4&2&&&&2&2 \\
5&4&&&&2&4 \\
\end{array}
\hspace{40pt}
\begin{array}{c|cccccc}
\multicolumn{6}{c}{q_\sright=-1} \\[4pt]
&\,0\,&\,1\,&\,2\,&\,3\,&\,4\,&\,5\, \\ \hline
0&4&4&2&&& \\
1&4&4&2&&& \\
2&2&2&2&&& \\
3&&&&2&2&2 \\
4&&&&2&4&4 \\
5&&&&2&4&4 \\
\end{array}
\label{eq:Gepnermarginaldef}
\end{align}
All $104$ states are neutral under the $\Lu(1)_\sleft\oplus\Lu(1)'_\sleft$ symmetry and, after factoring out the $\Le_8\oplus\Le_8$ sector, remain entirely within $\cC_{\text{int}}$.

Alternatively, decoupling the free non-compact sector allows us to interpret the CFT spectrum in terms of the resulting five-dimensional supergravity. Each of the marginal deformations above gives rise to a massless scalar in spacetime. Those obtained by combining a left-moving current with the free right-moving fermion transform in the adjoint representation of the five-dimensional gauge group and reside in vector multiplets. The remaining marginal deformations give rise to $26$ hypermultiplets. All of these are charged under the gauge group: while they are neutral under the $\Le_8\oplus\Le_8$ factor, they transform non-trivially under the $\so(4)\oplus\Lu(1)$ symmetry. Refining the partition function with respect to the $\Lu(1)^{\oplus 3}$ Cartan subalgebra of the enhanced factor allows us to determine their representations. We find that the hypermultiplet marginal deformations transform as the representation
\begin{align}
\mathcal{H} = \rep{4}_{+3}\oplus\rep{4}_{-3}\oplus\rep{9}_{+2}\oplus\rep{9}_{-2}
\label{eq:marginaldefenhanced}
\end{align}
under $\so(4)\oplus\Lu(1)$, where $\rep{9}$ denotes the symmetric traceless representation of $\so(4)$. Thus, at the Gepner point, the five-dimensional theory contains no massless gauge-neutral scalars.

\subsubsection*{Decompactification limit}
The orbifold CFT contains the marginal operator $J^\cir_\sleft J^\cir_\sright$ controlling the radius $r$ of the circle. Varying this modulus, the theory remains a solvable $\Z_6\times\Z_6$ orbifold of the product of the Gepner model and a compact scalar. Away from the special point $r^2=3$, however, the orbifold exhibits several qualitative changes. For a generic radius, several of the twisted ground states acquire a non-zero right-moving conformal weight $h_\sright$. Consequently, some of the currents responsible for the enhanced gauge symmetry cease to be holomorphic: the $\su(2)_\sleft\oplus\su(2)'_\sleft$ symmetry is broken to $\su(2)_\sleft\oplus\Lu(1)'_\sleft$, as the two broken currents carry non-zero winding. In the heterotic CFT, this breaks the second $\Le_8$ gauge factor to $\Le_7\oplus\Lu(1)'_\sleft$. The corresponding five-dimensional supergravity has gauge group
\begin{align}
G=\GE_8\times\GE_7\times\GU(1)^4~.
\end{align}
One of the abelian factors originates from the shift symmetry of the circle, while the remaining three arise from the $\Lu(1)^{\oplus 4}$ symmetry of the Gepner model, one linear combination of which enhances to $\su(2)_\sleft\subset\Le_7$.

The theory admits a decompactification limit in which the radius $r$ is sent to infinity. In a sector twisted by $g_1^{a_1}g_2^{a_2}$, the conformal weights of the twisted states of the circle CFT include ground state contributions of the form
\begin{align}
h_\sleft &\supset (a_1-a_2)^2\frac{(r^2+3)^2}{4r^2}~,&
h_\sright &\supset (a_1-a_2)^2\frac{(r^2-3)^2}{4r^2}~.
\end{align}
Thus, in the limit $r\to\infty$, all twisted sectors with $a_1\neq a_2$ become infinitely massive. The only sectors that remain at finite conformal weight are those twisted by the diagonal $\Z_6$ subgroup generated by $g=g_1g_2$, which acts trivially on the circle. The decompactification limit is therefore described by a $\Z_6$ orbifold of the LG model by this diagonal symmetry, which is a K3 Gepner model with $c_\sleft=c_\sright=6$. The corresponding heterotic CFT describes a six-dimensional (1,0) supergravity theory with one tensor multiplet and $\GE_8\times\GE_7\times\GU(1)^3$ gauge symmetry, with hypermultiplets transforming in $\rep{56}^{\oplus 10}\oplus\rep{1}^{\oplus 68}$ of $\GE_7$, all neutral under $\GE_8$ and carrying various charges under the abelian factors. Moving away from the Gepner point Higgses the $\Lu(1)$ factors, leaving an LGO model describing a heterotic compactification on a K3 surface at string-scale volume.

\subsubsection*{Deformations of the LG superpotential}
A subset of the SCFT deformations can be identified in the UV with deformations of the linear model. Specifically, the LG orbifold can be deformed by turning on additional monomials in the (0,2) superpotential. Dropping the massive fields $(\Phi_2,\Gamma_2)$ and $(\Phi_4,\Gamma_4)$ from the analysis, the most general potential compatible with the $\GU(1)^1_{\sleft} \times\GU(1)^2_{\sleft}\times\GU(1)_{\sright}$ symmetry takes the form
\begin{align}
\cW & = \Gamma_1 J_1 + \Gamma_3 J_3 +\Gamma_5 J_5 + \Gamma_6 J_6~,
\end{align}
where $J_{1,3,4,5}$ are arbitrary combinations of the monomials
\begin{align}
J_1 &\supset \Phi_1^5~,&
J_3 &\supset \Phi_3^5~,&
J_5,J_6 &\supset \Phi_5^2~,  \Phi_6^2~, \Phi_5 \Phi_6~,\Phi_5\Phi_1 \Phi_3 ~,  \Phi_6 \Phi_1 \Phi_3 ~,\Phi_1^2 \Phi_3^2~.
\label{eq:LGOdeformations}
\end{align}
This gives $14$ candidate deformations, but some are redundant. At the Gepner locus, characterized by the superpotential~\eqref{eq:Gepnersuperpotential}, the superspace equations of motion imply that the fields belonging to the ideal
\begin{align}
\cI = \la \Phi_1^5,\Phi_3^5, \Phi_5^2, \Phi_6^2, \Gamma_1 \Phi_1^4, \Gamma_3 \Phi_3^4, \Gamma_5 \Phi_5, \Gamma_6 \Phi_6 \ra
\end{align}
are $\cDb$--trivial. Hence, deformations in~\eqref{eq:LGOdeformations} that lie in this ideal do not correspond to non-trivial marginal operators, and only four independent deformations remain. This agrees with the counting~\eqref{eq:Gepnermarginaldef} of untwisted marginal deformations of the Gepner orbifold, which contains precisely four (0,2) marginal operators in the untwisted sector.

The matching can be refined by keeping track of the global symmetries. Although the deformations are neutral under the $\GU(1)_\sleft\times\GU(1)'_\sleft$ symmetry of the LGO model, they carry charges under the extra currents present at the Gepner point. Recall that the latter form an enhanced $\so(4)_\sleft\oplus\Lu(1)_\sleft''$ symmetry, with Cartan subalgebra generated by the currents $K_\sleft$, $K_\sleft'$, and $J_\sleft''$ defined in~\eqref{eq:enhancedsymmetrycurrents}. Since these currents are constructed from the left-moving $\Lu(1)$ symmetries of the four minimal models, they can be identified in the LGO model with the corresponding accidental global symmetries that emerge at the Gepner locus. Denoting the charges under the Cartan currents $K_\sleft,K_\sleft'$ of $\SO(4)$ by $[r_\sleft,r_\sleft']$, and the $\Lu(1)_\sleft''$ charge by $q_\sleft''$, the four non-trivial deformations carry
\begin{align}
\arraycolsep=10pt
\begin{array}{ccccc}
\Delta \cW_{\text{LG}} \supset & \Gamma_5 (\Phi_1\Phi_3)^2~, & \Gamma_6 (\Phi_1\Phi_3)^2~, & \Gamma_5\Phi_1\Phi_3\Phi_6~, & \Gamma_6\Phi_1\Phi_3\Phi_5~. \\
{[r_\sleft,r_\sleft']}_{q_\sleft''}& [2,2]_{-2} & [2,2]_{+2} & [1,1]_{-3} & [1,1]_{+3} 
\end{array}
\label{eq:LGOmarginal}
\end{align}
These charges precisely agree with those of the four (0,2) marginal deformations in the untwisted sector of the Gepner orbifold. The first two deformations in~\eqref{eq:LGOmarginal} arise in the TLSM from the superpotential couplings $\Gamma_5 f_1$ and $\Gamma_6 f_2$. Indeed, in the LGO phase, only the monomial $(\Phi_1\Phi_3)^2$ in $f_1$ and $f_2$ gives rise to a relevant deformation; all other monomials contain the superfields $\Phi_2$ and $\Phi_4$ which become massive in the IR.

The above description of untwisted marginal operators is specific to the Gepner point. Repeating the analysis for a generic superpotential leads to a different ideal, and only two of the superpotential deformations remain non-trivial in the quotient by this ideal. This result is consistent with the counting of physical parameters in the superpotential, namely, the number $N_{\text{LGO}}$ of parameters that cannot be eliminated by field redefinitions. The infinitesimal redefinitions of the superfields compatible with the global and gauge symmetries are
\begin{align}
\delta\Phi_1 &\supset \Phi_1~,&
\delta\Phi_3 &\supset \Phi_3~,&
\delta\Phi_5,\delta\Phi_6 &\supset \Phi_1\Phi_3~, \Phi_5~, \Phi_6~, \nonumber\\
\delta\Gamma_1 &\supset \Gamma_1~,&
\delta\Gamma_3 &\supset \Gamma_3~,&
\delta\Gamma_5,\delta\Gamma_6 &\supset \Gamma_5~, \Gamma_6~.
\end{align}
Of the $14$ corresponding parameters, two redefinitions act trivially on the superpotential, as they generate the $\GU(1)_\sleft^1\times\GU(1)_\sleft'$ symmetry of the LG superpotential. Consequently,
\begin{align}
N_{\text{LGO}} = 14-14+2 = 2~.
\end{align}
Thus, the generic LGO superpotential admits only two non-trivial marginal deformations in the untwisted sector. At the (2,2) locus of the superpotential, however, the number of untwisted marginal deformations jumps to four.

At first sight, this might suggest that the LGO theory can be continuously deformed away from the Gepner point by turning on an untwisted marginal coupling, or equivalently, by deforming the (0,2) superpotential by $\Delta\cW_{\text{LGO}}$. However, this interpretation is misleading: the deformation turns out to be marginally irrelevant. A simple diagnostic for this is that all four untwisted marginal operators carry charges of the same sign under a common $\Lu(1)$ factor of the enhanced gauge symmetry---indeed, this holds for any generator in the Cartan subalgebra of $\so(4)$. This provides an obstruction to exact marginality: in such a situation, conformal perturbation theory develops a logarithmic divergence, so that the corresponding deformation is not truly marginal~\cite{Bertolini:2014ela}. Instead, we expect the untwisted deformation to induce an RG flow that drives the theory back to the Gepner fixed point. Conversely, this suggests that the LGO model with a generic superpotential, and in particular with non-zero $f_1$ and $f_2$, flows in the IR to the same Gepner SCFT. Consequently, the LG phase of the TLSM really consists of a single point.

This obstruction can be also recovered from a spacetime perspective. To make this connection more transparent, we compactify the theory on an additional circle. The heterotic flux background $X\times S^1$ then gives rise to a four-dimensional supergravity theory preserving eight supercharges. Using an $\mathcal{N}=1$ superspace description as in~\cite{Argyres:1996eh,Argyres:1996hc}, a hypermultiplet in a representation $\rep{r}$ of the gauge group $G$ is described by two chiral multiplets $Q$ and $\widetilde{Q}$ transforming in $\rep{r}$ and $\brep{r}$, respectively. Their expectation values are constrained by the $\mathcal{N}=1$ potential, which gives rise to F- and D-term equations. On the Higgs branch the scalar components of the vector multiplets are set to zero, and the flat directions are determined by the simultaneous vanishing of
\begin{align}
\cF(M) &=\sum_{\rep{r}} \widetilde{Q}_{\brep{r}} M_{\rep{r}}Q_{\rep{r}}~,&
\cD(M) &=\sum_{\rep{r}} Q^\dagger_\rep{r} M_{\rep{r}}Q_{\rep{r}} -\widetilde{Q}_{\brep{r}} M_{\rep{r}}\widetilde{Q}_{\brep{r}}^\dagger ~,
\end{align}
for every generator $M$ of the gauge algebra, where the sum runs over the representations $\rep{r}$ carried by the hypermultiplets. The four untwisted marginal deformations~\eqref{eq:LGOmarginal} respectively reside in the four chiral multiplets $Q_{\rep{9}}^-$, $Q_{\rep{9}}^+$ $Q_{\rep{4}}^-$, and $Q_{\rep{4}}^+$. If we give expectation values only to these four complex scalars, the F-term constraints are automatically satisfied, since all scalar components of the $\widetilde{Q}$ multiplets vanish. The D-term constraints, however, cannot be satisfied for a non-zero expectation value: choosing a generator $M$ in the Cartan subalgebra of $\so(4)$ gives a non-vanishing $\cD(M)$. Thus, the four untwisted deformations alone cannot parametrize a Higgs branch.

On the other hand, the potential does admit flat directions. For a generic expectation value of the chiral matter in the $26$-dimensional representation~\eqref{eq:marginaldefenhanced} the F-term and D-term constraints can be solved simultaneously. In the Gepner SCFT such a deformation necessarily involves marginal states from the $\Z_6\times\Z_6$ twisted sectors. The expectation values completely break the $\SO(4)\times\GU(1)''$ gauge symmetry. The resulting four-dimensional supergravity has gauge group $\GE_8\times\GE_8$ and $19=26-7$ neutral hypermultiplets. This precisely matches the spacetime massless spectrum of the large radius phase of the TLSM, in which the hypers parametrize the K3 moduli, providing further evidence that the two SCFTs are connected.

\section{Discussion} \label{s:discussion}
In this work we presented a TLSM whose large radius phase realizes a (0,2) sigma model on a heterotic flux vacuum $X$ with trivial holomorphic bundle and $\GE_8\times\GE_8$ gauge symmetry. By studying the phases of the linear model, we accessed a stringy regime in which the volume of $X$ is of order $\alpha'$, and the compactification is described by a chiral orbifold of the product of a K3 Gepner model and a compact boson. The twisted sectors contain free fermions whose currents lead to the reassembly of the $\Le_8\oplus\Le_8$ current algebra, and the stringy point also exhibits enhanced gauge symmetry, which is Higgsed upon moving back to the large radius phase. We find compelling evidence that the two phases describe the same SCFT by comparing their dressed elliptic genera and spacetime massless spectra.

This explicit realization of a TLSM lends credence to the existence of SCFTs associated with (0,2) sigma models with flux. This is far from obvious from the supergravity perspective, since the corresponding geometries do not admit a large-radius limit with vanishing torsion: the $\alpha'$ expansion is therefore only formal, with no parametric control over higher-derivative corrections~\cite{Melnikov:2014ywa}. Avoiding these difficulties was one of the main successes of the UV completion of~\cite{Adams:2006kb}, and it is comforting that we find, in an explicit construction, a well-defined SCFT exhibiting the expected features of the flux compactification. This exact (0,2) theory provides a torsional analogue of the more familiar (2,2) Gepner models associated with Calabi--Yau compactifications.

Our study also reveals the intricate stratification of the (0,2) parameter space, which we expect to persist in more general settings. This structure stems from a basic difference between (0,2) and (2,2) theories: the space of holomorphic couplings in a (0,2) GLSM has a richer structure, with no invariant distinction between FI parameters and superpotential couplings, since the two can be mixed by field redefinitions. Consequently, the usual ``phase diagram'' of a (0,2) GLSM, viewed as the parameter space of FI couplings, must be interpreted with care. In particular, accidental anomalous $\GU(1)$ symmetries can render an FI parameter redundant, effectively collapsing a direction of the phase diagram by inducing an equivalence relation on the FI parameter space.
\\

The present analysis focused on models in which the left-moving fermions couple to an abelian bundle given by a sum of line bundles, and in particular to holomorphically trivial factors. This restriction was motivated primarily by simplicity: more general bundles with non-abelian structure group are realized in TLSMs as monad bundles over K3, whose stability is considerably harder to establish. For the branched cover geometries considered here, however, stability is easier to analyze for bundles pulled back from the base of the covering space. Combined with the generalized Hoppe criterion of~\cite{Jardim:2011jm}, this approach was used in~\cite{Festi:2025} to construct stable monad bundles on K3. It would be interesting to adapt this construction to (0,2) linear models and thereby realize a broader class of stable bundles.

The Gepner SCFTs uncovered in the stringy regime of TLSMs deserve further study, as they exhibit several unusual features. In the examples considered here, the LG phase realizes the heterotic flux background as a chiral orbifold $(\cC_{\text{LG}}\times S^1)/\Gamma$ of a (2,2) Gepner model and a circle-valued (0,2) chiral multiplet. It would be worthwhile to  develop a more detailed description of marginal deformations in the LGO phase, as in~\cite{Kachru:1993pg}, and to identify them with appropriatly twisted monomials in the cohomology of the right-moving supercharge dressed by holomorphic vertex operators of the compact boson. The chiral orbifold action also has a distinctive feature: decoupled free fermions emerge in the twisted sectors. It would be interesting to understand the origin of these fermionic degrees of freedom and the underlying structure of such chiral orbifolds. A useful perspective may be to describe the orbifold by $\Gamma$ in terms of simple current invariants. Indeed, as discussed in appendix~\ref{subsec:simplecurrents}, the left-moving chiral algebra of our model admits an extension by two simple currents of conformal weight $h_\sleft=\frac{1}{2}$, which can be identified with the two decoupled Weyl fermions. Generalizing this construction could lead to a new class of exactly solvable (0,2) backgrounds for the heterotic string.

Another natural direction for future work is to study orbifolds of TLSMs describing heterotic vacua with four spacetime supercharges. The circle  bundle $X$ admits a freely acting orientation-reversing isometry, that acts as the Enriques involution on the K3 base of $X$ accompanied with a reflection of the fibered circle. Combined with reflection of an additional circle, this induces an orientation-preserving $\Z_2$ action on the six-dimensional non-K\"ahler manifold $Y=X\times S^1$~\cite{Becker:2008rc}. The smooth quotient $Y/\Z_2$ admits an $\SU(2)\rtimes\Z_2$ structure and provides a flux analogue of the Enriques Calabi--Yau~\cite{Ferrara:1995yx}. The involution can be lifted to the TLSM as a $\Z_2$ orbifold acting on the (0,2) superfields~\cite{Israel:2023itj}. For the linear model analyzed in this paper, the action on chiral and fermi multiplets is\footnote{The underlying K3 geometry, realized as a biquadratic cover of $\P^1\times\P^1$, admits an Enriques involution exchanging the two $\P^1$ factors of the base together with a sign flip of the branch coordinates, provided $f_1$ and $f_2$ are invariant under this exchange. Combined with an exchange of the two factors of the $\cO\oplus\cO$ bundle and a reflection of the circle fiber, this gives an orientation-reversing $\Z_2$ symmetry of the heterotic background associated to $X$.}
\begin{align}
(\Phi_1,\Gamma_1) &\leftrightarrow (\Phi_3,\Gamma_3)~,&
(\Phi_5,\Gamma_5) &\to (-\Phi_5,\Gamma_5)~,&
\Phi_7 &\leftrightarrow \Phi_8~,
\nonumber\\
(\Phi_2,\Gamma_2) &\leftrightarrow (\Phi_4,\Gamma_4)~,&
(\Phi_6,\Gamma_6) &\to (-\Phi_6,\Gamma_6)~.
\end{align}
The action also reverses the sign of the shift multiplet and exchanges the two vector multiplets. In the LG phase, the theory is a $(\Z_6\times\Z_6)\rtimes\Z_2$ orbifold of the product of an LG model and two free (0,2) chiral multiplets. The $\Z_2$ factor acts as a permutation orbifold of the type studied in~\cite{Maio:2010eu}. It would be interesting to study the resulting $c_\sright=9$ SCFT and its relation to the non-linear sigma model on $Y/\Z_2$, as well as to investigate more general quotients $Y/\Z_N$ and their realization in TLSMs with an LG phase.

Finally, our analysis was guided by the close relation between heterotic flux vacua and compactifications with line bundles. This connection can be understood from the nine-dimensional decompactification limit and leads to Narain dualities relating compactification on the flux background $X$ to a compactification on $M\times S^1$ with a line bundle~\cite{Israel:2023tjw}. These dualities act non-trivially on the heterotic moduli, mixing the radius of the fibered circle with Wilson line parameters. In the absence of Wilson lines, the $r\to1/r$ duality can be derived exactly in the TLSM using a Buscher-like procedure~\cite{Israel:2013hna}. It would be interesting to understand how Wilson line moduli can be incorporated into the TLSM and whether topology-changing Narain dualities can likewise be established directly in the linear model.

\appendix

\section{Conventions for the shift multiplet} \label{app:shiftmultiplet}
In this short appendix, we collect some details about the Lagrangian for the shift multiplet. The relevant superfields are the shift multiplet $\Omega$ and the (0,2) vector multiplets $\cV_{0\alpha}$ and $\cV_{1\alpha}$ associated with the abelian gauge group of the GLSM. The contribution of the shift multiplet to the Euclidean action is
\begin{align}
S_{\text{shift}}=\frac{1}{2\pi}\int\dd^2 z\left\{\cD\cDb K_z^{\text{shift}}+\frac{1}{\sqrt{2}}\cD\cW^{\text{shift}}+\frac{1}{\sqrt{2}}\cDb\bar{\cW}^{\text{shift}}\right\}_{\theta,\thetab=0}~,
\end{align}
where $\dd^2 z=\ii\dd z\wedge\dd\zb$. The kinetic term and superpotential are
\begin{align}
K_z^{\text{shift}} &=\tfrac{1}{4}r^2(\Omega+\Omegab+2w^\alpha\cV_{0\alpha})(\p_z(\Omega-\Omegab)+w^\alpha\cV_{1\alpha})~,\nonumber\\
\cW^{\text{shift}} &=-\tfrac{1}{2}r^2 w^\alpha\Omega\Upsilon_\alpha~.
\end{align}
The superpotential contains an axial coupling between the shift multiplet and the chiral superfield strength $\Upsilon=\sqrt{2}\bar{\cD}(\p_z\cV_0+\frac{1}{2}\cV_1)$. The action is invariant under super-gauge transformations parametrized by chiral superfields $\Lambda_\alpha$, with anti-chiral conjugates $\Lambdab_\alpha$, acting as
\begin{align}
\cV_{0\alpha} &\to \cV_{0\alpha}-\tfrac{1}{2}(\Lambda_\alpha+\Lambdab_\alpha)~,&
\cV_{1\alpha} &\to \cV_{1\alpha}-\p_z(\Lambda_\alpha-\Lambdab_\alpha)~,&
\Omega &\to \Omega +w^\alpha\Lambda_\alpha~.
\end{align}
Thus, $\exp(\Omega)$ is a standard chiral superfield  of charge $w^\alpha$. After fixing to Wess--Zumino gauge, the superfields have the component expansions
\begin{align}
\Omega &= \omega+\sqrt{2}\theta\psi+\theta\thetab\pb_{\zb}\omega~,&
\bar{\Omega} &= \bar{\omega}-\sqrt{2}\thetab\bar{\psi}-\theta\thetab\pb_{\zb}\bar{\omega}~,
\nonumber\\
\cV_0 &= \ii\theta\thetab\bar{A}_{\zb}~,&
\cV_1 &= 2\ii A_z+\sqrt{2}(\thetab\upsilon-\theta\bar{\upsilon})+\theta\thetab D~,
\nonumber\\
\Upsilon&=\upsilon-\tfrac{1}{\sqrt{2}}\theta(D+2\ii F_{z\bar{z}})+\theta\thetab\pb_{\zb}\upsilon~,&
\bar{\Upsilon} &=\bar{\upsilon}-\tfrac{1}{\sqrt{2}}\thetab(D-2\ii F_{z\bar{z}})-\theta\thetab\pb_{\zb}\bar{\upsilon}~,
\end{align}
where  $F_{z\bar{z}}=\p_z\bar{A}_{\zb}-\pb_{\zb} A_z$. Substituting these expansions into the superspace action and discarding total derivatives gives
\begin{align}
S_{\text{shift}}=\frac{1}{2\pi}\int\dd^2 z \Big\{& r^2(\p_z\Img\omega+w^\alpha A_{z\alpha})(\pb_{\zb}\Img\omega+w^\alpha\bar{A}_{\zb\alpha}) -r^2w^\alpha F_{z\zb\alpha}\operatorname{Im}\omega
\nonumber \\
&+r^2\p_z\Rea\omega\,\pb_{\zb}\Rea\omega +\frac{r^2}{2}(\bar{\psi}\p_z\psi+\psi\p_z\bar{\psi})\Big\}~.
\end{align}
Under large gauge transformations, parametrized by $\Lambda_\alpha=2\pi\ii m_\alpha$, the action shifts by
\begin{align}
S \to S-\ii r^2 w^\alpha w^\beta m_\beta\int F_\alpha~,
\end{align}
where $\int F = -\ii \int\dd^2 z \,F_{z\bar{z}}$. Consequently, the Euclidean path integral weight $\ee^{-S}$ acquires a phase, which is trivial provided $r^2 w^\alpha$ is integral.

\section{Rational CFTs and characters}
In this appendix, we collect the definitions and conventions for the rational CFTs that appear throughout the text. Characters are defined on a genus-one worldsheet with complex structure parameter $\tau$ in the complex upper half-plane, and we use the notation $\tq=\ee^{2\pi\ii\tau}$ for the modular parameter. The Dedekind eta function is
\begin{align}
\eta(\tau) = \tq^{\frac{1}{24}}\prod_{n=1}^\infty(1-\tq^{n})~,
\end{align}
while the Jacobi theta function is
\begin{align}
\vartheta_1(\tau|z) &=-\ii\tq^{\frac{1}{8}}(\ty^{\frac{1}{2}}-\ty^{-\frac{1}{2}})\prod_{n=1}^\infty(1-\tq^n)(1-\tq^n\ty)(1-\tq^n\ty^{-1})~,
\end{align}
where $\ty=\ee^{2\pi\ii z}$. It satisfies the quasi-periodicity condition property
\begin{align}
\label{eq:theta_shift}
\vartheta_1(\tau|z+m\tau+n) &= (-1)^{m+n} \ee^{-2\ii\pi m z} \tq^{-m^2/2} \vartheta_1(\tau|z)~,&
m,n &\in \Z~,
\end{align}
and has a simple pole at the origin, with residue
\begin{align}
\label{eq:theta_resid}
\frac{1}{2\ii\pi}\oint_{u = 0} \frac{\dd u}{\vartheta_1 (\tau|u)} =  \frac{1}{2\pi \eta(\tau)^3}~.
\end{align}
The level $k$ theta functions are defined by
\begin{equation}
\Theta_{m,k} (\tau|z) = \sum_{n \in \Z } \ee^{2\ii\pi \tau k (n + \frac{m}{2k})^2 + 2\ii\pi k z (n+\frac{m}{2k})}~.
\end{equation}
They satisfy the holomorphic heat equation
\begin{align}
\label{eq:heateq}
\left(\frac{1}{2\pi\ii}\frac{\p}{\p\tau}+\frac{1}{4\pi^2 k}\frac{\p^2}{\p z^2}\right)\Theta_{m,k}(\tau|z)=0~.
\end{align}

\subsection{Compact boson conventions}
We first summarize the relevant conventions for the Narain CFT of a compact boson on a circle of radius $r$. The theory has a $\Lu(1)_\sleft\oplus\Lu(1)_\sright$ current algebra, normalized as
\begin{align}
J_\sleft(z_1)J_\sleft(z_2) &\sim\frac{1/2r^2}{z_{12}^2}~,&
J_\sright(\zb_1)J_\sright(\zb_2) &\sim\frac{1/2r^2}{\zb_{12}^2}~.
\end{align}
For quantized radius $r^2=k$, the CFT admits two extended holomorphic operators of weight $h_\sleft=k$ and charges $q_\sleft=\pm1$. The representations of the extended algebra generated by these operators together with $J_\sleft$ are labeled by $m\in\Z_{2k}$, with characters
\begin{align}
\chi^{\Lu(1)}_m(\tau) = \frac{1}{\eta(\tau)} \Theta_{m,k} (\tau|0)~.
\end{align}
Under the modular $\tau\to-1/\tau$ transformation these characters obey
\begin{align}
\chi^{\Lu(1)}_m(-1/\tau) = \frac{1}{\sqrt{2k}} \sum_{m\in\Z_{2k}} \ee^{-\ii\pi\frac{mm'}{k}}\chi^{\Lu(1)}_{m'}(\tau)~.
\label{eq:Stransformationu1}
\end{align}
An analogous extended algebra exists in the right-moving sector, with characters $\overline{\chi}^{\Lu(1)}_m$. The circle CFT has the diagonal partition function
\begin{align}
Z_\cir = \sum_m \chi^{\Lu(1)}_m \overline{\chi}^{\Lu(1)}_m ~.
\end{align}

\subsection{Aspects of (2,2) minimal models}
\label{app:minimalmodels}
We collect here some standard facts about N=2 minimal models relevant to our study. We denote by $\cA_k$ the (2,2) A-type minimal model with central charges
\begin{align}
c_\sleft(\cA_k)=c_\sright(\cA_k)=\frac{3(k-2)}{k}~.
\end{align}
It arises as the IR fixed point of a LG model~\cite{Martinec:1988zu,Vafa:1988uu} with a single chiral multiplet $\Phi$, a fermi multiplet $\Gamma$, and (0,2) superpotential
\begin{align}
\cW = \Gamma\Phi^{k-1}~.
\end{align}
The theory flows in the IR to a unitary (2,2) SCFT. The left- and right-moving N=2 superconformal algebras contain $\Lu(1)_\sleft$ and $\Lu(1)_\sright$ current algebras. In the LG description, these arise from a global $\GU(1)_\sleft\times\GU(1)_\sright$ symmetry. The charges of the superfields are given by the following table, with the right-moving superspace coordinate $\theta$ itself charged under the $\GU(1)_\sright$ symmetry.
\begin{align*}
\arraycolsep=4pt
\begin{array}{cccc}
& \Phi & \Gamma & \theta \\
q_\sleft & \frac{1}{k} & \frac{1-k}{k} & 0 \\
q_\sright & \frac{1}{k} & \frac{1}{k} & 1
\end{array}
\end{align*}
The (0,2) superpotential is neutral under $\GU(1)_\sleft$ and carries unit charge under $\GU(1)_\sright$.

The minimal model $\cA_k$ can be described as the coset
\begin{align}
\frac{\mathfrak{su}(2)_{k-2}\oplus\mathfrak{u}(1)_{2}}{\mathfrak{u}(1)_{k}}~,
\end{align}
which makes the spectrum and modular properties particularly transparent. The left-moving primary fields of the coset are labeled by triples $(l,m,s)$ valued in
\begin{align}
(l,m,s)&\in\{0,1,\dots,k-2\}\times\Z_{2k}\times\Z_4~,&
l+m+s&=0\mod2~,
\end{align}
and subject to the field identification
\begin{align}
(l,m,s)\simeq(k-2-l,m+k,s+2)~.
\end{align}
The conformal weight $h_\sleft$ and left-moving R-charge $q_\sleft$ of a primary labeled by $(l,m,s)$ satisfy
\begin{align}
h_\sleft &=\frac{l(l+2)-m^2}{4k}+\frac{s^2}{8}\mod 1~,&
q_\sleft &=-\frac{m}{k}+\frac{s}{2}\mod 2~.
\end{align}
We denote the corresponding characters by
$\chi^l_m{}^{(s)}$. Their modular transformation follows from the coset construction and takes the form
\begin{align}
\chi^l_m{}^{(s)}(-1/\tau) = \frac{1}{2k} \sum_{l',m',s'} \sin\left(\frac{\pi (l+1)(l'+1)}{k}\right) \ee^{\ii\pi\frac{mm'}{k}-\ii\pi\frac{ss'}{2}}\chi^{l'}_{m'}{}^{(s')}(\tau)~.
\label{eq:Stransformationminmodel}
\end{align}
In this expression we have not imposed the field identification, so that each representation appears twice in the sum. We also set
$\chi^l_m{}^{(s)}=0$ whenever $l+m+s$ is odd.

The characters of the N=2 superconformal algebra are obtained by combining the coset primaries into supermultiplets. The NS sector corresponds to $s=0,2$, while the R sector corresponds to $s=1,3$. In the NS sector we write
\begin{align}
\chi^l_m = \chi^l_m{}^{(0)}+\chi^l_m{}^{(2)}~.
\end{align}
The corresponding supermultiplets are labeled by $(l,m)\simeq(k-2-l,m+k)$. Using the field identification, an NS multiplet may always be represented with
\begin{align}
|m|\leq l~.
\end{align}
For such a representative, the lowest component of the supermultiplet has conformal weight and R-charge
\begin{align}
h_\sleft &=\frac{l(l+2)-m^2}{4k}~,&
q_\sleft &=-\frac{m}{k}~.
\end{align}
The chiral and anti-chiral representations are characterized by $|m|=l$. The (NS,NS) partition function of the minimal model $\cA_k$ is equal to
\begin{align}
Z_{\cA_k}=\tfrac{1}{2}\sum_{l,m}\chi^l_m\overline{\chi}^l_m~,
\end{align}
where the overall factor accounts for the field identification. The (a,c) ring of $\cA_k$ is trivial, whereas the (c,c) ring contains $k-1$ states. Their left- and right-moving R-charges agree and take the values $q_\sleft=q_\sright\in\{0,\frac{1}{k},\dots,\frac{k-2}{k}\}$.

\section{Gepner partition function}
\subsection{Modular properties}
\label{app:modularinvariance}
The (NS,NS) states of the Gepner orbifold in the sector twisted by $a\in\Z_6\times\Z_6$ are counted by the partition function
\begin{align}
Z^{(a,0)} &=\frac{1}{2^4}\sum_{\boldsymbol{l},\boldsymbol{m},\boldsymbol{\varepsilon},m} 
\chi^{\boldsymbol{l}}_{\boldsymbol{m}+\boldsymbol{\mu}_a}{}^{(2\boldsymbol{\varepsilon}+\boldsymbol{\nu}_a)} \chi^{\Lu(1)}_{m+a_1-a_2}\overline{\chi}^{\boldsymbol{l}}_{\boldsymbol{m}}\overline{\chi}^{\Lu(1)}_{m}~.
\end{align}
The characters appearing in this expression carry definite charges, modulo integers, under the left- and right-moving $\Lu(1)^{\oplus 5}$ current algebra. These charges are
\begin{align}
q_\sleft^{\mm,1} &=-\tfrac{1}{6}(m_1+2a_1)\mod1~,&
q_\sright^{\mm,1} &=-\tfrac{1}{6}m_1\mod1~,\nonumber\\
q_\sleft^{\mm,3} &=-\tfrac{1}{6}(m_3+2a_2)\mod1~,&
q_\sright^{\mm,3} &=-\tfrac{1}{6}m_3\mod1~,\nonumber\\
q_\sleft^{\mm,5} &=-\tfrac{1}{3}m_5+\varepsilon_1+\tfrac{1}{6}(a_1+a_2)\mod 2~,&
q_\sright^{\mm,5} &=-\tfrac{1}{3}m_5\mod1~,\nonumber\\
q_\sleft^{\mm,6} &=-\tfrac{1}{3}m_6+\varepsilon_2+\tfrac{1}{6}(a_1+a_2)\mod 2~,&
q_\sright^{\mm,6} &=-\tfrac{1}{3}m_6\mod1~,\nonumber\\
q_\sleft^\cir &=+\tfrac{1}{6}(m+a_1-a_2)\mod1~,&
q_\sright^\cir &=+\tfrac{1}{6}m\mod1~,
\end{align}
and hence the corresponding states carry  $\mathfrak{u}(1)_\sleft^1\oplus\mathfrak{u}(1)_\sleft^2\oplus\mathfrak{u}(1)_\sright$ charges
\begin{align}
q_\sleft^1&=-\tfrac{1}{6}(m_1+m_5+m_6-m)+\tfrac{1}{2}(\varepsilon_1+\varepsilon_2)\mod1~,
\nonumber\\
q_\sleft^2&=-\tfrac{1}{6}(m_3+m_5+m_6+m)+\tfrac{1}{2}(\varepsilon_1+\varepsilon_2)\mod1~,
\nonumber\\
q_\sright&=-\tfrac{1}{6}(m_1+m_3)-\tfrac{1}{3}(m_5+m_6)\mod1~.
\label{eq:qLgepner}
\end{align}
The contributions to the left-moving conformal weight are
\begin{align}
h_\sleft^{\mm,1} &=\tfrac{1}{24}\left(l_1(l_1+2)-(m_1+2a_1)^2\right)\mod \tfrac{1}{2}~,
\nonumber\\
h_\sleft^{\mm,3} &=\tfrac{1}{24}\left(l_3(l_3+2)-(m_3+2a_2)^2\right)\mod \tfrac{1}{2}~,
\nonumber\\
h_\sleft^{\mm,5} &=\tfrac{1}{12}\left(l_5(l_5+2)-(m_5+a_1+a_2)^2\right)+\tfrac{1}{8}(2\varepsilon_1+a_1+a_2)^2\mod 1~,
\nonumber\\
h_\sleft^{\mm,6} &=\tfrac{1}{12}\left(l_6(l_6+2)-(m_6+a_1+a_2)^2\right)+\tfrac{1}{8}(2\varepsilon_2+a_1+a_2)^2\mod 1~,
\nonumber\\
h_\sleft^\cir &=\tfrac{1}{12}\left(m+a_1-a_2\right)^2\mod 1~,
\end{align}
while those to the right-moving weight are
\begin{align}
h_\sright^{\mm,1} &=\tfrac{1}{24}\left(l_1(l_1+2)-m_1^2\right)\mod \tfrac{1}{2}~,
\nonumber\\
h_\sright^{\mm,3} &=\tfrac{1}{24}\left(l_3(l_3+2)-m_3^2\right)\mod \tfrac{1}{2}~,
\nonumber\\
h_\sright^{\mm,5} &=\tfrac{1}{12}\left(l_5(l_5+2)-m_5^2\right)\mod \tfrac{1}{2}~,
\nonumber\\
h_\sright^{\mm,6} &=\tfrac{1}{12}\left(l_6(l_6+2)-m_6^2\right)\mod \tfrac{1}{2}~,
\nonumber\\
h_\sright^\cir &=\tfrac{1}{12}m^2\mod 1~.
\end{align}
Combining these contributions, one finds that the conformal weights
\begin{align}
h_\sleft &= h_\sleft^{\mm,1}+h_\sleft^{\mm,3}+h_\sleft^{\mm,5}+h_\sleft^{\mm,6}+h_\sleft^\cir~,
\nonumber\\
h_\sright &= h_\sright^{\mm,1}+h_\sright^{\mm,3}+h_\sright^{\mm,5}+h_\sright^{\mm,6}+h_\sright^\cir~,
\end{align}
satisfy the relation
\begin{align}
h_\sleft-h_\sright = a_1 q_\sleft^1 + a_2 q_\sleft^2 \mod\tfrac{1}{2}~.
\label{eq:orbifoldspin}
\end{align}
The partition function of the orbifold is obtained by summing over twisted sectors and projecting onto invariant states,
\begin{align}
Z=\frac{1}{6^2}\sum_{a,b\in\Z_6\times\Z_6}Z^{(a,b)}
\end{align}
where $Z^{(a,b)}$ denotes the contribution of the sector twisted by $g_1^{a_1}g_2^{a_2}$, with an insertion of the symmetry $g_1^{b_1}g_2^{b_2}$. Explicitly,
\begin{align}
Z^{(a,b)} &=\frac{1}{2^4}\sum_{\boldsymbol{l},\boldsymbol{m},\boldsymbol{\varepsilon},m} 
\chi^{\boldsymbol{l}}_{\boldsymbol{m}+\boldsymbol{\mu}_a}{}^{(2\boldsymbol{\varepsilon}+\boldsymbol{\nu}_a)} \chi^{\Lu(1)}_{m+a_1-a_2}\overline{\chi}^{\boldsymbol{l}}_{\boldsymbol{m}}\overline{\chi}^{\Lu(1)}_{m}\ee^{2\pi\ii (b_1q_\sleft^1+b_2 q_\sleft^2)}~,
\end{align}
where $q_{\sleft}^1$ and $q_{\sleft}^2$ are given as in~\eqref{eq:qLgepner}. It follows from~\eqref{eq:orbifoldspin} that, after the $\Z_6\times\Z_6$ projection, the operators that survive in the orbifold theory have half-integral spin: $h_\sleft-h_\sright\in\frac{1}{2}\Z$, or equivalently
\begin{align}
Z(\tau+2,\bar{\tau}+2)=Z(\tau,\bar{\tau})~.
\end{align}
The $\SL(2,\Z)$ transformation 
\begin{align}
Z^{(a,b)}(-1/\tau,-1/\bar{\tau}) &= Z^{(b,-a)}(\tau,\bar{\tau})
\end{align}
follows instead from the modular transformations~\eqref{eq:Stransformationu1} and~\eqref{eq:Stransformationminmodel} of the individual characters.

\subsection{Relation to simple currents}
\label{subsec:simplecurrents}
An ordinary (4,4) Gepner model for K3 is constructed from a product of minimal models
\begin{align}
\cC=\cA_{k_1} \times\dots\times\cA_{k_n}~,
\end{align}
with $c_\sleft=c_\sright=6$, orbifolded by a non-chiral $\Z_k$ symmetry, where $k$ is the least common multiple of the levels $k_i$. The GSO-projected heterotic partition function can then be described as a simple current invariant of the product of minimal models. Recall that, in a rational CFT, a simple current is a primary operator whose fusion with any primary produces a single primary. Simple currents form a finite abelian group under the fusion product, called the center of the CFT. For the product $\cC$ this group has a simple description. Restricting to the left-moving chiral algebra, the simple currents are in one-to-one correspondence with the primary representations with $l_1=\dots=l_n=0$. We denote the corresponding simple currents by
\begin{align}
\text{J} = \text{J}_{(m_1,\dots,m_n|s_1,\dots,s_n)}~.
\end{align}
The monodromy charge of a primary operator labeled by $(\boldsymbol{l}',\boldsymbol{m}',\boldsymbol{s}')$ with respect to $\text{J}$ is
\begin{align}
q^{\text{J}} = \sum_{i=1}^n\left(\frac{m_im'_i}{2k_i}-\frac{s_is'_i}{4}\right) \mod 1~.
\end{align}
When $\text{J}$ is a simple current of the left-moving chiral algebra with integral conformal weight $h_\sleft(\text{J})$, extending the chiral algebra by $\text{J}$ has a natural interpretation. The resulting CFT contains the holomorphic operator associated with $\text{J}$ in its spectrum, and only operators with trivial monodromy charge $q^{\text{J}}=0$ appear in the theory.

The (4,4) SCFT obtained as the Gepner orbifold $\cC/\Z_k$ has its left-moving chiral algebra extended by several simple currents. First, fermion alignment is implemented by a $(\Z_2)^{n-1}$ subgroup generated by
\begin{align}
\text{J}^{\text{align}}_{ij}&=\text{J}_{(0,\dots,0|0,\dots,0,2,0,\dots,0,2,0,\dots,0)}~,
\end{align}
where $i<j$ and $s_i=s_j=2$. These currents have $h_\sleft(\text{J}^{\text{align}}_{ij})=3$ and correspond to the products $G_{\mm,i}^{\pm}G_{\mm,j}^{\pm}$ of the minimal model supercurrents. Their presence projects onto states satisfying $s_i=s_j\mod 2$. In addition, the Gepner SCFT has a chiral spectral flow operator associated to the simple current
\begin{align}
\text{J}^{\text{sf}}&=\text{J}_{(1,\dots,1|1,\dots,1)}~,
\end{align}
with $h_\sleft(\text{J}^{\text{sf}})=\frac{1}{4}$ and $q_\sleft(\text{J}^{\text{sf}})=1$. Its square
\begin{align}
\text{J}&=(\text{J}^{\text{sf}})^2=\text{J}_{(2,\dots,2|2,\dots,2)}~,
\end{align}
has $h_\sleft(\text{J})=1$ and $q_\sleft(\text{J})=2$. The corresponding holomorphic (NS,NS) state is the image of the identity under chiral spectral flow and furnishes one of the two currents that enhance the left-moving $\Lu(1)_\sleft$ R-symmetry to $\su(2)_\sleft$; the other is associated to $\text{J}^{-1}$. Extending by $\text{J}$ projects onto states of integral R-charge, $q_\sleft\in\Z$. The right-moving sector has an analogous structure, with the corresponding anti-holomorphic operators.

In our case, the center of $\cC_{\text{LG}}\times S^1$ contains, in addition, the simple currents of the circle CFT, with each $m\in\Z_6$ defining a simple current. We thus label a left-moving simple current by
\begin{align}
\text{J} = \text{J}_{(m|m_1,m_3,m_5,m_6|s_1,s_3,s_5,s_6)}~.
\end{align}
The orbifold partition function exhibits an extension by the two simple currents
\begin{align}
\text{J}^1&=\text{J}_{(+1|2,0,1,1|2,0,1,1)}~,&
\text{J}^2&=\text{J}_{(-1|0,2,1,1|0,2,1,1)}~,
\end{align}
which appear in the (NS,NS) spectrum and have $h_\sleft(\text{J}^1)=h_\sleft(\text{J}^2)=\frac{1}{2}$. Their products
\begin{align}
\text{J}&=\text{J}^1\text{J}^2=\text{J}_{(0|2,2,2,2|2,2,2,2)}~,&
\text{J}'&=\text{J}^1(\text{J}^2)^{-1}=\text{J}_{(2|2,-2,0,0|2,2,0,0)}~,
\end{align}
generate a $\Z_6\times\Z_6$ subgroup of the CFT center. The currents $\text{J}$ and $\text{J}'$ are mutually local and have integral conformal weight $h_\sleft(\text{J})=h_\sleft(\text{J}')=1$. The extension projects onto states of integer charges $q_\sleft$ and $q_\sleft'$. In contrast, not all fermion alignment currents $\text{J}^{\text{align}}_{ij}$ appear in the spectrum. This misalignment, visible directly in the partition function~\eqref{eq:LGOpartitionfunction}, reflects the breaking of (2,2) supersymmetry to (0,2).

\bibliographystyle{./utphys}
\bibliography{./newref}

\end{document}